\documentclass[fleqn,usenatbib]{mnras}

\usepackage{newtxtext,newtxmath}

\usepackage[T1]{fontenc}

\DeclareRobustCommand{\VAN}[3]{#2}
\let\VANthebibliography\thebibliography
\def\thebibliography{\DeclareRobustCommand{\VAN}[3]{##3}\VANthebibliography}

\usepackage{graphicx}   
\usepackage{amsmath}    
\usepackage{multirow}
\usepackage[dvipsnames]{xcolor}
\usepackage{float}
\usepackage{subfloat}
\usepackage{colortbl}
\usepackage{subcaption}
\usepackage{orcidlink}
\usepackage{soul}

\definecolor{bad}{HTML}{D50000}
\definecolor{DS}{HTML}{5055EB}
\definecolor{SS}{HTML}{117733}
\definecolor{CBS}{HTML}{1A8CFF}
\definecolor{ADD}{HTML}{D50000}

\def\kms{\ifmmode{~{\rm km~s^{-1}}}\else{~km s$^{-1}$}\fi}
\def\cc{\ifmmode{~{\rm cm^{-3}}}\else{~cm$^{-3}$}\fi}
\def\fesc{\ifmmode{{f_{esc}}}\else{$f_{\rm esc}$}\fi}
\def\fstar{\ifmmode{{f_\star}}\else{$f_\star$}\fi}
\def\lsim{\lower0.3em\hbox{$\,\buildrel <\over\sim\,$}}
\def\gsim{\lower0.3em\hbox{$\,\buildrel >\over\sim\,$}}

\def\cubecm{\ifmmode{~{\rm cm^{-3}}}\else{cm$^{-3}$}\fi}
\def\Ms{\ifmmode{~{\rm M_\odot}}\else{M$_\odot$}\fi}
\def\Zs{\ifmmode{~{\rm Z_\odot}}\else{Z$_\odot$}\fi}
\def\h2{H$_2$}

\def\nat{Nature}
\def\apj{ApJ}

\def\apjl{ApJL}

\def\mnras{MNRAS}
\def\araa{ARA\&A}
\def\aap{A\&A}

\def\pasa{PASA}

\newcommand{\hii}{H {\sc ii}}

\title[High-z galaxy properties from JWST]{Inferred galaxy properties during Cosmic Dawn from early JWST photometry results}

\author[Brummel-Smith \& Skinner et al.]{Corey Brummel-Smith$^{\orcidlink{0000-0001-6204-5181}}$,$^{1}$\begin{NoHyper}\thanks{E-mail: cdbs3@gatech.edu (CBS)}\thanks{These authors contributed equally to this work.}\end{NoHyper}
    Danielle Skinner$^{\orcidlink{0000-0002-5346-1308}}$,$^{1}$\footnotemark[2]
    Snigdaa S. Sethuram$^{\orcidlink{0000-0001-7908-3934}}$,$^{1}$
    John H. Wise$^{\orcidlink{0000-0003-1173-8847}}$,$^{1}$\newauthor
    Bin Xia$^{\orcidlink{0000-0002-7077-9836}}$,$^{1}$
    Khushi Taori$^{\orcidlink{0000-0003-0460-5333}}$$^{1}$
\\
$^{1}$Center for Relativistic Astrophysics, School of Physics, Georgia Institute of Technology, Atlanta, GA 30332, USA\\
}

\pubyear{2023}

\begin{document}
\label{firstpage}
\pagerange{\pageref{firstpage}--\pageref{lastpage}}
\maketitle

\begin{abstract}
Early photometric results from JWST have revealed a number of galaxy candidates above redshift 10. The initial estimates of inferred stellar masses and the associated cosmic star formation rates are above most theoretical model predictions up to a factor of 20 in the most extreme cases, while this has been moderated after the recalibration of NIRCam and subsequent spectroscopic detections. Using these recent JWST observations, we use galaxy scaling relations from cosmological simulations to model the star formation history to very high redshifts, back to a starting halo mass of $10^{7} \ \mathrm{M}_{\odot}$, to infer the intrinsic properties of the JWST galaxies.  Here we explore the contribution of supermassive black holes, stellar binaries, and an excess of massive stars to the overall luminosity of high-redshift galaxies. Despite the addition of alternative components to the spectral energy distribution, we find stellar masses equal to or slightly higher than previous stellar mass estimates. Most galaxy spectra are dominated by the stellar component, and the exact choice for the stellar population model does not appear to make a major difference.  We find that four of the 12 high-redshift galaxy candidates are best fit with a non-negligible active galactic nuclei component, but the evidence from the continuum alone is insufficient to confirm their existence.  Upcoming spectroscopic observations of $z>10$ galaxies will confirm the presence and nature of high-energy sources in the early universe and will constrain their exact redshifts.
\end{abstract}




\section{Introduction}

JWST has provided us with data from a time never seen before. We are now seeing galaxies as they first form when the universe was only a few hundred million years old. The conclusions that we draw as a community will continue to inform our models for many years to come, and thus, the early JWST results act as our initial conditions for this new era of astronomy. The first observations of these high-redshift galaxies are coming from two major surveys: the Cosmic Evolution Early Release Survey \citep[CEERS;][]{CEERS} and the Grism Lens-Amplified Survey from Space \citep[GLASS;][]{GLASS}. CEERS is a photometric and spectroscopic survey covering 100 square arcminutes of sky utilizing the NIRSpec, NIRCam and MIRI instruments. One of CEERS's specific goals is to discover galaxies between $9 < z < 13$ and observe their spectra to constrain details on early galaxy formation. One of the first and more striking results coming from the CEERS survey is the detection of a galaxy at $z = 11.8$ \citep{Finkelstein22}. The GLASS survey is aimed at answering questions related to reionization and baryon usage and recycling within faintly magnified galaxies by utilizing the NIRISS, NIRSpec, and the NIRCam instruments. Some of the first results in this survey showed lensed low-metallicity and star forming galaxies at $z \geq 7$ \citep{Roberts-Borsani22} and higher redshift galaxies at $z \sim 9 - 15$ \citep{Castellano22}, demonstrating the power of high redshift observations that can be done by JWST.

These first galaxies began to form approximately 100 Myr after the Big Bang as small dark matter halos (minihalos) merged to form larger halos, hierarchically building a home for future galaxies. While the term ``first galaxy'' is not well-defined, it is commonly agreed that the first galaxies were ones that were able to form stars and withstand their feedback \citep{Bromm11}, in order to continue to merge and grow. At early times, the first stars to form were metal-free (Population III; Pop III) stars that are thought to have been very massive from the lack of efficient cooling \citep{Yoshida03}. Simulations have shown that Pop III stars primarily formed in minihalos with masses $\sim$$10^{5-6} \mathrm{M}_{\odot}$ at $z > 10$ \citep{Yoshida03, Bromm11, Schauer19, Skinner20}. If we consider these dark matter minihalos to be the hosts of the first galaxies, their initial formation would have occurred around $z \sim 30$ and continued to grow and merge with one another until they reach the atomic cooling limit at $\mathrm{M}_{\rm halo} \simeq 10^8 \Ms$ \citep[e.g.][]{Wise07}. These halos are then able to continuously form stars in a less bursty fashion, and if massive enough, can withstand the feedback. These galaxies are ones that we may be more familiar with in the present day. This process of going from small dark matter halos, to the first generation of bursty star formation, to atomically cooling halos drove the early stages of reionization \citep{Wise14_Galaxy}. Observing these first galaxies with JWST is a massive step forward in understanding the importance of these sources in relation to reionization.

Still, at such early times, far from the views of any telescopes that have looked to the cosmos so far, simulations remain the only tool informing our understanding of galaxy formation in the very early universe \citep[see][for reviews]{Somerville15, Wechsler18}. The Renaissance simulations \citep{Xu13, Xu14, Chen14, OShea15} are a suite of hydrodynamical adaptive mesh refinement zoom-in simulations focusing on different regions of the universe; namely overdense, normal, and low density regions. These simulations were initially used to study Pop III star formation, but have since been used to study reionization, first galaxy formation, and black hole (BH) growth.  In particular, \citet{Chen14} used the Renaissance simulations to provide various scaling relations for galaxies forming at $z \geq 15$. They found that halos below the atomic cooling threshold show bursty star formation as opposed to halos above the atomic cooling threshold, where the star formation becomes more efficient.

The SPHINX simulations are another suite of cosmological hydrodynamical adaptive mesh refinement simulations studying the effects of various astrophysical processes on reionization. In their first paper, \citet{Rosdahl18} studied the effects of binary systems on reionization. They found that binary stars are necessary to include if the simulation volume is to be reionized by $z \sim 6$. Including binary stars leads to higher escape fractions as compared to the single star counterparts. The Obelisk simulation followed up on the SPHINX simulations, finding that stellar radiation predominately drove reionization \citep{Trebitsch21}.  More recently, \citet{Katz22} used the SPHINX simulations to study the C and O abundances in galaxies during reionization and found that the stellar population within galaxies at $z > 6$ may need to be under the influence of a top-heavy initial mass function (IMF) in order to reproduce the observed abundances. At these high redshifts, this is somewhat expected since the first generation of stars are expected to come from a more top-heavy IMF. FLARES (First Light and Reionization Epoch Simulations) is also a cosmological suite following galaxy formation in the early universe based on the EAGLE simulations, a smoothed particle hydrodynamics (SPH) simulation suite \citep{Schaye15}. \citet{Vijayan21} used FLARES to predict photometric data for high redshift galaxies, in preparation for JWST. They presented a UV luminosity function (LF) that matched well with the observed data at that time. More recently, \citet{Wilkins22} used FLARES to study the stellar histories of galaxies in the early universe, within the redshift range that JWST aims to cover. They found a stellar mass -- metallicity relationship at very high redshifts, which can be probed and verified by results from JWST, and that the environment does not play a huge role in affecting the stellar masses, metallicities, star formation rates, or ages. As more data comes from JWST, the results from such simulations will be further constrained and will help us understand galaxy formation at these early times. 

As the first galaxies begin to grow into the large galaxies we see today, so do the BHs that reside in their hosts and over time, will migrate towards their centers. Beginning in the 1960s, quasars were first discovered as radio bright sources in the sky \citep{Schmidt63}, only later to be discovered and labelled as quasars, or supermassive BHs (SMBHs) growing rapidly from their surroundings. Since 2001, nearly 100 quasars have been discovered with masses of $\gsim 10^{9} \Ms$ at $z > 6$ \citep[see][for a collection of these sources]{Inayoshi22}. The existence of these SMBHs at such early times requires an explanation for how they grew so quickly in such a short period of time. There are many possible origins of these BHs, including stellar mass remnants from Pop III stars, intermediate mass BHs formed from stellar collisions, or direct collapse BHs formed from collapsing gas clouds \citep{Valiante17}. From here, models of active galactic nuclei (AGN) have helped answer some questions regarding how efficient accretion can be at growing these BHs to be very massive. Assuming that the galaxies that are observed by JWST are some of the brightest and thus most massive galaxies, they are likely to have a SMBH at their centers. 

\section{The first six months of JWST results} \label{sec:recent_history}

As soon as data from JWST became publicly available in early July 2022, preprints started to be posted to the arXiv. Bright, high redshift galaxies were being reported \citep[e.g.][]{Finkelstein22, Castellano22, Labbe22} and along with it, a growing concern that these galaxies were not abiding by the laws of $\Lambda$CDM cosmology. But with new telescopes comes the natural growing pains of calibration and data re-analysis. Calibrating the instruments aboard JWST is a critical step to ensure that we are able to interpret the data as accurately as possible \citep[e.g.][]{Rigby22, Bagley22}. At the time of writing, calibration is still ongoing which will require the reprocessing of already published data. An example of the effects of recalibration and reanalysis comes from \citet{Adams22}. Using post-launch calibrations, they search for galaxies at $z > 9$ and compare with results from other studies. They find that compared to the studies using the early July 2022 calibration, a large percentage of the reportedly high-redshift galaxies are actually at much lower redshifts (see their Table 4 and references therein), and there is not a lot of overlap of sources between studies of the same fields. The reasons for these issues have not been fully explored, but they do find that some sources are affected by the post-flight calibration, in that a lower redshift is obtained when using the new calibrations. \citet{Boyer22} found that there is a large flux offset in the NIRCam filters resulting in magnitude differences of $\sim 0.01 - 0.2$ mag, and reported new zeropoints for the NIRCam filters. Calibration will take place throughout Cycle 1 of JWST, and will continue to be updated, requiring reanalysis of already published data. While calibration is an important issue, improved reanalysis of these objects as time goes on may lead to different results. For example, \citet{Finkelstein22} initially reported finding a highly star forming galaxy at $z \sim 14$ with a high stellar mass. After improving their astrometry methods, certain filters were better aligned with each other and with the aperture, resulting in an overall brighter flux. This resulted in the galaxy's redshift being decreased to $z \sim 11.8$. While this is still a very high redshift galaxy, this is an example of how initial measurements of these early galaxies will likely change with the improved calibration and analysis of JWST data.

Alongside the rush of reports on these high redshift galaxies, theoretical explorations on the abundance and limits of these galaxies had begun to be reported as well. \citet{Boylan-Kolchin22} presented a straightforward calculation of the comoving number density of halos and the maximum amount of stellar mass contained within a halo as a function of redshift assuming maximum star formation efficiency (SFE). They found that some early measurements of JWST galaxies with high stellar masses are found in a volume much smaller than expected, and it appears that there are two galaxies whose stellar mass density lies above the comoving stellar mass density assuming maximum SFE, i.e. their stellar mass is larger than the amount of baryons available to their host halos. The analysis presented in \citet{Boylan-Kolchin22} provided some of the first examples of the possible tension with $\Lambda$CDM cosmology. Relieving this tension, \citet{Mason22} investigated the upper limit of the UV LF by modeling the star formation rate (SFR) and including a simple stellar spectrum as a function of time. They found that when the SFE is at a maximum, the upper limit to the UV LF is about four orders of magnitude higher than what is currently detected, showing that these detections are still in the realm of $\Lambda$CDM cosmology. They also found that when the SFE is significantly decreased to a more realistic value and is calculated as a function of the halo mass, and if dust attenuation is negligible at high redshifts, the UV LF more closely matches currently observed values. \citet{Mason22} go on to show that younger star forming galaxies that have rapidly evolved are the main galaxies that are detectable due to their increased UV magnitude, implying that we are selectively only seeing the brightest galaxies, and they may not represent the entire population of galaxies at that time. Another possible explanation comes from \citet{Ferrara22}, who constructed a minimal physical model of the UV LF as a function of time. They modelled the SFR using the classical Schmidt-type expression and altered the SFE by including feedback from supernova (SN). They then fit their UV LF to the ALMA REBELS survey to calibrate for dust attenuation, and find that some high redshift bright detections at $z > 11$ are best described with a UV LF that does not include dust attenuation, supporting results from \cite{Mason22}. \citet{Inayoshi22} also determined the upper bound of the LF at three different redshift ranges and compare to the LF with varying SFEs. They found that current detections fit well with SFEs that closely match SFEs in starburst galaxies. They also inspected the product of the SFE and the UV radiative efficiency, providing another constraint. Meeting this constraint could mean a combination of periods of very efficient metal-poor and relatively efficient metal-free star formation. They concluded that weakly accreting BHs and quasars could also contribute to the UV continuum. Results from \citet{Inayoshi22} agree with results from \citet{Mason22} in that these high redshift massive galaxies do not violate $\Lambda$CDM cosmology, and they also provide us with constraints on their stellar population.

Groups have also been reanalyzing the JWST data to provide more insight and tools for future JWST observations. \citet{Furtak22} detailed their spectral energy distribution (SED) fitting process for a number of weakly-lensed galaxies, and provided their galactic best fit parameters. They found that the galaxies at $z \geq 10$ are young and highly star forming, which aligns with results from \citet{Mason22}. They calculated the mass-to-light ratios of each galaxy and found no redshift dependent evolution of this relation. Importantly, the galaxies they analyze are all weakly magnified by gravitational lensing, and thus they are only looking at the brightest galaxies in the UV, echoing concerns about selection bias from \citet{Mason22}. Higher magnification is needed to observe galaxies on the fainter end of the UV spectrum. \citet{Furtak22} concluded that spectra for these galaxies are needed in order to confirm their redshift, as they find that for some galaxies that do not have strong Balmer breaks, a low-redshift SED can also describe the data. Finally, they say that their data remains within the upper limits $\Lambda$CDM cosmology. 

Some results are showing the importance of considering other sources contributing to the SED. \citet{Brinchmann22} analyzed JWST NIRSpec data to compare high redshift sources with local analogues. They found that a few of the high redshift sources they analyze are likely affected by an AGN due to emission lines associated with AGN activity at lower redshifts. This introduces the important idea that the UV flux  may be coming from more than just a stellar population, and additional sources should be included. \citet{Steinhardt22} generated new photometric templates in order to more accurately fit the JWST photometry at high redshifts, rather than using templates based on IMFs calibrated to the local universe. This is important due to a major environmental difference at early times, namely, the increased temperature of the CMB, resulting in a bottom-light IMF \citep{Britton09}. They find that a change in the IMF produces a 0.2 -- 0.3 dex increase in stellar mass and that a change in the gas temperature produces a much stronger effect. Since their work is based on gas temperature being heated by the CMB, this is a lower bound on the gas temperature, and in reality, the gas temperature could be much higher, resulting in an even bottom-lighter IMF. They state that this could mean that fitting high redshifts galaxies with the local IMF could be overestimating the stellar mass by $\geq 1$ dex. \citet{Steinhardt22} assert that the ``early galaxy problem'' is solved by using fits that account for the CMB heating gas at higher redshifts.  \citet{Volonteri22} found that the unobscured AGN emission in these JWST high-z galaxy candidates is generally fainter than the stellar component and will be difficult to detect in color-color selections; however, overmassive SMBHs may be detectable that could point toward it being seeded by a massive BH.

Given the theory discussions so far, the ``early galaxy problem'' may not be as big a problem as initially thought. A top-heavy IMF may help explain the higher UV magnitudes and may be a more appropriate choice to model high redshift galaxies, as explained by \citet{Steinhardt22}. Including an AGN and more luminous sources may be another missing piece to the puzzle, since there have been some emission lines detected that are indicative of an AGN in some galaxies so far \citep{Brinchmann22}. Adjusting our models for each of these pieces may provide explanations to the high stellar masses found in the high redshift galaxies. To tackle this problem, we investigate how an AGN, binary stars, and top-heavy IMFs may affect the UV magnitude of a set of JWST galaxies by modeling the halo mass and the star formation history. From the star formation history, we determine the stellar spectrum given an IMF choice and add to that to an AGN spectrum. We vary a set of free parameters in order to find the set that best fit the observed rest-frame UV and optical magnitudes. We describe our model in section \ref{sec:methods}, detailing the star formation history in section \ref{sec:SFH}, the stellar spectrum in section \ref{sec:SS} and the AGN spectrum in section \ref{sec:SMBH}. Details about dust attenuation, computing the magnitudes, and tests are described in Sections \ref{sec:dust}, \ref{sec:mag}, and \ref{sec:testing} respectively. We vary our model using a Monte Carlo Markov Chain (MCMC), described in section \ref{sec:emcee}. Our results are described and discussed in sections \ref{sec:results} and \ref{sec:discussion} respectively. We conclude with section \ref{sec:conclusions}.

\section{Methods} \label{sec:methods}

\begin{figure*}
    \includegraphics[width=\textwidth]{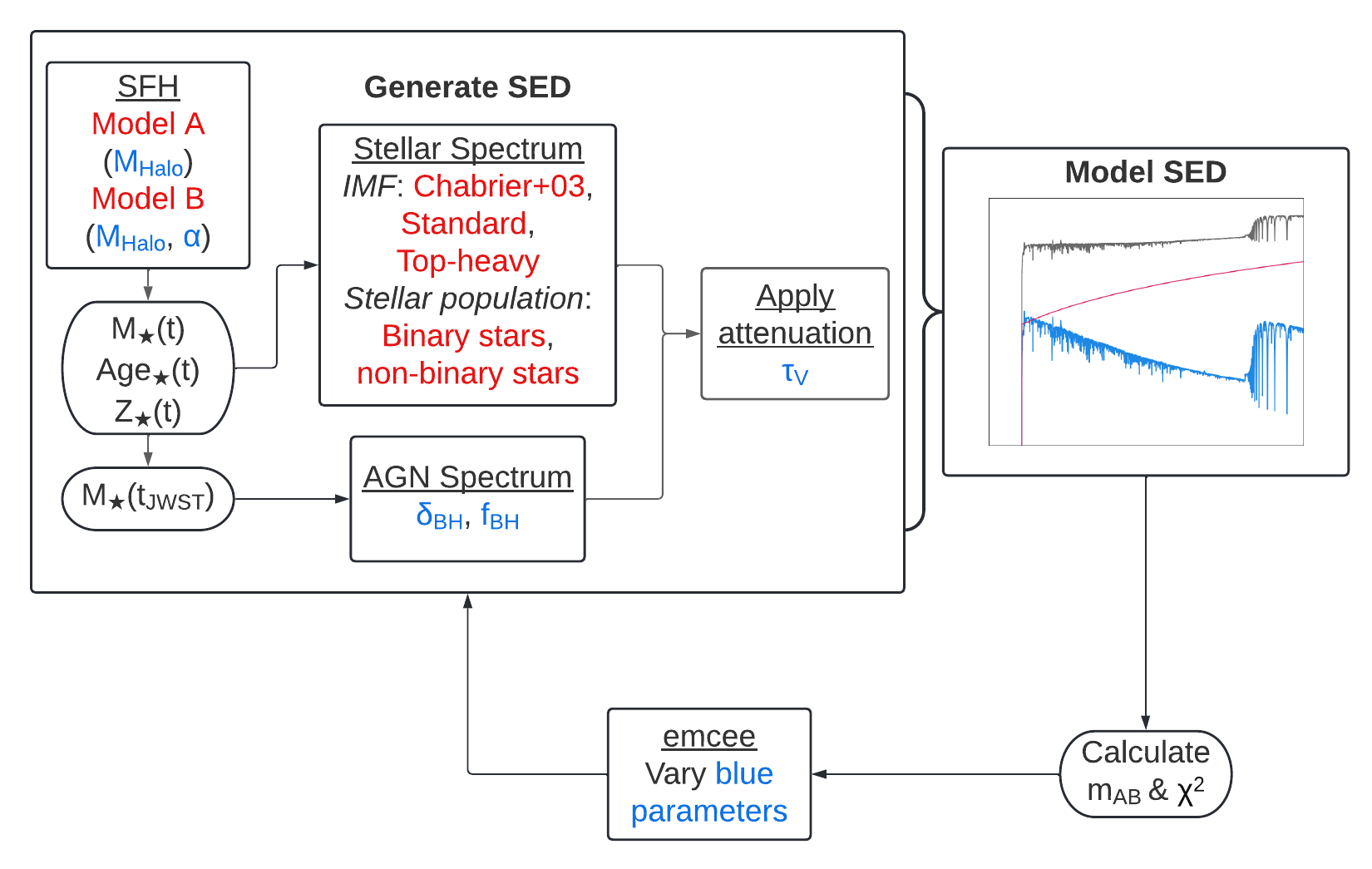}
    \centering
    \caption{Flowchart of the analysis pipeline. The red parameters indicate individual model choices, while the blue parameters indicate parameters that are continuously varied by {\sc emcee}. To generate a model SED, model A or B is chosen to model the star formation history. The outputs are then fed to the stellar and AGN spectrum models. The stellar spectrum is controlled by choosing one IMF and one stellar population. The AGN spectrum is controlled by two parameters varied by {\sc emcee}. Once the model SED (the gray line shows the total SED, the pink line the AGN component, and the blue line the stellar component) is generated and attenuation is applied, the AB magnitudes and $\chi^{2}$ are calculated. {\sc emcee} then runs the analysis again for a set of red parameters and varies the blue parameters.}
    \label{fig:analysis_pipeline}
\end{figure*}

To explore how the JWST photometry is affected by different properties of a galaxy, we create a model made of three discrete parts, varying the star formation history, the stellar population, and the central SMBH. We then use {\sc emcee} \citep{emcee}, a MCMC ensemble sampler, to vary certain free parameters in order to determine the most probable set of characteristics that would explain the observed photometry. We run our analysis pipeline on each of the JWST galaxies listed in Table \ref{table:jwst_galaxies}, coming from \citet[][hereafter A22]{Adams22} and \citet[][hereafter H22]{Harikane22}, where we only consider the candidates in the latter that are also selected by other works. The complete analysis pipeline can be seen in Figure \ref{fig:analysis_pipeline} and the model choices and free parameters are listed in Tables \ref{table:fixed_params} and \ref{table:free_params} respectively. Throughout our calculations, we use the cosmological parameters from \citet{Planck18_Cosmo}: $\Omega_{M} = 0.3153$, $\Omega_{\Lambda} = 0.6847$, $\Omega_{b} = 0.04923$, $\sigma_{8} = 0.8111$, and $h = 0.6736$, where the parameters are their usual definitions. 

\begin{table}
	\caption{Sample JWST galaxies} \label{table:jwst_galaxies}
	\renewcommand{\arraystretch}{1.5}
	\resizebox{\columnwidth}{!}{
		\begin{tabular}{|c|c|c|c|} 
			\hline
			ID & $z$ & log$_{10}(M_{\star}/\Ms)$ & Other detections (alternative ID) \\
			\hline\hline
			\multicolumn{4}{|c|}{\citet{Adams22} (A22)}\\
			\hline
			1514 & 9.85 & $9.8$ & -- \\
			1696 & 9.59 & $8.7$ & At23 (SMACS-z10c) \\ 
			2462 & 9.50 & $9.5$ & At23 (SMACS-z10b) \\ 
			2779 & 9.51 & $8.7$ &-- \\
			6115 & 10.94 & $8.4$ & -- \\ 
			6878 & 9.59 & $9.3$ & At23 (SMACS-z10a) \\ 
			10234 & 11.42 & $7.8$ & -- \\
			\hline\hline
			\multicolumn{4}{|c|}{\citet{Harikane22} (H22)}\\
			\hline
			GL-z9-1 & 10.49 & $9.04^{+0.61}_{-0.74}$ & C22, C22B (GHZ1), D23 (17487), N22 (GLz11) \\ 
			GL-z9-2 & 10.46 & $8.64^{+0.16}_{-1.04}$ & C22, C22B (GHZ4) \\ 
			GL-z12-1 & 12.28 & $8.36^{+0.90}_{-0.28}$ & C22 (GHZ2), D23 (1698), N22 (GLz13) \\ 
			CR2-z12-1 & 11.63 & $7.88^{+0.95}_{-0.26}$ & F22 (Maisie's Galaxy) \\ 
			CR2-z16-1 & 16.25 & $9.20^{+1.06}_{-0.72}$ & D23 (93316) \\
			\hline
		\end{tabular}
		}
		\centering
        \parbox[t]{\columnwidth}{\vspace{1em} \textit{Notes:} Numeric and alphanumeric IDs are taken from A22 and H22, respectively. Photometric redshifts and stellar masses are shown from these two works with the exception of GL-z9-2 where the stellar mass is from C22. Sources: At23 \citep{Atek22}, C22 \citep{CastellanoDec22}, C22B \citep{Castellano22}, D23 \citep{Donnan22}, F22 \citep{Finkelstein22}, N22 \citep{Naidu22}.}
\end{table}

\begin{table}
	\caption{Fixed model choices} \label{table:fixed_params} 
\resizebox{\columnwidth}{!}{%
	\begin{tabular}{|c|c|c|}
		\hline
		 & Name & Explanation \\
		\hline
		\multirow{2}{*}{Growth Model} 	& A & Press-Schechter \\ 
								& B & Exponential Growth \\
		\hline
		\multirow{6}{*}{IMF} 	& \multirow{2}{*}{Chabrier} 		& $\alpha_{1} = \mathrm{exp\ cutoff} ;\ \alpha_{2} = -2.3$ \\
								& 								& $\mathrm{M}_1 = 1.0\ \Ms;\ M_{\mathrm{max}} = 100\ \Ms$\\  \cline{2-3}				 
								
								& \multirow{2}{*}{Standard} 		& $\alpha_{1} = -1.30 ;\ \alpha_{2} = -2.35$ \\
								& 								& $\mathrm{M}_1 = 0.5\ \Ms;\ M_{\mathrm{max}} = 100\ \Ms$\\  \cline{2-3} 
								
								& \multirow{2}{*}{Top-heavy} 	& $\sigma = 1\ \Ms ;\ \ M_{\rm c} = 10\ \Ms$ \\
								& 								& $\mathrm{M}_{\mathrm{min}} = 1\ \Ms;\ M_{\mathrm{max}} = 500\ \Ms$\\
		\hline
		\multirow{2}{*}{Stellar Population} & Binary & binary \& non-binary population spectra \\ 
											& Non-binary & only single stellar population spectra \\
		\hline
	\end{tabular}
	}
\parbox[t]{\columnwidth}{\vspace{1em} \textit{Notes:} A single model consists of a choice of growth model, IMF and stellar population. Note that the Chabrier and Standard IMF models come from BPASS \citep{bpassPaper} and the Top-Heavy IMF model comes from Yggdrasil \citep{yggdrasil}. {\sc emcee} is run separately for each fixed model.}
\end{table}

\begin{table}
	\caption{Free parameter choices} \label{table:free_params}
	\resizebox{\columnwidth}{!}{%
		\begin{tabular}{|c|c|c|} 
			\hline
			& \textbf{Definition} & \textbf{Range} \\ 
			\hline
			M$_{\rm Halo}$ & Mass of the halo in $\Ms$ & 9 < $\mathrm{log}_{10}(\mathrm{M} / \Ms)$ < 12 \\
			\hline
			$\alpha$ & Controls halo growth rate \citep{Wechsler02} & 0.4 < $\alpha$ < 0.8 \\
			\hline
			$\delta_{\rm BH}$ & Alters AGN slope \citep{Yang22} & $-1 \leq \delta_{\rm BH} \leq 1$ \\ 
			$f_{\rm BH}$ & Controls BH mass in terms of stellar mass   & $-4 \leq \textrm{log}_{10}(f_{\rm BH}) \leq 0$ \\
			\hline
			$\tau_{\textrm V}$ & Optical depth in the visual band & 0 < $\tau_{\textrm V}$ < 2 \\ 
			\hline
		\end{tabular}
	}
	\parbox[t]{\columnwidth}{\vspace{1em} \textit{Notes:} Parameter choices. These values are varied by {\sc emcee} to map out parameter space. The $\alpha$ parameter is only required for Model B.}
\end{table}

\subsection{Star formation history} \label{sec:SFH}

\begin{figure}
    \includegraphics[width=\columnwidth]{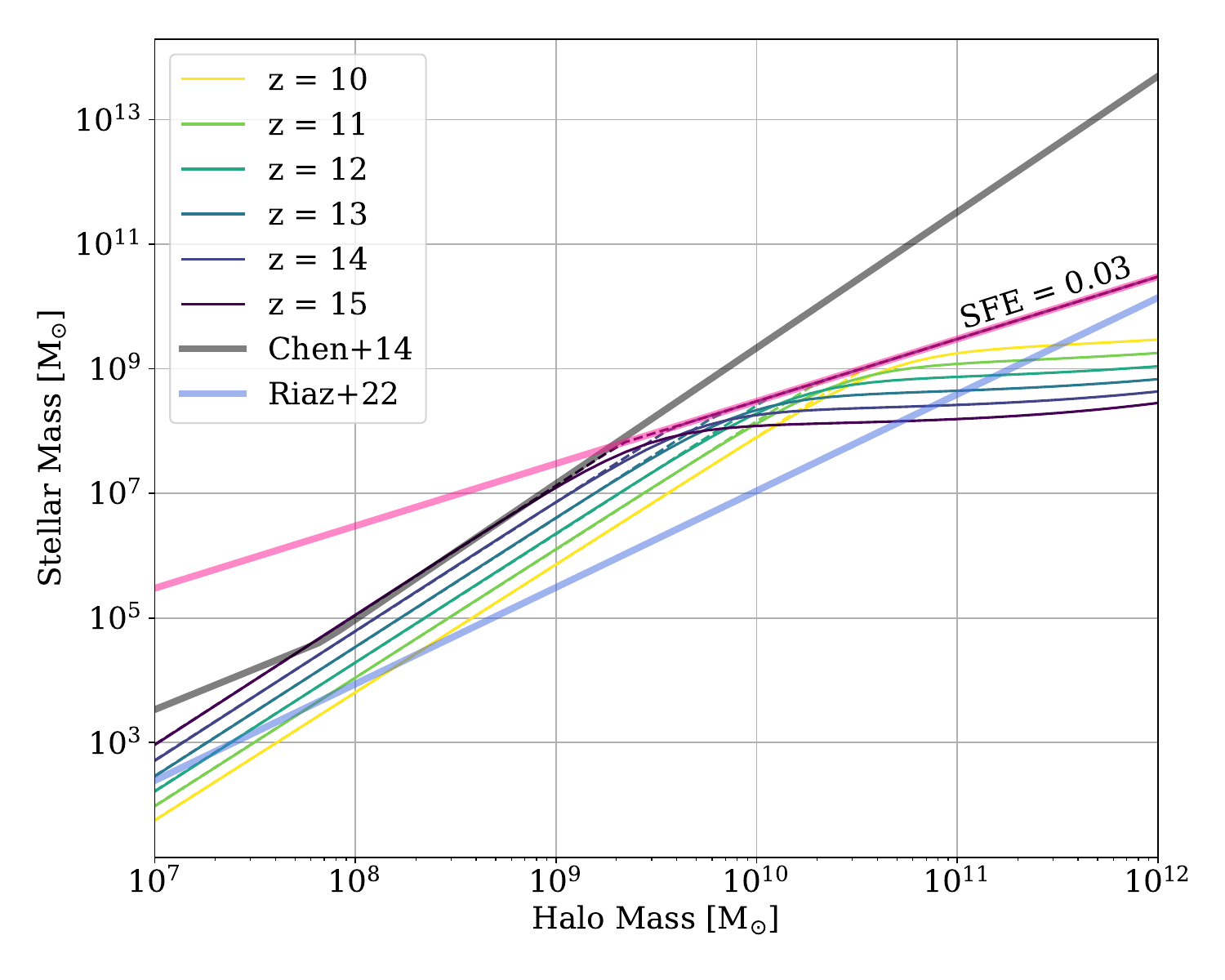}
    \centering
    \caption{The SMHM relation implemented in our model compared to different sources. The solid lines labelled by a redshift come from the fit from \citet{Behroozi13}. The similarly colored dashed lines are our linear fits to the low mass power laws at the noted redshifts. They converge to a stellar mass of $0.03 \mathrm{M}_{\rm Halo}$ since we limit the stellar mass to a maximum SFE, indicated by the opaque pink line. The opaque black line shows the fit from \citet{Chen14}. Notice how this relation matches well with the \citet{Behroozi13} fit at $z = 15$. The opaque blue line shows the relation from \citet{Riaz22}.}
    \label{fig:SMHM}
\end{figure}

In order to produce a model SED that best represents the observed photometry for a given JWST galaxy, the history of the galaxy must be determined. This includes the halo mass, stellar mass, stellar ages, and stellar metallicity throughout time. These components are inputs to the spectrum models in Section \ref{sec:SS} and Section \ref{sec:SMBH}. As described in Section \ref{sec:emcee}, the halo mass at the observed redshift will be varied to determine the most probable set of parameters to fit the observed luminosity. The stellar mass is computed from the fits in \citet[their Equation 3]{Behroozi13}, given the host halo mass. While this equation is fit only for halos between $0 < z < 8$, \citet{Chen14} found that these relations fit well with their results for $z \leq 15$, thus we use the $z=15$ fit for redshifts above that value. This fit comparison can be seen in Figure \ref{fig:SMHM}. In order to better model halos at higher masses, we extrapolate the low mass power law to higher masses. We also limit the stellar mass to a maximum SFE of 0.03 \citep{Pillepich18}, as shown in Figure \ref{fig:SMHM}. This relation is also consistent with \citet{Riaz22}, who determined the stellar mass - halo mass (SMHM) relation from a semi-analytic model of halo masses at high redshifts, including the contribution of Pop III stars.

The stellar-mass-halo-mass relation is not well studied for massive halos at such high redshifts. \cite{Behroozi13} only modelled this relation for halos from redshift 0 to 8. At redshift 8, the largest halo mass that they modelled was $10^{11} \mathrm{M_\odot}$, which is reasonable given that a halo of that mass is very rare at $z = 8$. Although comparison with \cite{Chen14} shows that extending the SMHM relation to redshift 15 is consistent and matches simulations very well at lower masses, the maximum halo mass in \cite{Chen14} was $10^9 \mathrm{M_\odot}$. Equation 3 in \cite{Behroozi13}, is not meant to be used for halo masses larger than $10^{11} \mathrm{M_\odot}$ at redshift 8. This means the plateau seen in Figure \ref{fig:SMHM} is unrealistic. The plateau is an artifact of applying their equation beyond the intended range of halo mass and redshift. The turnover describes the transition to elliptical galaxies which occurs at lower redshifts. Therefore, we cannot simply use the \cite{Behroozi13} equation as-is for large halo masses. The SMHM relation is often modelled as a double power law \citep{Yang12, Moster13} so in the absence of a better model, we argue that the most sensible choice is to transition to a high-mass power law with a physically motivated maximum slope, determined by the maximum star formation efficiency. Our SFE limiting causes the low-mass mass \cite{Behroozi13} relation to transition to a shallower slope which describes the high-mass regime.

We explore two models to generate a halo growth history based on the halo mass of the JWST galaxy calculated previously at the observed redshift. The first model comes from the ellipsoidal variant of Press-Schechter formalism \citep{Sheth01}. Given the halo mass and observed redshift, we directly calculate the rarity of the halo ($\nu$) and the halo mass history beginning at the observed redshift. The second model is an exponential growth history from \citet{Wechsler02}: $M(z) = M_{0}e^{-\alpha z}$, where $M_{0}$ is the present-day halo mass and $\alpha$ is a free parameter controlling the rate of halo growth. The higher the $\alpha$ value, the more rapid the halo growth. For both models, the halo growth history is calculated back to when M$_{\mathrm{Halo}}$ = 10$^{\mathrm{7}} \Ms$ in steps of 5 Myr. Given the halo mass growth history, we now directly calculate the stellar mass history as given by \citet{Behroozi13} and shown in Figure \ref{fig:SMHM}. This allows us to calculate the SFR, stellar ages, and metallicities as a function of time. The SFR is calculated simply by taking the difference in the stellar mass between two time steps: SFR = $(M_{\star, i} - M_{\star, i-1}) / (t_{i} - t_{i-1})$. The stellar age is simply the lookback time from the photometric redshift estimate. The stellar metallicities are calculated from \citet{Torrey19} (T19), where they describe the mass-metallicity relation from the IllustrisTNG simulation suite. They present the $\textrm{Log(O/H)} + 12$ galactic metallicity value for a series of stellar mass bins -- $\log_{10} M_{\star}/\Ms = [8,\ 8.5,\ 9, \ 9.5, \ 10, \ 10.5]$ -- from $0 < z < 10$ (see their Figure 7). We take the y-intercept points and interpolate between these lines for our given stellar mass and redshift. We then convert this into a total metallicity (Z$_{\mathrm{Tot}}$) in units of Z$_{\odot}$. To calculate the stellar metallicity ($\mathrm{Z}_{\star}$) of a single stellar population (SSP) at each timestep, we multiply the change in total metallicity by the mass fraction of new stellar mass created: $\mathrm{Z}_{\star} = (Z_{\mathrm{Tot, i}} - Z_{\mathrm{Tot, i-1}}) \times (\mathrm{M}_{\star, i} / (\mathrm{M}_{\star, i} - \mathrm{M}_{\star, i-1}))$. Below $10^{8} \Ms$ and above $10^{10.5} \Ms$, we use the minimum and maximum of the T19 relationship at $10^{8} \Ms$ and $10^{10.5} \Ms$, respectively. While this relation is fit only to a particular mass range and up to $z \sim 10$, it generally matches well with simulations. The T19 metallicity differs slightly from metallicity relations from other simulations for the same stellar mass, and redshift. Here we present comparisons with three other simulations: \cite{Wilkins22} (W22), \cite{Abe21} (Abe21), and \cite{Jeon15} (J15). We note that the MZR we use from T19 is a galactic gas-phase MZR, not specifically a stellar MZR. However we use this as an approximation for the stellar MZR. J15 and Abe21 also use a gas-phase metallicity, however W22 use a stellar metallicity from a smoothing kernel applied to the star particles in their simulation. In each of these comparisons, we measure the difference in UV magnitude of a single stellar population at a given redshift, age, and stellar mass, only differing in the metallicity.

Compared to W22, with $\mathrm{M}_{\star} = 10^9 \Ms$, $z = 10$, Age = $45$ Myr, $\mathrm{Z_{Wilkins}}$ = 0.48 $\mathrm{Z_\odot}$, $\mathrm{Z_{Torrey}}$ = 0.16 $\mathrm{Z_\odot}$, using the T19 metallicity produces brighter UV flux by 0.15 mag.

In the following comparisons, the simulated stellar mass quoted did not form from a single stellar population at one age, however, since the authors do not present a full mass-metallicity relation as a function of redshift, the best we can do is compare the spectra of SSPs at various ages.

We compared to the fiducial simulation from Abe21, with $\mathrm{M}_{\star} = 7.5 \times 10^4 \Ms/h$, $z = 9$, $\mathrm{Z_{Abe}}$ = $7.4 \times 10^{-3} \mathrm{Z_\odot}$, $\mathrm{Z_{Torrey}}$ = 0.12 $\mathrm{Z_\odot}$, for SSPs with ages ranging from 10--200 Myr. Using the T19 metallicity resulted in a dimmer UV magnitude than when using the metallicity from Abe21. The range of the difference in magnitudes was 0.14--0.35 mag. The difference monotonically increased with increasing stellar age.

We performed the same comparison using parameters from J15. Namely, $\mathrm{M}_{\star} = 3 \times 10^4 \Ms$, $z = 10.5$, $\mathrm{Z_{Jeon}} = 10^{-4} \mathrm{Z_\odot}$, $\mathrm{Z_{Torrey}}$ = 0.095 $\mathrm{Z_\odot}$, for SSPs with ages ranging from 10--200 Myr. Using the much larger T19 metallicity resulted in a much brighter UV magnitude than when using the metallicity from J15. The range of the difference in magnitudes was 0.60--1.93 mag. Again, the difference monotonically increased with increasing stellar age.

This shows that although the T19 MZR shows different metallicities than other simulations, the change in the UV magnitude compared to W22 and Abe21 is small. However, the difference in metallicity between T19 and J15 is large and results a large difference in flux, especially for old stellar populations. However, given that T19 has the most robust MZR for high-redshift with stellar masses spanning many orders of magnitude, it is an ideal model for generating spectra for our SED fitting pipeline.

From each growth history model, the stellar masses, ages and metallicities are passed to the next two components of our pipeline in order to generate a stellar and AGN spectrum. 

\subsection{Stellar spectrum} \label{sec:SS}

Given the stellar mass, age, and metallicity histories for each galaxy from the star formation history model, we calculate a stellar spectrum at the observed redshift by summing the stellar spectra of SSPs forming at 5 Myr intervals. We obtain the stellar spectra of each SSP from Binary Population and Spectral Synthesis (BPASS) \citep{bpassPaper} and Yggdrasil \citep{yggdrasil}. 

As the BPASS name suggests, it is used for simulating stellar populations as well as hosting a library of spectra generated from simulations using a set of IMFs and stellar populations at various stellar ages and metallicities. BPASS specializes in synthesizing binary stellar populations, creating spectra with similar binary system distributions compared to the local universe. The Yggdrasil spectra instead use single stellar populations and focus on Pop III and low-metallicity stars for a better comparison with the observations of the first galaxies. The Yggdrasil SED database models Pop III (characteristic masses of 100~\Ms), Pop III.2 (characteristic masses of 10~\Ms), and metal-poor Pop II stars.

In our parameter exploration, we use the \citet{chab03} IMF and a ``standard'' IMF from BPASS and a ``Pop III.2'' (top-heavy) IMF from Yggdrasil. The SMHM relation used to generate our star formation history is based on the Chabrier IMF, but we explore the effect of different IMFs on the stellar spectrum here. We include the top-heavy IMF in our analysis since the first galaxies formed while the CMB temperature was higher and the interstellar medium was predominantly metal-poor, cooling less efficiently than solar-metallicity gas and possibly reducing fragmentation into low-mass stars, leading to a larger number of more massive stars than seen in the local Universe.

The IMF functions modelled in BPASS follow a broken power law:
\begin{align}
    N(M < M_{\mathrm{max}}) \propto \int_{0.1}^{M_1} \left( \frac{M}{\Ms}\right)^{\alpha_ 1} \, dM \\ 
    + \ M_{1}^{\alpha_1} \int_{M_{1}}^{M_{\mathrm{max}}} \left( \frac{M}{\Ms} \right)^{\alpha_2} \, dM. \nonumber
\end{align}
Here each IMF modelled in BPASS varies the $\alpha$ parameters as well as $\mathrm{M}_1$ \& $\mathrm{M}_{\mathrm{max}}$ that are listed in Table \ref{table:fixed_params}. The Chabrier IMF in BPASS is modelled with an exponential cutoff in the low-mass regime to recover the smooth transition from low to high-mass stars that is typical of the Chabrier IMF. Our chosen ``standard'' IMF will have a more discontinuous drop in $N(M)$ from low to high-mass stellar mass ranges. 

The IMF function for Yggdrasil adopts a log-normal IMF taking the form
\begin{equation}
\mathrm{ln} \left(\frac{dN}{d\mathrm{ ln}(M)} \right) = A - \frac{1}{2\sigma^2} \left[ \mathrm{ln} \left( \frac{M}{M_\mathrm{c}} \right) \right]^2,
\end{equation}
where $\sigma$ is the width of the distribution, $M_\mathrm{c}$ is the characteristic mass, and $A$ is an arbitrary normalization \citep{Tumlinson06}, whose values are also listed in Table \ref{table:fixed_params}.

Each SED available in BPASS is organized into a specific IMF, metallicity, stellar age, and stellar population -- binary or non-binary. Available absolute stellar metallicities in BPASS are 0.00001, 0.0001, 0.001, 0.002, 0.003, 0.004, 0.006, 0.008, 0.01, 0.014, 0.02, 0.03, and 0.04, where the stellar ages range from $10^{6-11}$~yr in increments of 0.1 dex. Since the Yggdrasil spectra assume a fixed metallicity, they are organized only by stellar age.  All BPASS spectra are calculated between $10^{-3}$\,--\,10 $\mu$m, and all Yggdrasil spectra are calculated between 0.0091 -- 9~$\mu$m.  The stellar spectra for each burst of star formation are chosen by taking the age and metallicity of each SSP and selecting the nearest available neighbor in the database for the appropriate IMF and stellar population model. The total stellar SED is the sum of the SEDs from the individual SSPs that form every 5 Myr, starting when M$_{\mathrm{Halo}}$ = 10$^{\mathrm{7}} \Ms$ and ending at the observed redshift.

\subsection{AGN spectrum} \label{sec:SMBH}

To include the effects of an AGN on our model galaxy, we implement a disk continuum from the SKIRTOR model \citep{Stalevski16} with a modification from \citet{Yang22}. To better match observations, \citet{Yang22} introduced a free parameter, $\delta_{\rm AGN}$, in order to let the optical spectral slope deviate from the intrinsic shape in the piecewise powerlaw
\begin{equation} \label{eq:agn_SED}
    \lambda L(\lambda) \propto
    \begin{cases}
        \lambda^{1.2} & 0.001 \leq \lambda \leq 0.01\ \ (\mu m) \\
        \lambda^{0} & 0.01 < \lambda \leq 0.1\ \ (\mu m) \\
        \lambda^{-0.5 + \delta_{\rm AGN}} & 0.1 < \lambda \leq 5\ \ (\mu m) \\
        \lambda^{-3} & 5 < \lambda \leq 1000\ \ (\mu m) \\
    \end{cases}.
\end{equation}
We take the AGN luminosity to be Eddington-limited $L_{\textrm{AGN}} = 1.26 \times 10^{38}\ f_{\rm Edd} \ f_{\textrm{BH}}\ (M_{\star}/\Ms) \; \textrm{erg} \ \mathrm{s^{-1}}$. The above continuum represents the face-on accretion disk luminosity. Dust obscuration and inclination angle are not included in this spectrum. We include $\delta_{\rm AGN}$ as a free parameter in our study. We normalize the AGN SED with the bolometric luminosity $L_{\rm AGN}$ using the BH mass as a percentage of the stellar mass ($f_{\textrm{BH}} \equiv M_{\textrm{BH}} / M_{\star}$) determined from the star formation history at the observed redshift, where we vary $f_{\textrm{BH}}$ in our MCMC analysis.  Because of the degeneracy of the Eddington and BH-stellar mass ratios, we do not vary $f_{\rm Edd}$, and the variations in $f_{\rm BH}$ will account for any variations in the product of these two parameters.

\subsection{Dust Attenuation} \label{sec:dust}
Once we compute both the stellar SED and the AGN SED, we sum them to obtain the total SED of the galaxy. This spectrum is still the intrinsic, unattenuated spectrum in the rest-frame of the galaxy. In reality, as UV photons travel through the (dusty) interstellar medium, they are absorbed and reemitted in various emission lines and in the infrared. Following the same method used in H22, we apply two models of dust attenuation; however, we do not consider nebular emission. The effects of this will be discussed in Section \ref{sec:discussion}. First, absorption from the neutral IGM from \cite{Madau95} and second, extinction via the starburst reddening curve from \cite{Calzetti00}. To attenuate the spectrum, we compute an effective optical depth, $\tau_\mathrm{eff}$, and scale the intrinsic luminosity by $e^{-\tau_\mathrm{eff}}$ to compute the attenuated luminosity. There are three contributions to the IGM absorption from \cite{Madau95}:

\begin{enumerate}
    \item Metal-line blanketing:
\begin{equation}
\tau_\mathrm{metal}(\lambda) = A_\mathrm{metal}\bigg( \frac{\lambda_\mathrm{obs}}{\lambda_\alpha} \bigg)^{1.68}\ \ (\lambda_\mathrm{obs} < \lambda_\alpha)
\end{equation}
where $\lambda_\mathrm{obs} = \lambda (1 + z)$ is the observer frame wavelength of the SED, $A_\mathrm{metal}$ = 0.0017, and $\lambda_\alpha$ = 1215.67 \AA.

\item Lyman series line blanketing:
\begin{equation}
\tau_\mathrm{LB}(\lambda) = \sum_{j=2,i} A_j \bigg( \frac{\lambda_\mathrm{obs}}{\lambda_j} \bigg)^{3.46} \ \ (\lambda_{i+1} < \lambda_\mathrm{obs}/(1+z) < \lambda_i)
\end{equation}
where $\lambda_j$ are the wavelengths of photons emitted by neutral hydrogen from an electron transition from excited state $j$ to the first excited state. We compute 17 terms for $\tau_\mathrm{LB}$ for $j = 2$ to $j = 18$ given the coefficients $A_j$\footnote{$A_j$ = (0.0036, 0.0017, 0.0011846, 0.0009410, 0.0007960, 0.0006967, 0.0006236, 0.0005665, 0.0005200, 0.0004817, 0.0004487, 0.0004200, 0.0003947, 0.000372, 0.000352, 0.0003334, 0.00031644)} and $1/\lambda_j = R_\mathrm{H} (1 - 1/j^2)$ with $R_\mathrm{H} \approx 1.0968 \times 10^{-3}\, \mathrm{nm}^{-1}$.

\item Photoelectric absorption by intervening systems:
\begin{align}
    \begin{split}
    \label{eq:photoelectric_absorption}
    \tau_\mathrm{PA}(\lambda) \simeq\  & 0.25 x_\mathrm{c}^3 (x_\mathrm{em}^{0.46} - x_\mathrm{c}^{0.46}) + 9.4x_\mathrm{c}^{1.5} (x_\mathrm{em}^{0.18} - x_c^{0.18}) \\
     - &0.7x_c^3 (x_\mathrm{c}^{-1.32} - x_\mathrm{em}^{-1.32}) - 0.023 (x_\mathrm{em}^{1.68} - x_\mathrm{c}^{1.68}),
    \end{split}
\end{align}
where $x_\mathrm{c} \equiv \lambda_\mathrm{obs}/\lambda_\mathrm{L}$, $\lambda_L$ = 911.75 \AA\ and $x_\mathrm{em} \equiv 1 + z$.  We use the approximation\footnote{Note Madau's definition of $x_\mathrm{c}$ in footnote 3 contains a typo. We give the correct definition here.} of Equation 16 in \cite{Madau95}.
\end{enumerate}

We also apply dust extinction to the spectrum using the starburst reddening curve from \cite{Calzetti00}, adding another contribution to the total optical depth
\begin{equation}
\tau_\mathrm{red}(\lambda) = \frac{k'(\lambda)}{R_V'}\, \tau_\mathrm{V} 
\end{equation}
with
\begin{equation}
k'(\lambda) =
\begin{cases}
 2.659(-2.156 + 1.509/\lambda \\ \ \quad- 0.198/\lambda^2 + 0.011/\lambda^3) + R'_\mathrm{V} & [0.12,\, 0.63)\, \mu\mathrm{m}  \\ \\
2.659(-1.857 + 1.040/\lambda) + R'_\mathrm{v} & [0.63,\, 2.20]\, \mu\mathrm{m}
\end{cases}
\end{equation}
and $R'_\mathrm{V} = 4.05$. $\tau_\mathrm{V}$ is the V-band optical depth and is a free parameter in our model. Finally, we calculate the dust attenuated luminosity from the intrinsic luminosity as $L_\mathrm{obs}(\lambda) = L(\lambda) \exp[-\tau_\mathrm{eff}(\lambda)]$ with $\tau_\mathrm{eff}(\lambda) = \tau_\mathrm{metal} + \tau_\mathrm{LB} + \tau_\mathrm{PA} + \tau_\mathrm{red}$.

\begin{figure}
    \includegraphics[width=\columnwidth]{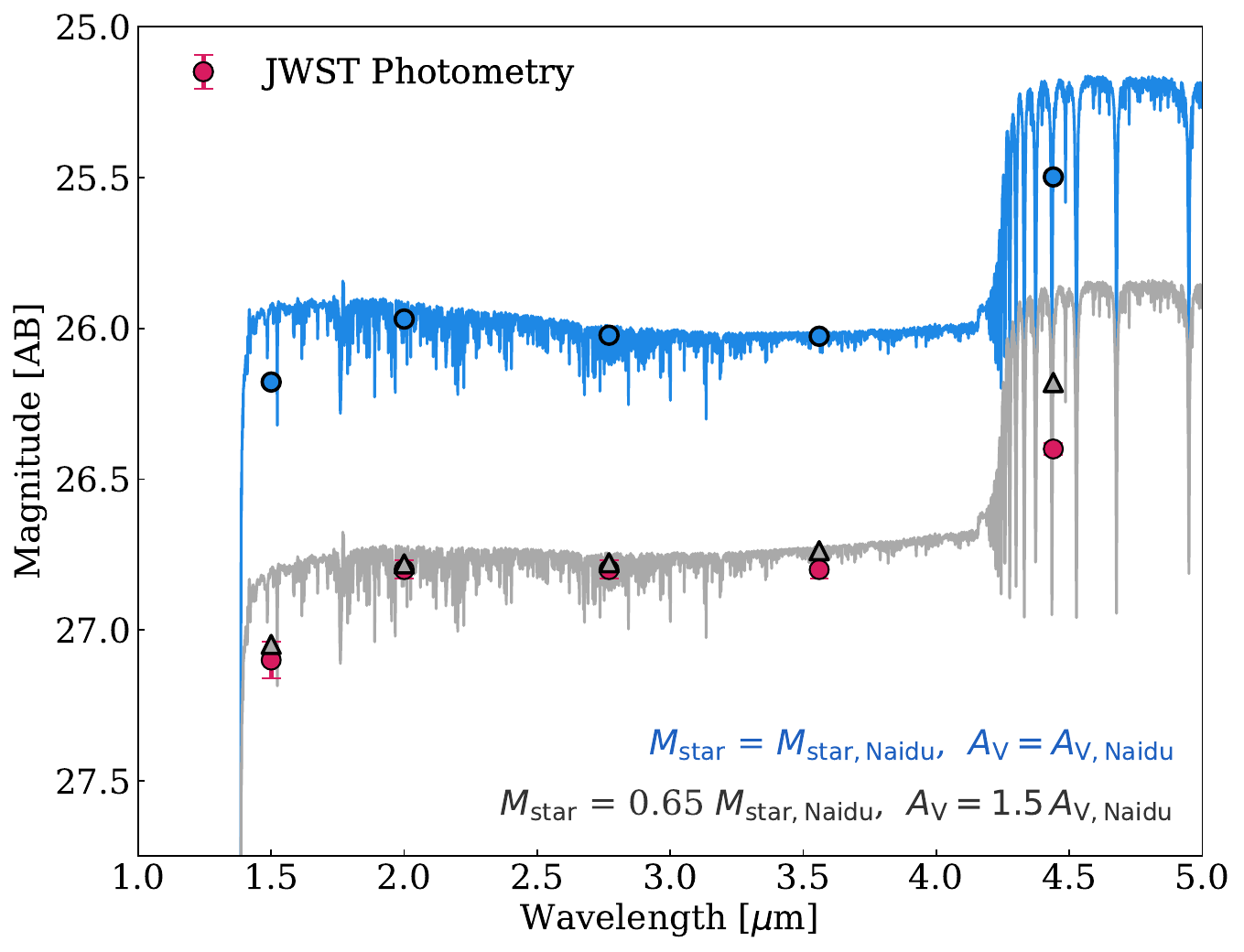}
        \centering
        \caption{Model stellar spectra for GL-z10 from our initial test assuming no AGN, no stellar binaries, a chab100 IMF, and exponential halo growth with $\alpha$ = 0.65. The blue line shows the spectra using the stellar mass and dust attenuation from N22: $\log_{10}(\textrm{M}_\star/\Ms)$ = 9.6 and $A_\mathrm{V}$ = 0.3. The grey line shows the spectrum when decreasing the stellar mass by a factor of 0.65 while increasing the dust attenuation by a factor of 1.5. The red points with error bars show the magnitudes in NIRCam filters F150W, F200W, F277W, F356W, and F444W. The flux in each filter was taken from Naidu et al. (2022).}
        \label{fig:spectra_jwst_mstar}
\end{figure}

\subsection{Computing magnitudes from the SED} \label{sec:mag}

We convert the rest-frame attenuated luminosity density to a flux density given the redshift $z$ and luminosity distance $D_{\rm L}$ to the galaxy. Using the JWST NIRCam filters, we convolve the throughput with the flux density and integrate the spectrum to obtain the total flux in each filter. All of our apparent magnitudes are computed in the AB system using
\begin{equation}
m_{\mathrm{AB}} = -2.5 \log_{10} \Bigg(\frac{\int f_\nu(\nu) T(\nu) \frac{d\nu}{\nu} }{f_{\mathrm{AB}} \int T(\nu) \frac{d\nu}{\nu}} \Bigg),
\end{equation}
where
\begin{equation}
f_\nu(\nu) = \frac{(L_\lambda^\mathrm{stellar} + L_\lambda^\mathrm{AGN})/\nu}{4\pi D_L^2 (1+z)}.
\end{equation}
Here $f_\nu$ is the flux density per unit frequency; $T$ is the throughput for a given NIRCam filter; $f_{\mathrm{AB}}$ = 3630.78 Jy is the zero-point for the AB magnitude system, and $L_\lambda$ is the luminosity density (per unit wavelength). For the absolute UV magnitude, we follow \cite{Donnan22} and convolve the flux density with a top-hat filter centered on 1500 \AA\ with a width of 100 \AA.

\subsection{Validating the SED Model} \label{sec:testing}

Before running a full MCMC analysis, we validate our model spectra against the photometry of GL-z10 from \citet[][hereafter N22]{Naidu22}, using the same SED fitting parameters as them. We chose GL-z10 because this is the same object analyzed by H22, named GL-z9-1, which we analyze and discuss in this paper. The reason we used the SED fitting parameters from N22 rather than H22 is because N22 include the extinction $A_\mathrm{V}$ and the median stellar age, while H22 do not. The goal with this test is to confirm that our dust attenuation model is similar to the N22 model. We used star formation model B, the chab100 IMF, and we did not include an AGN or stellar binaries, i.e. we used a stellar-only SED. Rather than choosing a halo mass and computing the stellar mass from the \cite{Behroozi13} SMHM relation, we used the stellar mass of GL-z10 directly from N22. This eliminates three of the five free parameters from our model, $M_\mathrm{halo}$, $f_\mathrm{BH}$, and $\delta_\mathrm{AGN}$. This leaves two free parameters: the exponential growth rate $\alpha$ and the V-band optical depth $\tau_\mathrm{V}$. We chose $\alpha = 0.65$ such that the median stellar age is 165 Myr, similar to the value of $163^{+20}_{-133}$ Myr for the same object analyzed by N22. We used the dust attenuation $A_{5500\text{\AA}}$ = 0.3 which corresponds to an optical depth $\tau_\mathrm{V} \approx 0.28$, where $\tau_\mathrm{V} = 0.4 A_{5500\text{\AA}} / \log_{10}(e)$. We found that when using the SED parameters from N22, our SED was brighter than the photometry by approximately 1 magnitude and the slope was slightly flatter. However, we were able to match the slope and normalization when we decreased the stellar mass by a factor of 0.65 and increased the dust attenuation by a factor of 1.5 (see Figure \ref{fig:spectra_jwst_mstar}).

\begin{figure}
    \includegraphics[width=\columnwidth]{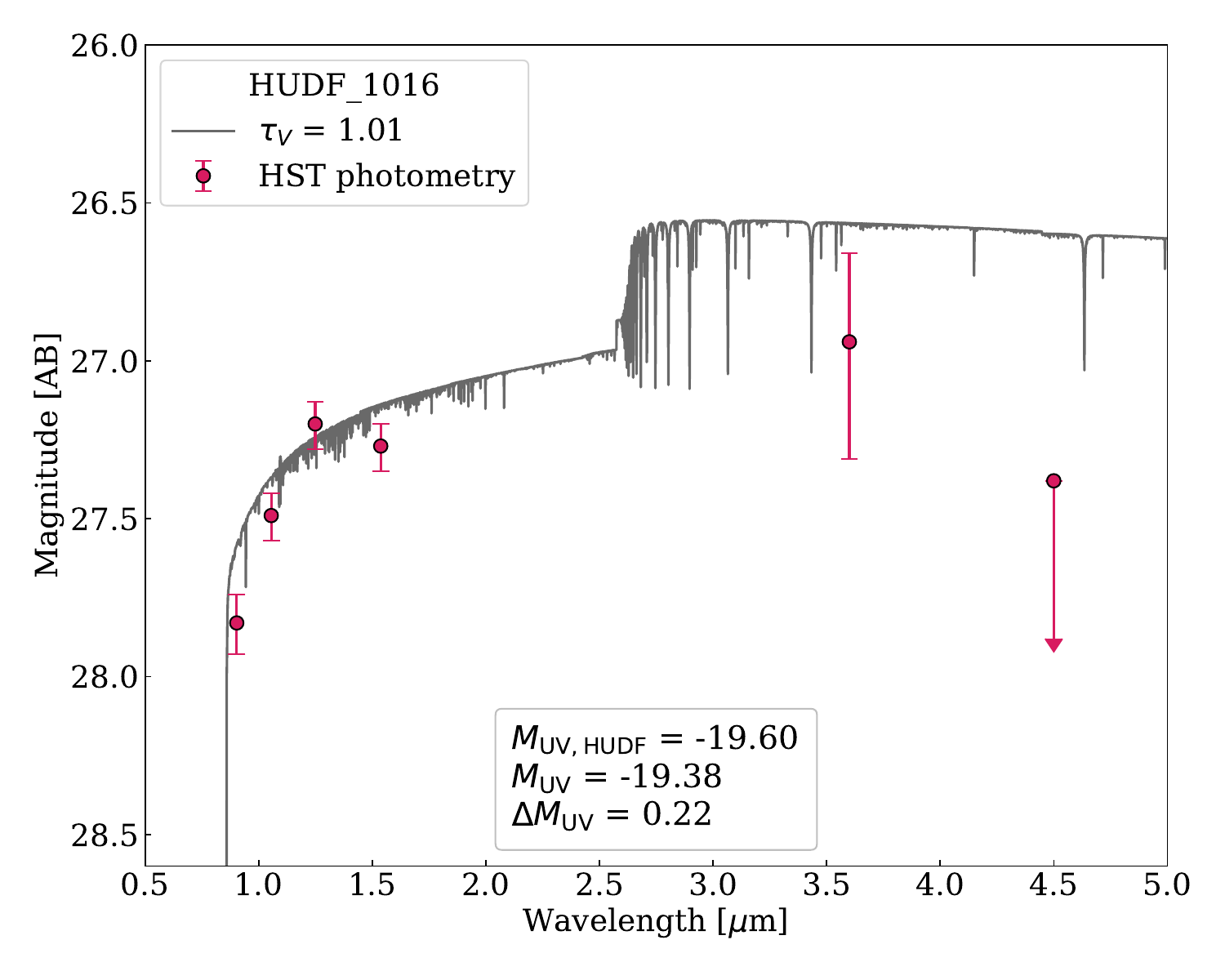}
    \centering
    \caption{Model stellar spectra for HUDF\_1016 from our test assuming no AGN, no stellar binaries, a chab100 IMF, and a single burst of star formation. This is only one of the eight galaxies we compared with and it has neither the best nor the worst fit.}    
    \label{fig:spectra_hubble}
\end{figure}

\begin{figure*}
    \includegraphics[trim={2cm 1cm 3.5cm 0.5cm},width=\textwidth]{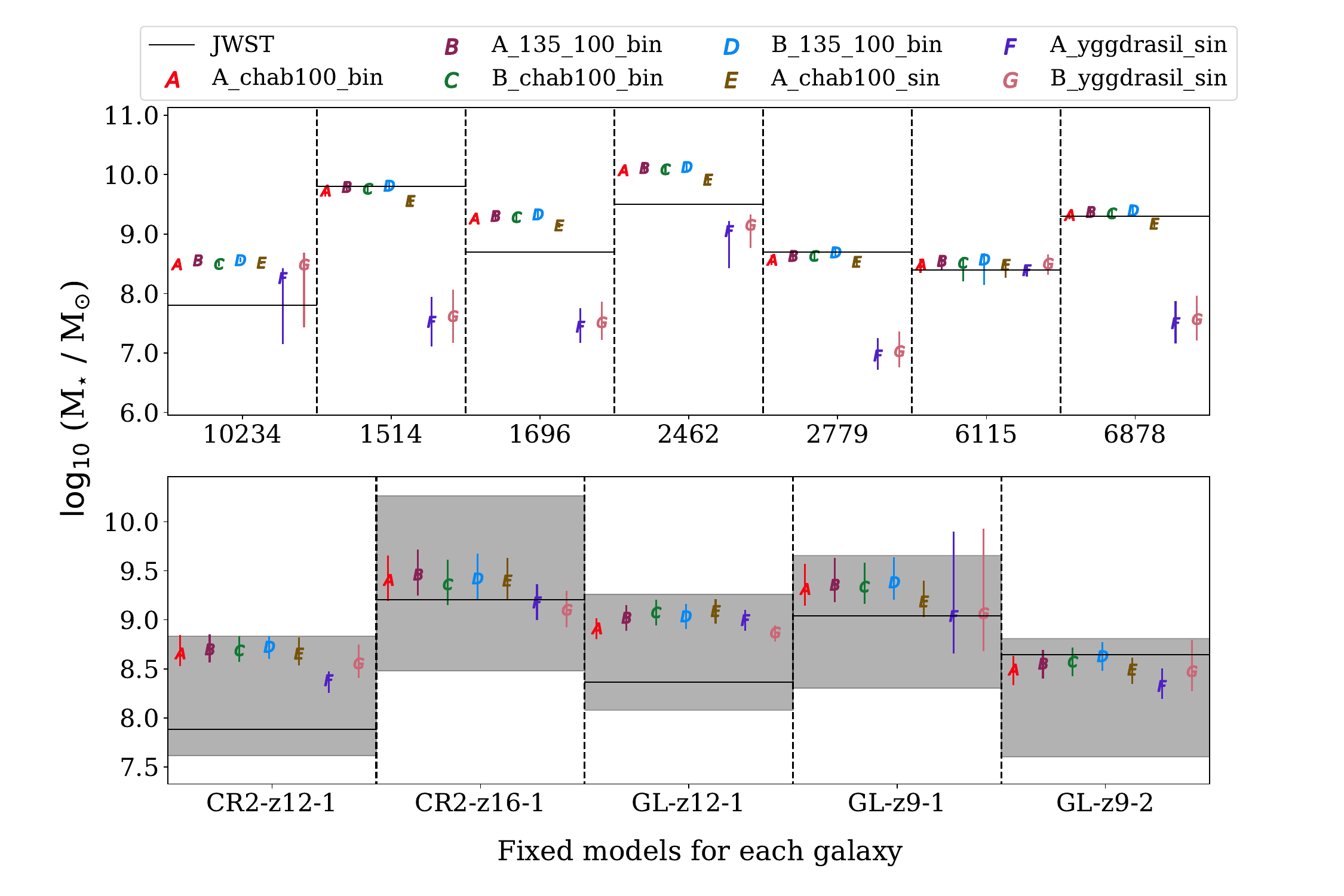}
    \centering
    \caption{The {\sc emcee}-predicted median stellar mass for each galaxy according to our seven fixed models is shown in with lettered markers. The error bars represent the 16th and 84th percentiles of the {\sc emcee} distribution after discarding any data prior to three times the integrated autocorrelation time, which indicates when the walkers begin converging. The black line is the stellar mass determined by A22 and H22, and the black shaded region shows their error if provided. Note that the stellar mass and error shown in black for galaxy GL-z9-2 is predicted by \citet{CastellanoDec22}, not A22 or H22. It is evident that a significant number of our predictions overestimate the stellar mass as compared to those predicted by H22 and A22. The AGN-dominated Yggdrasil spectra, which are used in models F and G, recover significantly lower stellar masses as expected when there is contribution from a central black hole.}
    \label{fig:Mstar_median}
\end{figure*}

Ideally, using the fitted parameters from N22 with our stellar-only spectra would match the photometry perfectly, but our model produces a higher luminosity for a given stellar mass. However, the stellar mass and dust attenuation needed to match the photometry are within the uncertainties in stellar mass and dust attenuation from N22. For reference, these values are $\log_{10}(\textrm{M}_\star/\Ms) = 9.6^{+0.2}_{-0.4}$ and $A_\mathrm{V} = 0.3^{+0.4}_{-0.2}$ and our values are $\log_{10}(\mathrm{M}_\star/\Ms) = 9.41$ and $A_\mathrm{V} = 0.45$. From these results, it was unclear whether this discrepancy was due to a problem without our model or differences between the models. It is not a perfect "apples-to-apples" comparison because we use a different star formation history model and different SED libraries. 

To investigate this further, we compared our model spectra with fitted spectra of galaxies from the Hubble Ultra Deep Field (HUDF) \citep{McLure11}. The goal of the HUDF analysis was to compare our model to theirs in the most similar way possible. To do this, we turned off the AGN in our model and we chose the standard chab\_100 IMF. The models we compared with also use a \cite{chab03} IMF and stellar-only spectra without binaries. We only compared with HUDF galaxies that were modelled using a single burst of star formation. This eliminates the star-formation history from our model, i.e. we use neither SFH model A or B, we just use a spectrum from a single stellar population given the stellar mass, age, metallicity, and extinction in the V-band from \cite{McLure11}. This means that the only differences between our model and theirs are the SED templates and the treatment of dust. Since we trust that BPASS provides high quality SED templates, the primary difference is only our treatment of dust. If there was a systematic error in our dust attenuation code, we would expect this to be apparent because our SEDs of the HUDF galaxies would look much different than the SEDs from \cite{McLure11}.

Using the redshift, stellar mass, stellar age, metallicity, and V-band extinction from \cite{McLure11} as inputs to our model, we found that the model spectra matched the HUDF galaxies reasonably well (see Figure \ref{fig:spectra_hubble} for an example of one model), particularly at lower wavelengths. To assess the quality of our spectra, we computed the difference in absolute UV magnitude between our model and theirs, $\Delta M_\mathrm{UV}$. For example, for object HUDF\_1173 with $\tau_\mathrm{V} = 0$, $\Delta M_\mathrm{UV} = 0.05$, and for object HUDF\_1016 with $\tau_\mathrm{V} = 1.01$, $\Delta M_\mathrm{UV} = 0.22$. We chose to show HUDF\_1016 in Figure \ref{fig:spectra_hubble} because although our spectra do not fit the photometry as well as some of the other examples, HUDF\_1016 has the largest optical depth and the primary reason for this test was to verify that our dust attenuation was working properly.

Figure \ref{fig:spectra_hubble} shows that the spectra does not fit the 4.5 micron flux for HUDF\_1016. We tested seven other HUDF/ERS galaxies and all but one of the others fit the 4.5 micron flux better than this one. Three of the objects fit the 4.5 micron flux well. For these three objects, the difference between the observed IRAC2 (4.5 micron) magnitudes and the model magnitudes were 0.05, 0.26, and 0.29 mag. Overall, HUDF\_1016, is neither the best nor the worst example from the objects we tested. Four of the of the seven other objects match the UV flux better than this one.

Since we are able to match the photometry of HUDF\_1016 which has a large optical depth and the other objects, this confirms that the discrepancy between our model of GL-z10 and the model from N22 is not due to a problem with our dust attenuation. These results, combined with the fact that we can match the photometry of GL-z10 using N22 fitting parameters (within their uncertainty) validates our model. We now describe how we determine the posterior distributions of our input parameters for each object in our sample.

\subsection{Exploring the parameter space} \label{sec:emcee}

\begin{figure*}
    \includegraphics[width=0.9\textwidth]{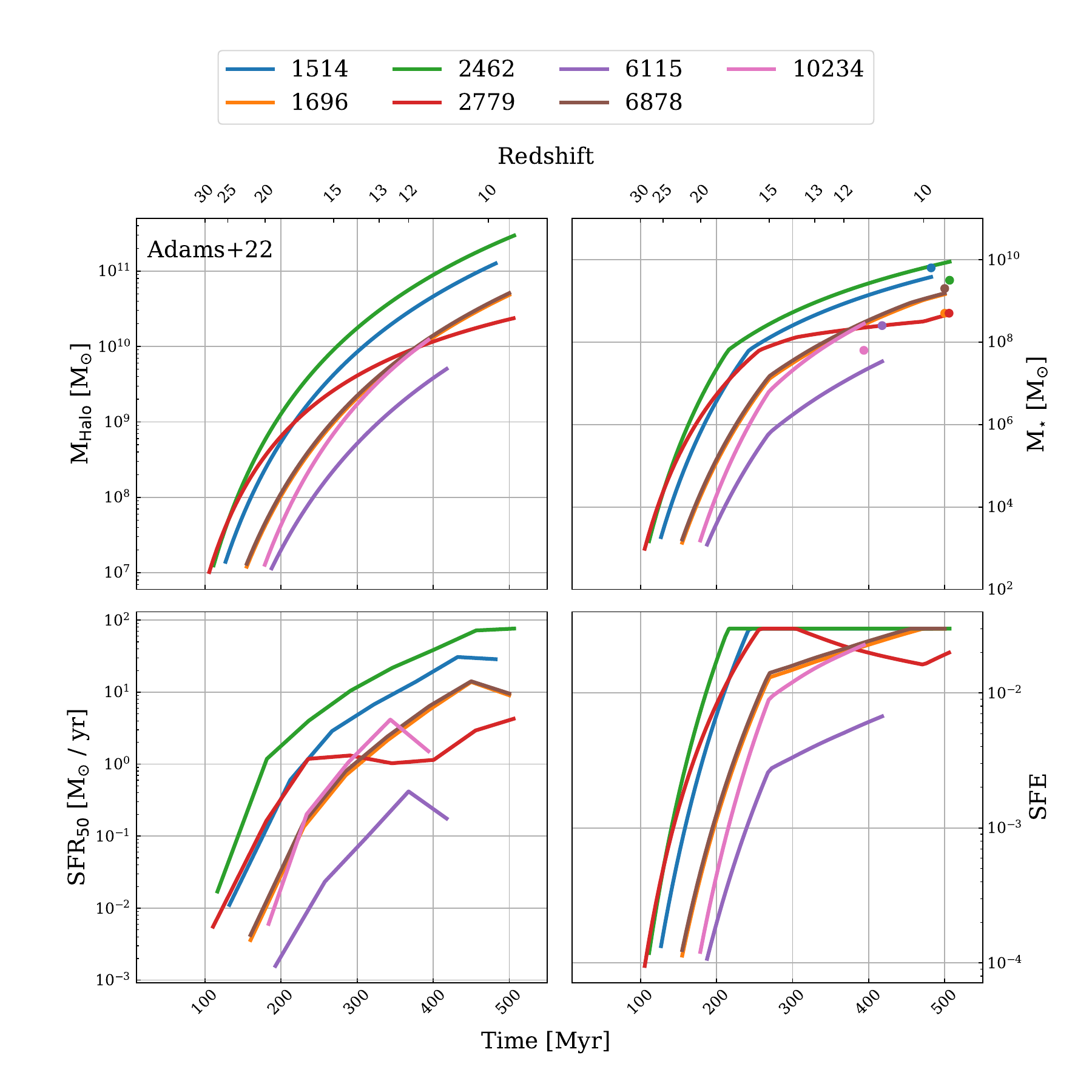}
        \centering
        \caption{The halo growth and star formation histories for each galaxy from A22 given the best-fit parameters. Upper left: Halo mass history. Upper right: Stellar mass history. The scatter points indicate the predicted stellar mass from JWST. Lower left: SFR history averaged over the last 50 Myr. Lower right: SFE history.}
        \label{fig:adams_history}
\end{figure*}

\begin{figure*}
    \includegraphics[width=0.9\textwidth]{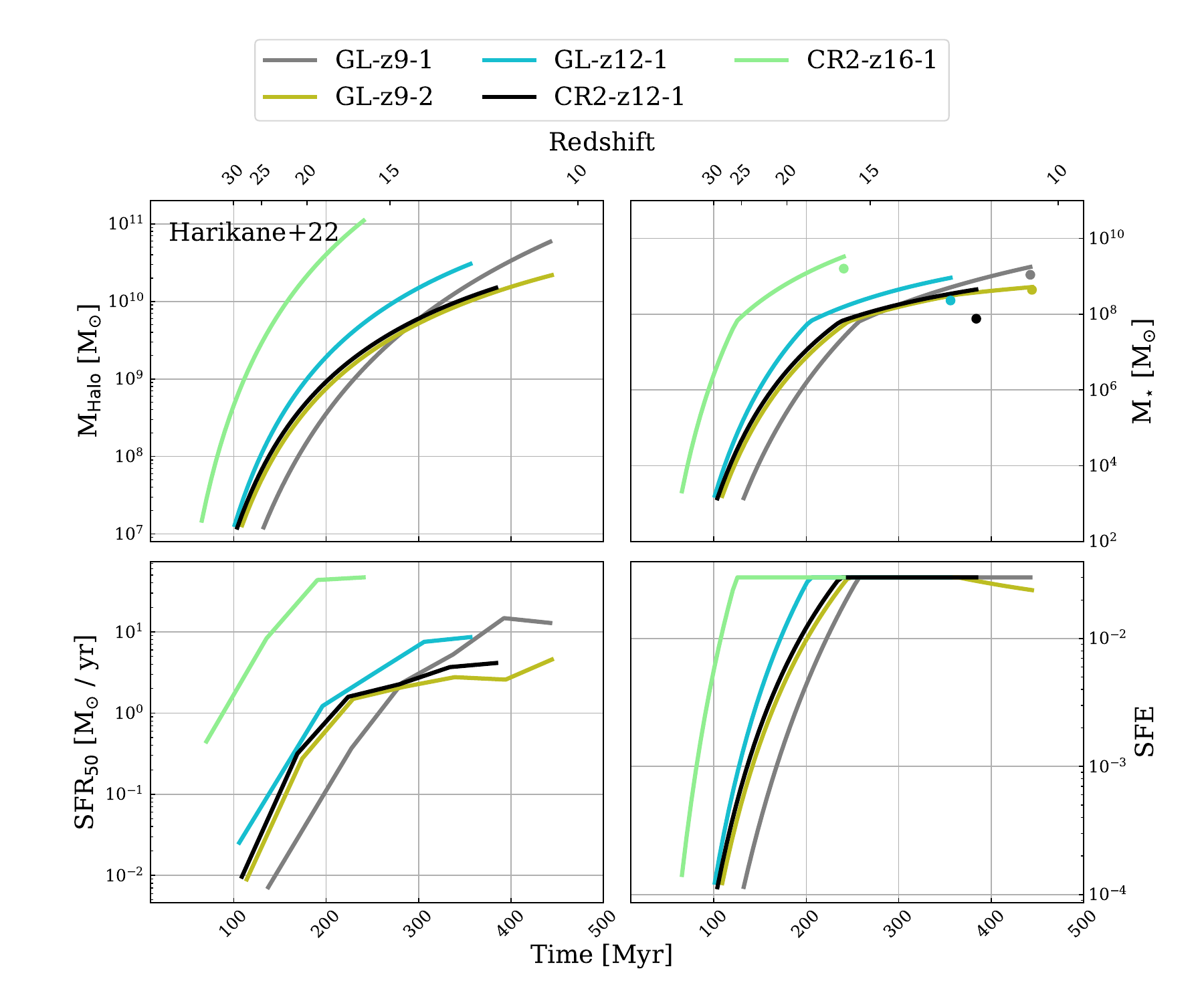}
        \centering
        \caption{The halo growth and star formation histories for each galaxy from H22 given the best-fit parameters. Upper left: Halo mass history. Upper right: Stellar mass history. The scatter points indicate the predicted stellar mass from JWST. Lower left: SFR history averaged over the last 50 Myr. Lower right: SFE history.}
        \label{fig:harikane_history}
\end{figure*}

\begin{figure*}
    \includegraphics[width=\textwidth]{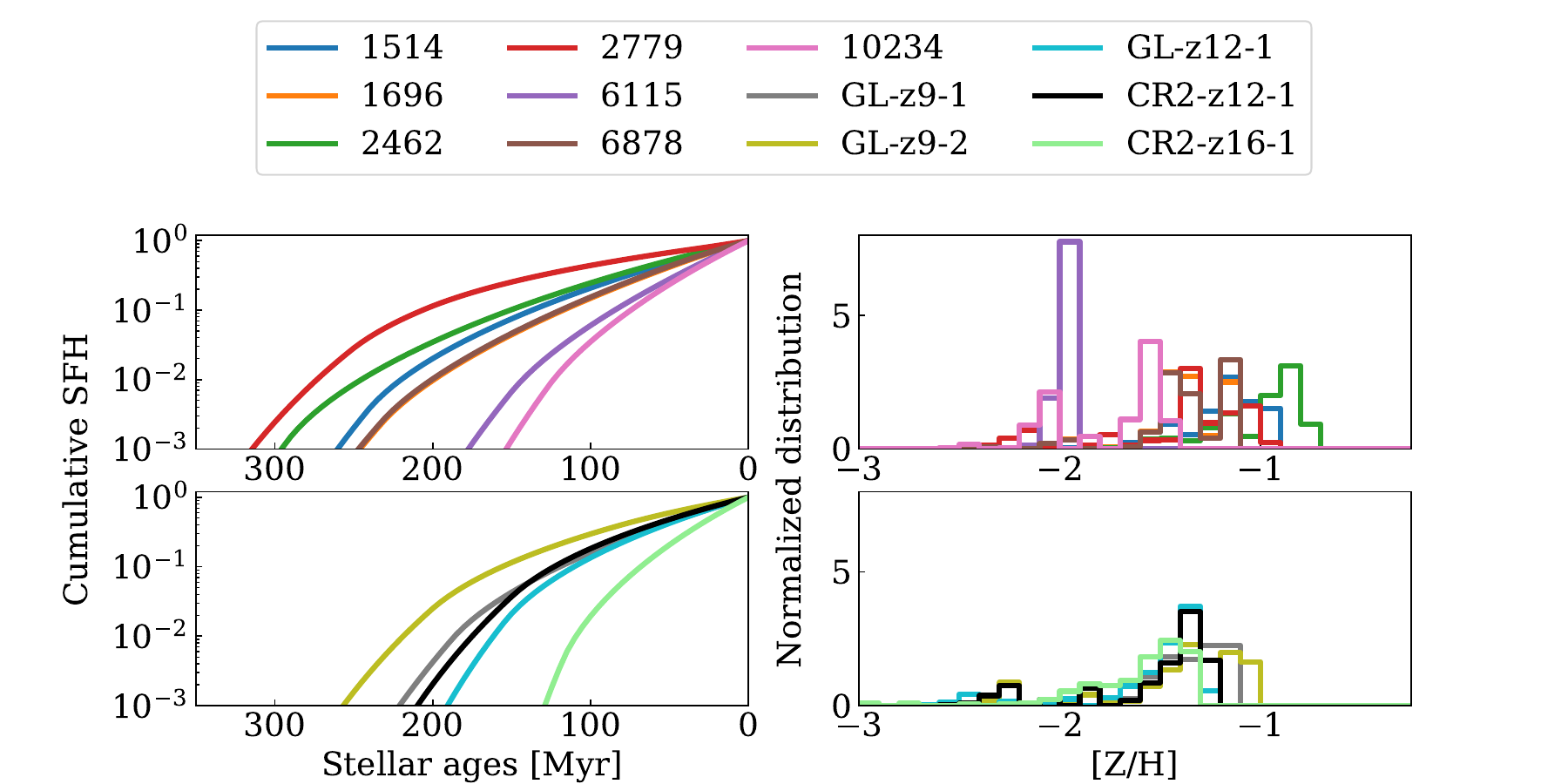}
        \centering
        \caption{The stellar population from A22 (top panels) and H22 (bottom panels) given the best-fit parameters. Left panels: The cumulative star formation history as a function of stellar age. Right panels: The normalized distribution of stellar metallicities.}
        \label{fig:all_stellar}
\end{figure*}

To find the parameters that produce a spectrum with the photometry that best matches the observed JWST values, we use the MCMC ensemble sampler {\sc emcee} \citep{emcee}. The MCMC algorithm uses Bayesian inference to determine the probability distributions of the model parameters which also allows us to quantify the most likely values and their uncertainties. For each model permutation, we use {\sc emcee} to vary four or five continuous parameters: the halo mass $M_{\mathrm{halo}}$, the BH mass to stellar mass ratio $f_{\mathrm{BH}}$, the AGN spectrum power law slope offset $\delta_{\mathrm{AGN}}$, the V-band optical depth $\tau_\mathrm{V}$, and for models using exponential halo growth, we also vary the growth rate $\alpha$. {\sc emcee} works by sending out ``walkers'' to explore the parameter space. As the walkers move, we compute the photometric magnitudes associated with the walker's position in the parameter space and compute the posterior probability of the parameters given the JWST observations. The posterior probability is given by 
\begin{align}
p(\Theta|O_{\mathrm{JWST}}) = \frac{p(O_{\mathrm{JWST}}|\Theta) p(\Theta)}{ p(O_{\mathrm{JWST}}) }
\end{align}
where $\Theta$ is the set of model parameters \{$M_{\mathrm{halo}}, f_{\mathrm{BH}}, \delta_{\mathrm{AGN}}, \tau_\mathrm{V}, \alpha$\}; $O_{\mathrm{JWST}}$ is the set of JWST observed magnitudes in each filter; $p(O_{\mathrm{JWST}}|\Theta)$ is the likelihood of a JWST observation given our parameters; $p(\Theta)$ are the parameter priors, and $p(O_{\mathrm{JWST}})$ is the probability of the JWST observations. In practice, we do not need to know $p(O_{\mathrm{JWST}})$ because we are only interested in maximizing the likelihood of the model parameters and $p(O_{\mathrm{JWST}})$ is a constant that does not affect this process. We assume flat priors in the ranges for the parameters described in Table \ref{table:free_params}.

\begin{align} 
p(\Theta) = 
\left  \{
\begin{array}{ll} 
1 \ , & \mathrm{if}\ \theta \in [\theta_{\mathrm{min}},\, \theta_{\mathrm{max}}]\ \ \forall\ \  \theta \in \Theta  \\
0 \ , & \mathrm{otherwise} 
\end{array}
\right.\
\end{align}

We define the log of our likelihood function as $-\chi^2/8$ between the model magnitudes and the observed magnitudes because {\sc emcee} seeks to maximize the log likelihood function. In our case, this means minimizing $\chi^2$,
\begin{align}
\mathrm{ln}\  p(\Theta|O_{\mathrm{JWST}}) =
    -\frac{1}{2}\sum_{i} \biggl(\frac{m_i^{\mathrm{model}} - m_i^{\mathrm{obs}}}{\sigma_i^{\mathrm{obs,+}} - \sigma_i^{\mathrm{obs,-}}}\biggl)^2,
\end{align}
where we sum over each filter $i$. Here $m_i$ are the magnitudes in each filter, and $\sigma_i^+$ and $\sigma_i^-$  are the upper and lower uncertainties in the observed magnitudes respectively. Therefore the reduced $\chi^2$ is

\begin{align}
\chi^2 = \sum_{i} \biggl(\frac{m_i^{\mathrm{model}} - m_i^{\mathrm{obs}}}{\sigma_i^{\mathrm{obs}}}\biggl)^2 = -8\, \mathrm{ln}\  p(\Theta|O_{\mathrm{JWST}})
\end{align}
where $\sigma_i = 0.5(\sigma_i^+ - \sigma_i^-)$.

We ran {\sc emcee} for each of the seven model choices for each of the 12 galaxies in our sample for a total of 84 different runs. For each run we used 288 chains of walkers with a maximum of 20,000 steps, though not all models made it to 20,000 steps in the allocated computing time. Each model was run in parallel on a cluster using 144 cores each for a maximum of eight hours. To verify that the Markov chains had converged to a steady state distribution, we checked that the chains were at least 50 times longer than the integrated autocorrelation time, $\mathrm{\hat{\tau}}$, and that the autocorrelation time had not changed by more than 1 per cent between steps at the end of the run. The autocorrelation time is the number of steps it takes for the walkers to settle into the true distribution of the parameter space. Eight out of the 84 models did not meet these standards. These were the Yggdrasil models for object 10234, and six other models for object 6115. However, the best fitting models had chains longer than 50$\mathrm{\hat{\tau}}$, and all but three of the best models had chains longer than 100$\mathrm{\hat{\tau}}$. The best model for each object -- except 6115 and GL-z9-2 -- also met the condition of $\mathrm{\hat{\tau}}$ not changing by more than 1 per cent by the time the runs finished. For the two objects where this condition was not met, the fractional change in $\mathrm{\hat{\tau}}$ was only marginally larger than 1 per cent. For details on how the autocorrelation time is calculated and its relevance to convergence, see \cite{emcee}.

At each step in the chain, there is a certain chance that a walker will move to a new position in the parameter space. In general, a walker has a lower -- although non-zero -- chance to move to a new position if the posterior probability $p(\Theta|O_{\mathrm{JWST}})$ at the new position is very low. When the move is rejected, the walker will remain in place and the next step in the chain will be a duplicate of the previous step. This often occurs when a walker reaches an edge of the parameter space and the next step would take it outside the range of the priors. For this reason, we remove the duplicate samples and only compute statistics for the distributions using the unique samples. We also discard the initial 3$\mathrm{\hat{\tau}}$ steps as ``burn-in'' to ensure we only work with parameters that are sampled from the true distribution. We calculate the median and the uncertainty in the distributions of each parameter from the 16th, 50th, and 84th percentiles of the distributions.

\section{Results} \label{sec:results}

For most galaxies, we find that our analysis pipeline determines a stellar mass this is consistent with A22/H22 within their uncertainties.  For half of the objects, the best-fit parameters are higher than their best-fit values but still within their uncertainties.  Figure \ref{fig:Mstar_median} shows the median stellar mass from each model for each galaxy. The black lines indicate the stellar mass from A22/H22, except for galaxy GL-z9-2 which comes from \citet{CastellanoDec22}. For most galaxies, the BPASS models lie above the A22/H22 stellar mass, whereas the Yggdrasil models lie well below the black line and tend to have much larger error bars. While we set out to see if we could determine stellar masses that are lower than what the literature predicts by including other luminous sources, we found the opposite. We discuss the exact reasons for this overestimate in Section \ref{sec:discussion}.

\begin{figure*}
    \begin{subfigure}{\columnwidth}
      \centering
      \includegraphics[width=\columnwidth]{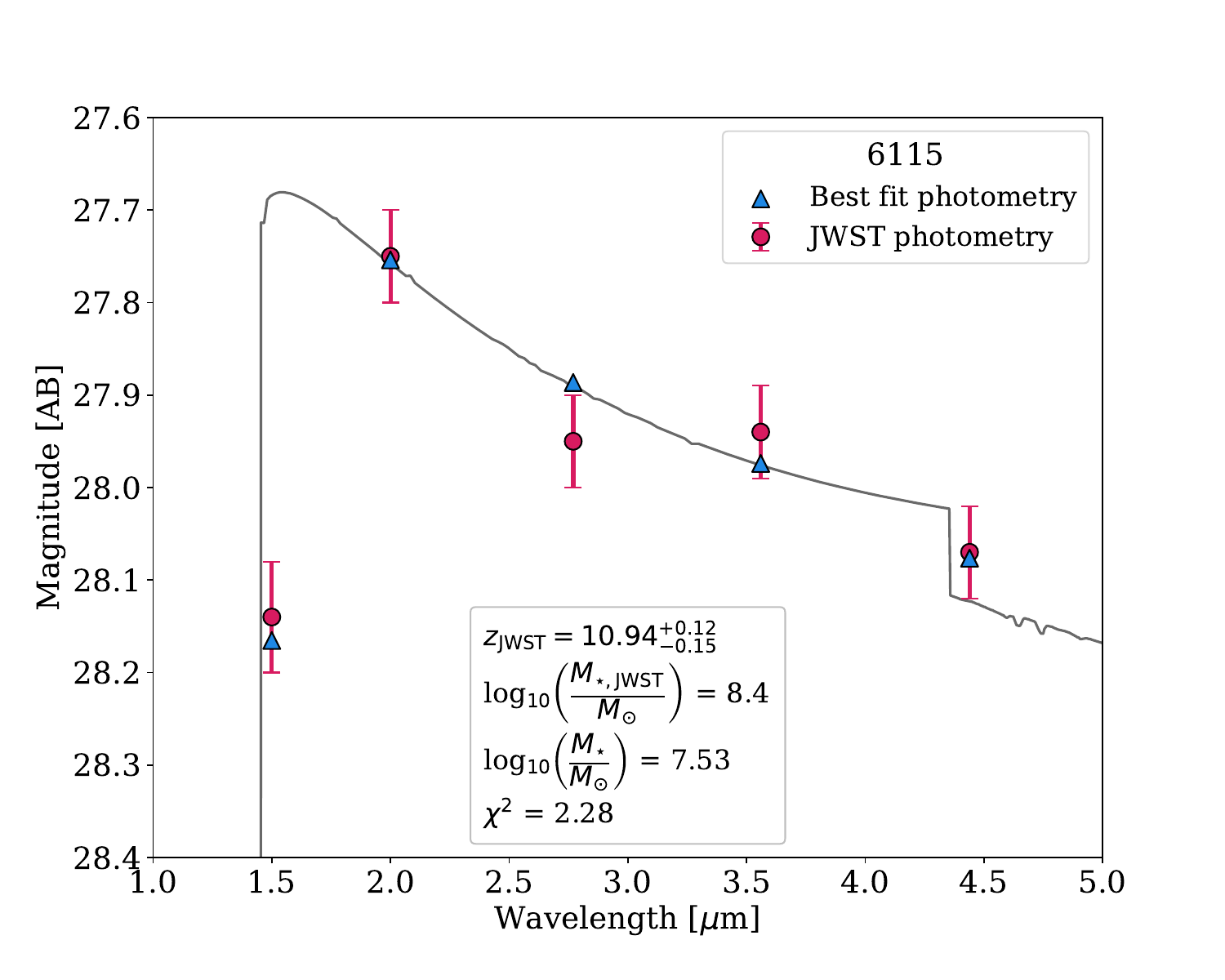}
      \caption{}
      \label{fig:6115}
    \end{subfigure}
    \begin{subfigure}{\columnwidth}
      \centering
      \includegraphics[width=\textwidth]{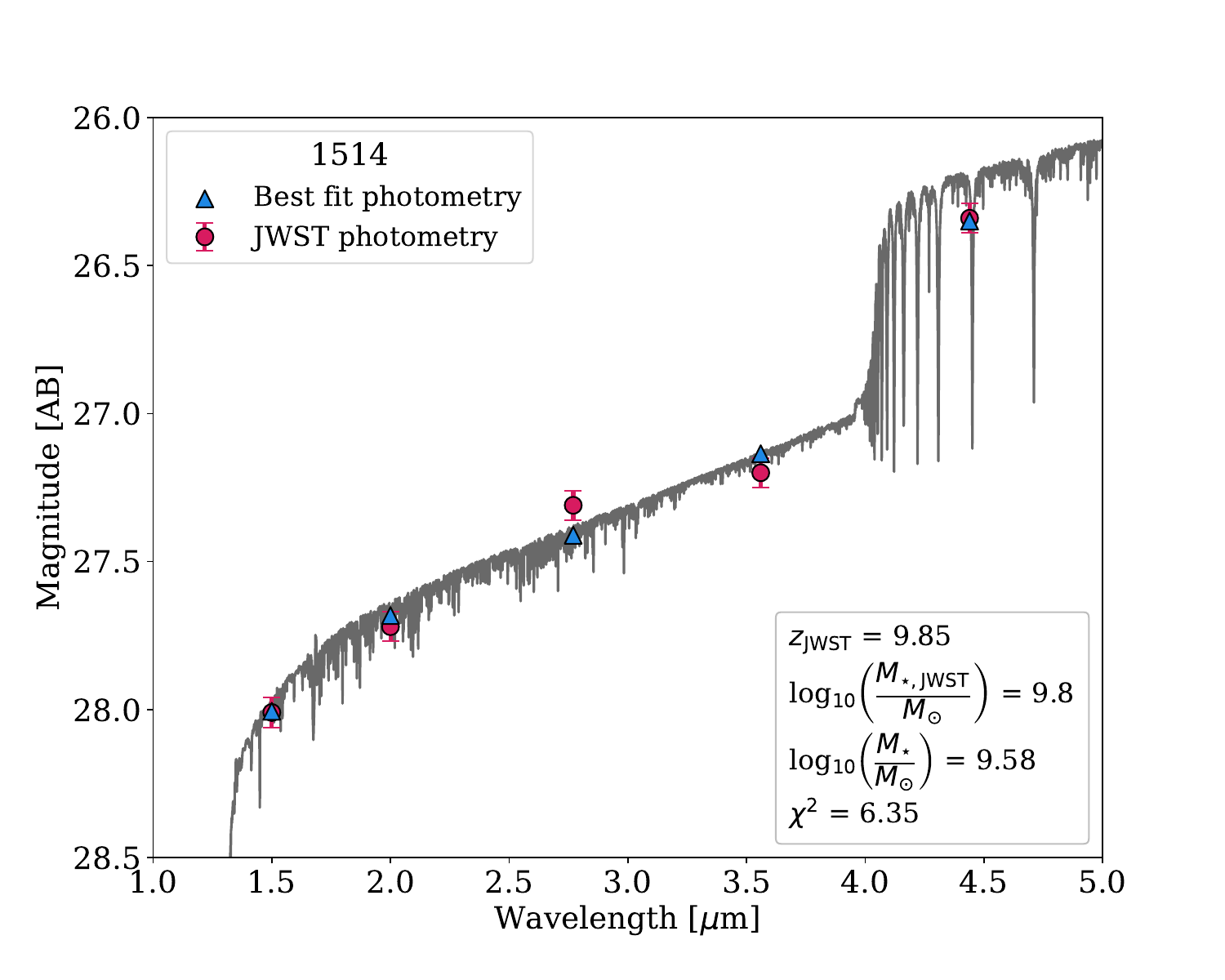}
      \caption{}
      \label{fig:1514}
    \end{subfigure}
    \begin{subfigure}{\columnwidth}
      \centering
      \includegraphics[width=\columnwidth]{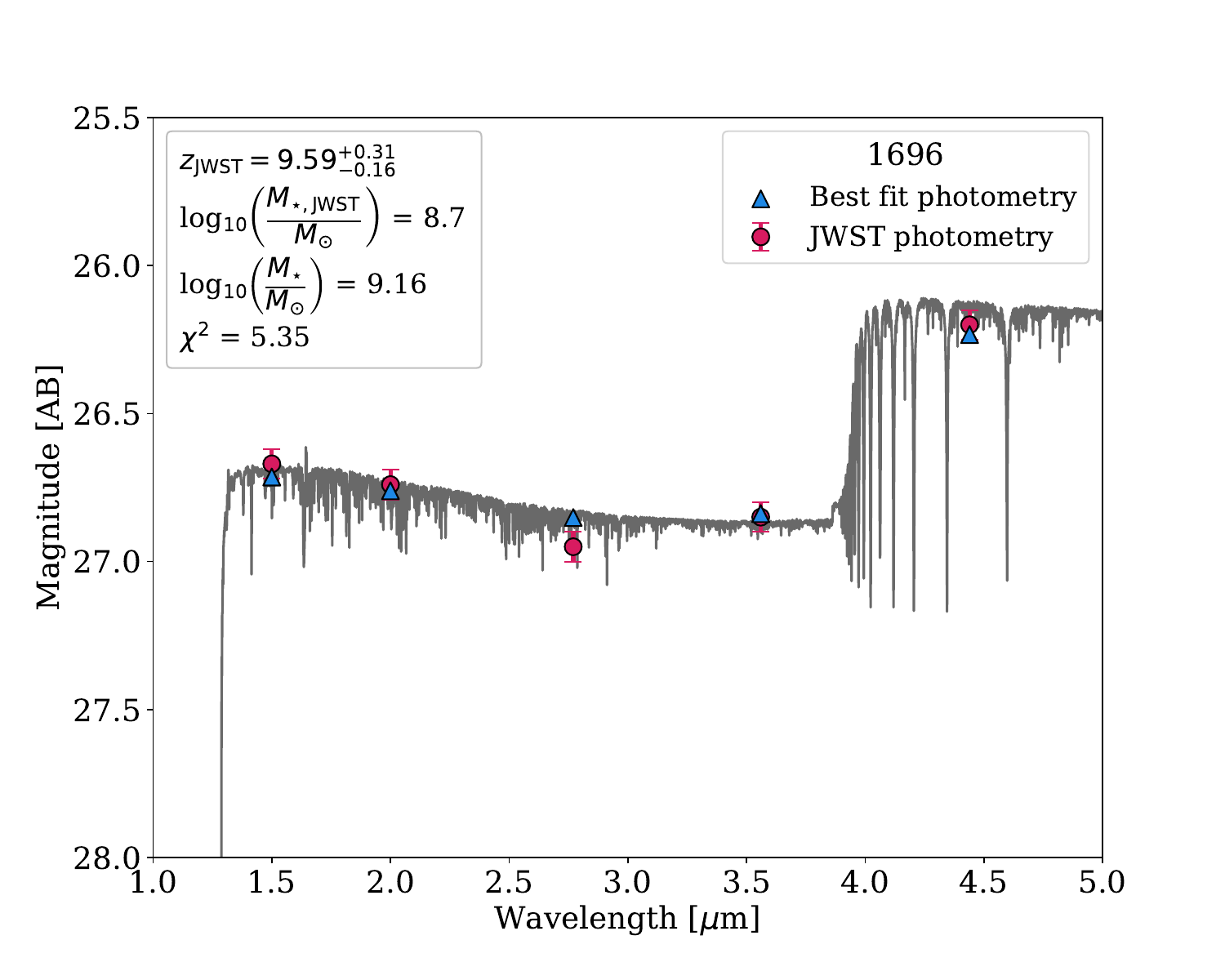}
      \caption{}
      \label{fig:1696}
    \end{subfigure}
    \begin{subfigure}{\columnwidth}
      \centering
      \includegraphics[width=\columnwidth]{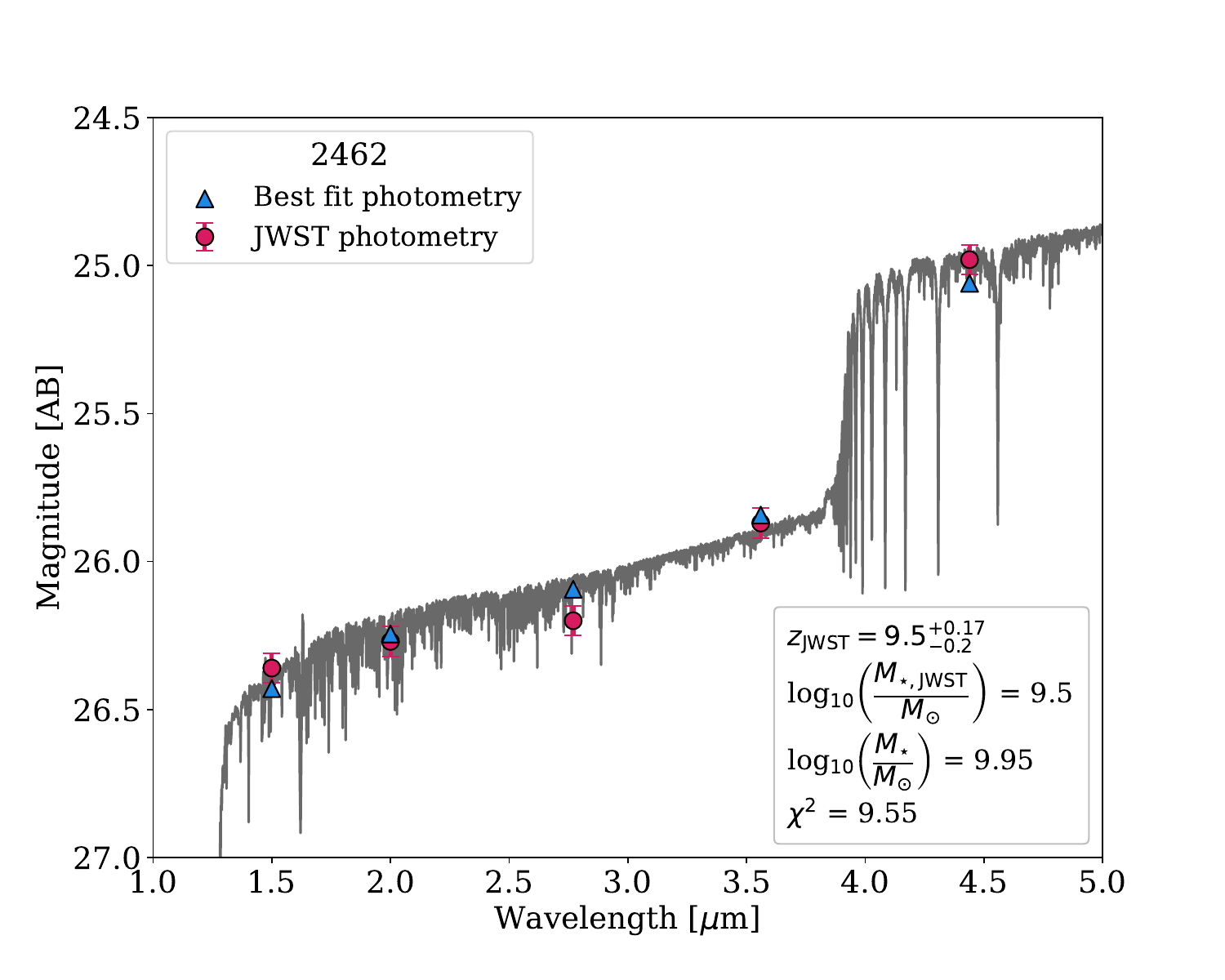}
      \caption{}
      \label{fig:2462}
    \end{subfigure}
    \caption{Fitted SED given the best-fit models for galaxies (a) 6115, (b) 1514, (c) 1696, and (d) 2462. Blue triangles indicate the measured photometry from our fitted spectra and the pink circles are the observed photometry values and errors. Annotated are the measured redshifts and stellar masses from A22, as well as our stellar mass given the best-fit halo mass and our $\chi^{2}$ value.}
    \label{fig:best_spectra_1}
\end{figure*}

\begin{figure*}
    \begin{subfigure}{\columnwidth}
      \centering
      \includegraphics[width=\columnwidth]{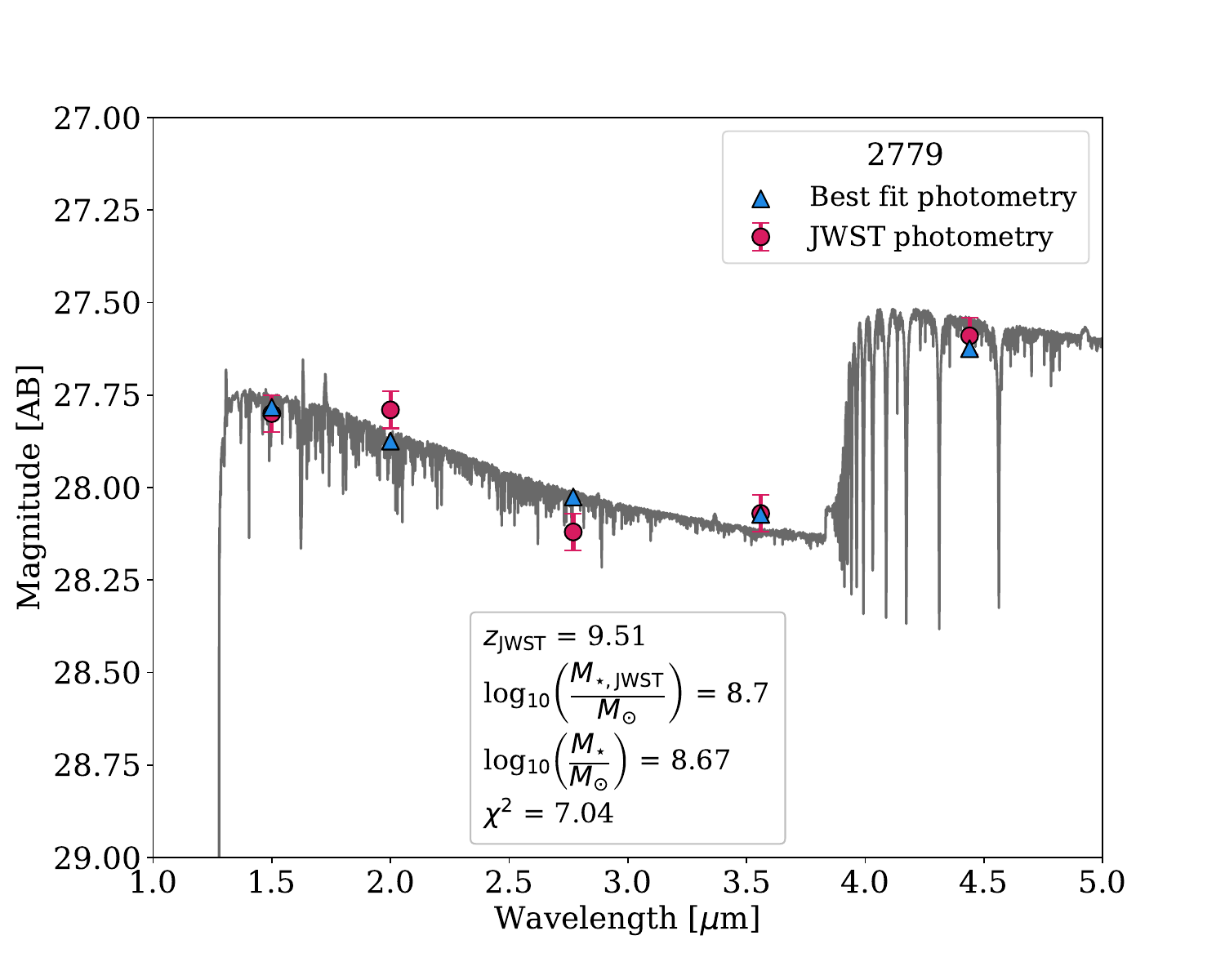}
      \caption{}
      \label{fig:2779}
    \end{subfigure}%
    \begin{subfigure}{\columnwidth}
      \centering
      \includegraphics[width=\columnwidth]{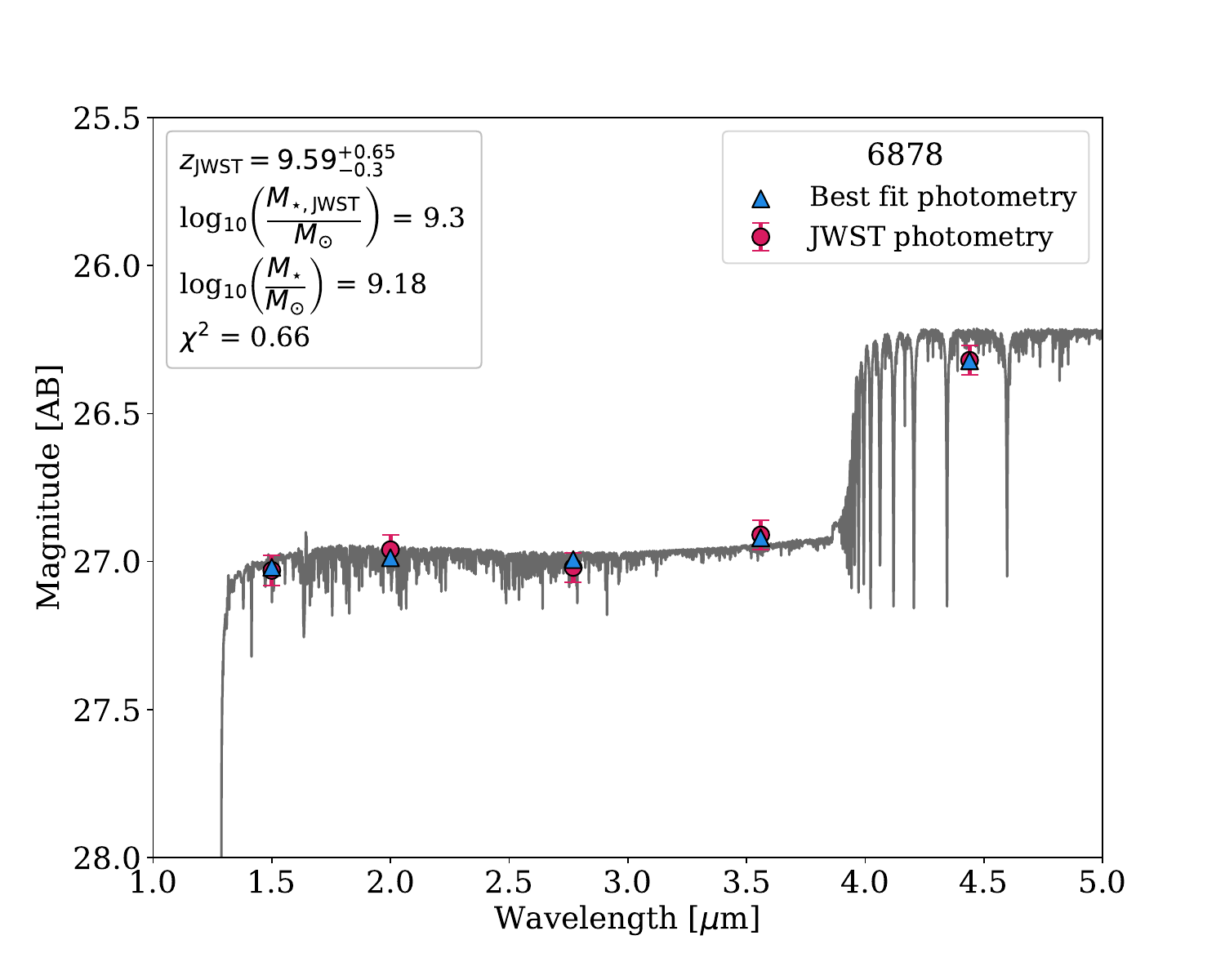}
      \caption{}
      \label{fig:6878}
    \end{subfigure}
    \begin{subfigure}{\columnwidth}
      \centering
      \includegraphics[width=\columnwidth]{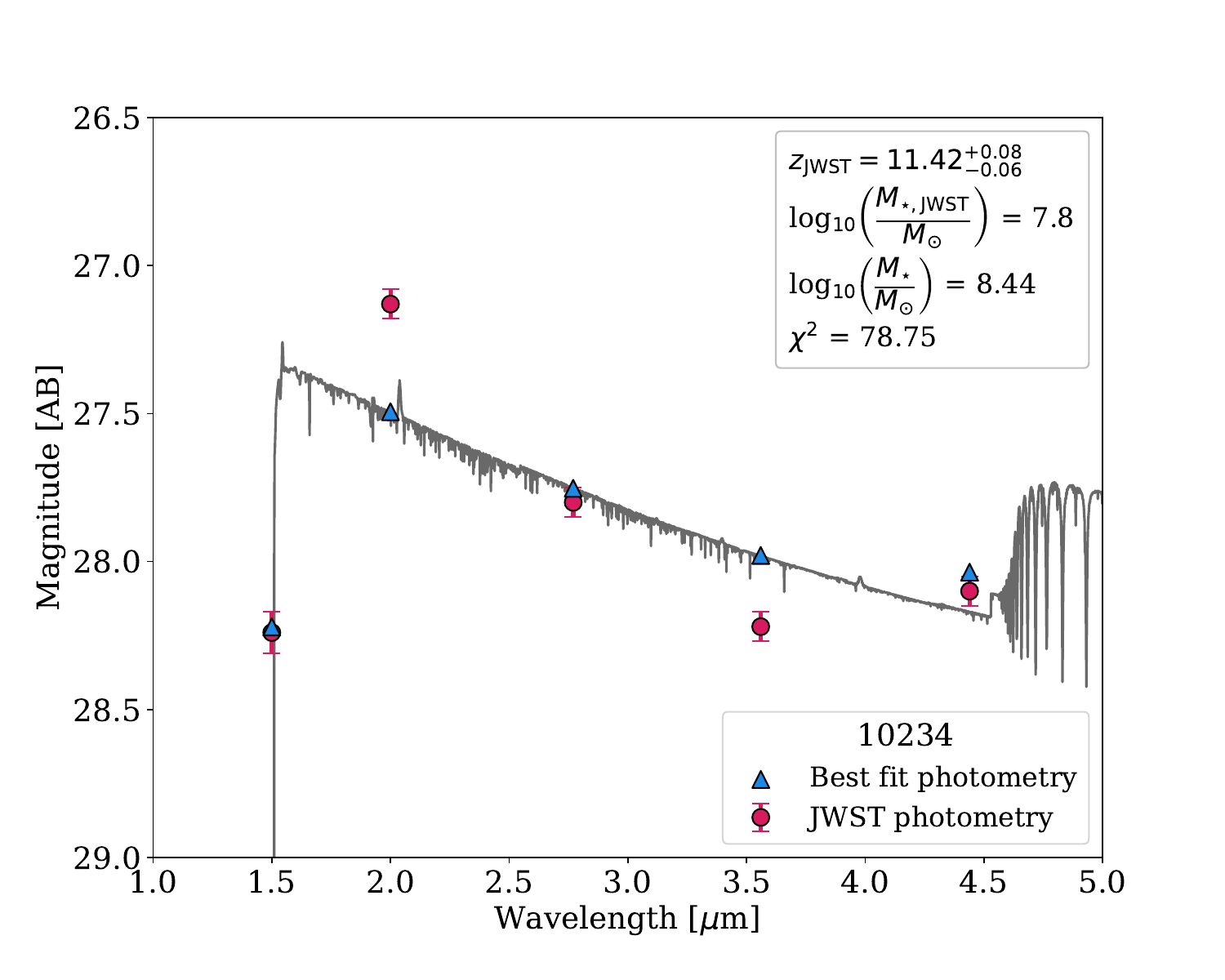}
      \caption{}
      \label{fig:10234}
    \end{subfigure}
    \begin{subfigure}{\columnwidth}
      \centering
      \includegraphics[width=\columnwidth]{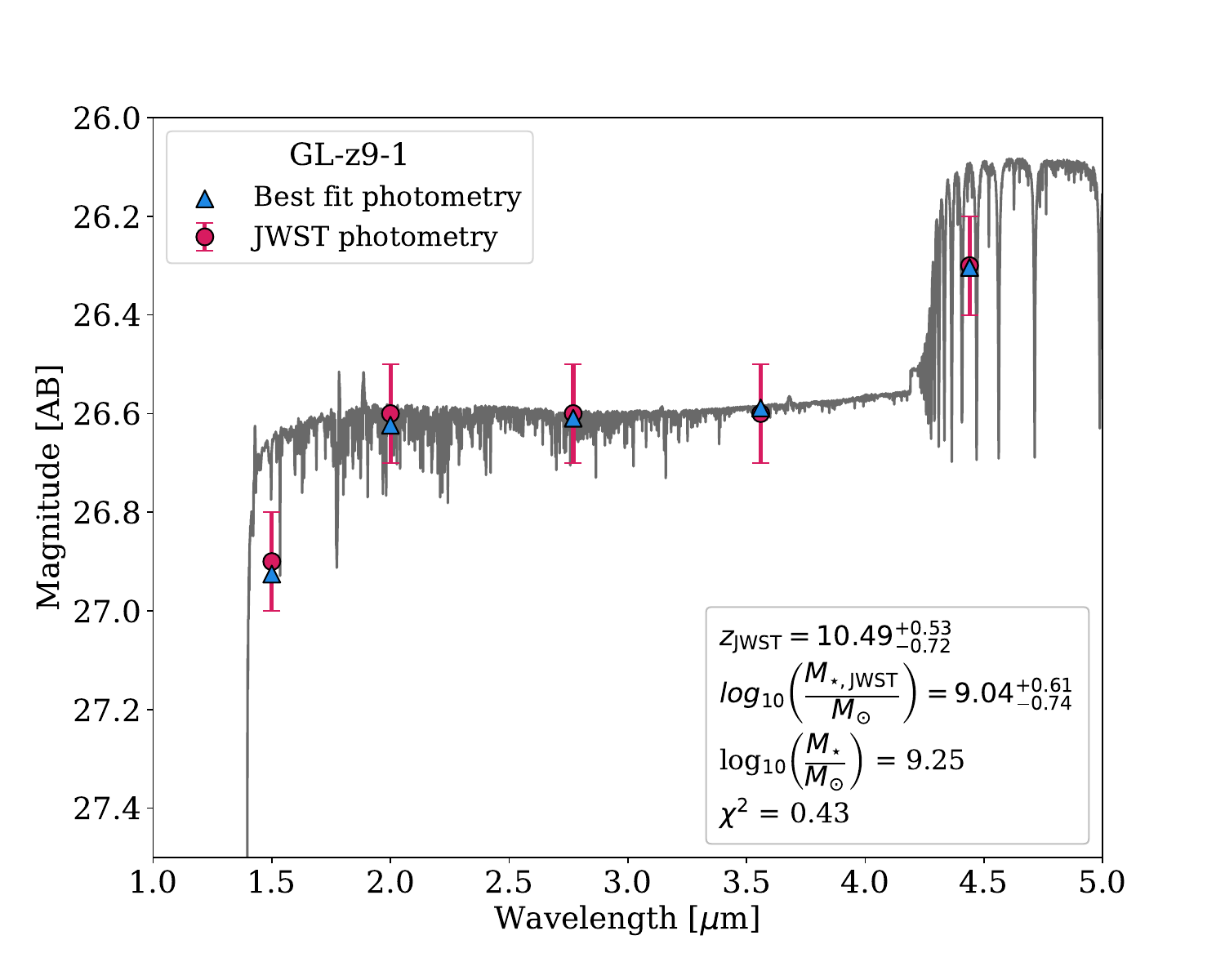}
      \caption{}
      \label{fig:GL-z9-1}
    \end{subfigure}
    \caption{Fitted SED given the best-fit models for galaxies (a) 2779, (b) 6878, (c) 10234, and (d) GL-z9-1. Blue triangles indicate the measured photometry from our fitted spectra and the pink circles are the observed photometry values and errors. Annotated are the measured redshifts and stellar masses from A22 and H22, as well as our stellar mass given the best-fit halo mass and our $\chi^{2}$ value.}
    \label{fig:best_spectra_2}
\end{figure*}

\begin{figure*}
    \begin{subfigure}{\columnwidth}
      \centering
      \includegraphics[width=\columnwidth]{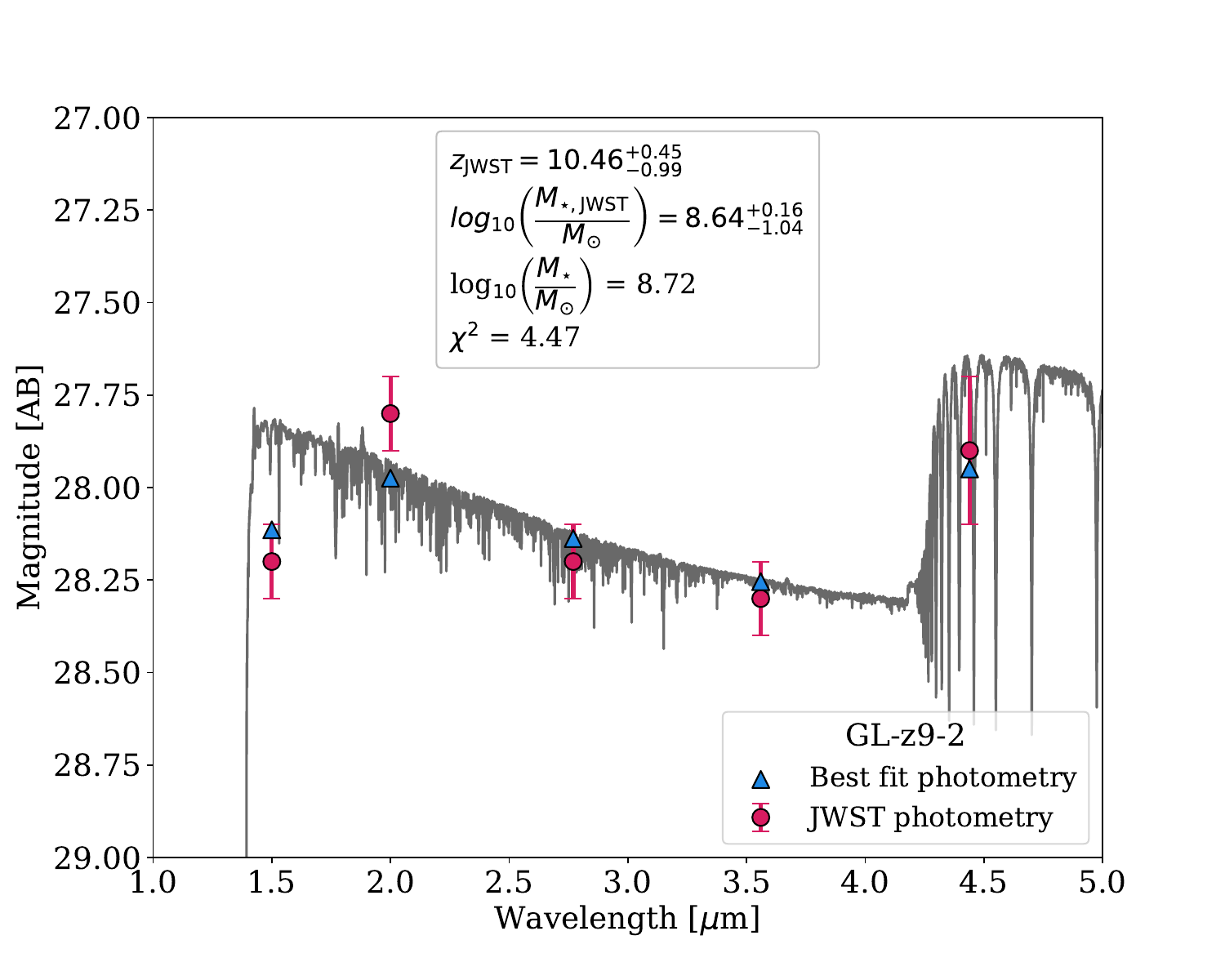}
      \caption{}
      \label{fig:GL-z9-2}
    \end{subfigure}%
    \begin{subfigure}{\columnwidth}
      \centering
      \includegraphics[width=\columnwidth]{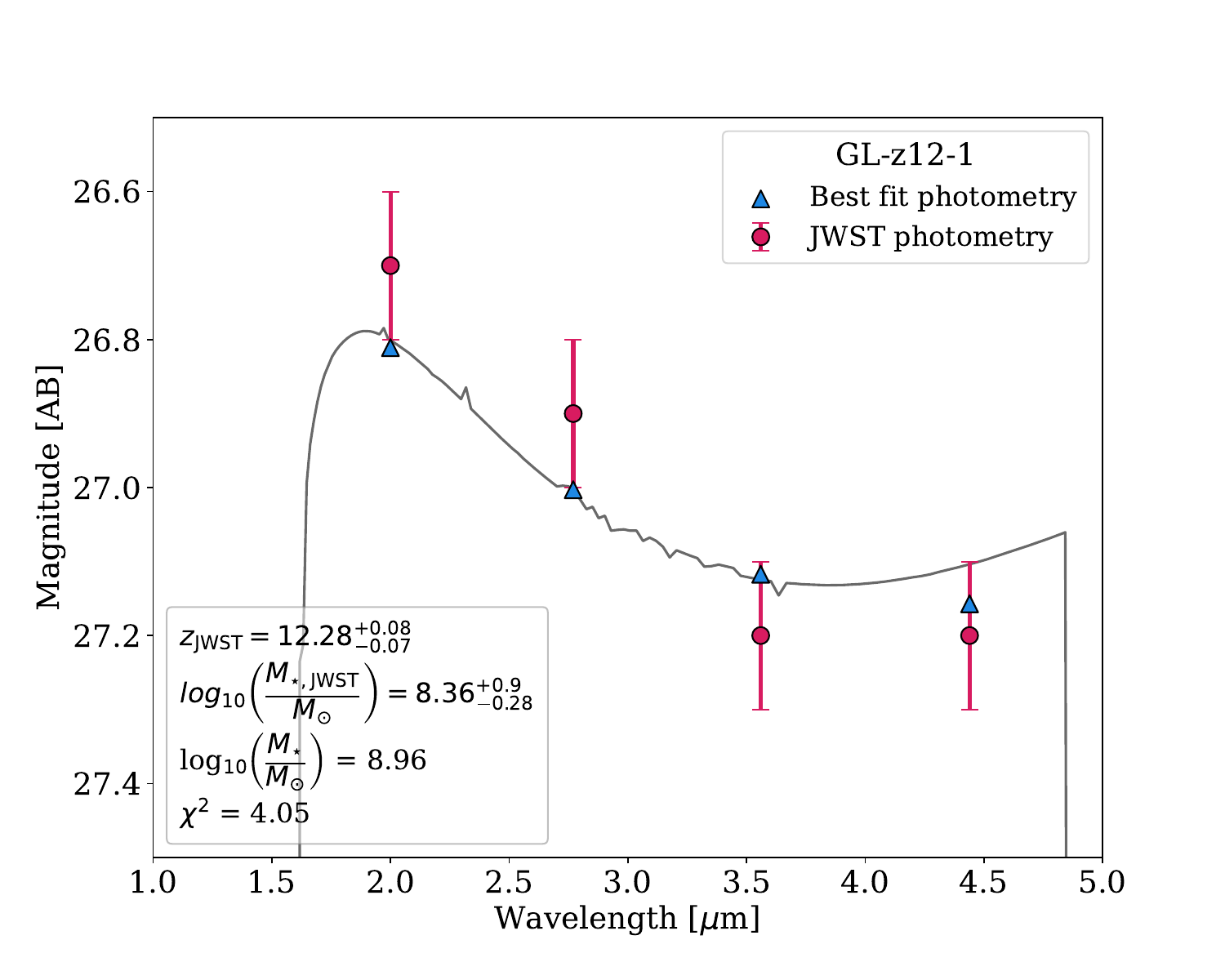}
      \caption{}
      \label{fig:GL-z12-1}
    \end{subfigure}
    \begin{subfigure}{\columnwidth}
      \centering
      \includegraphics[width=\columnwidth]{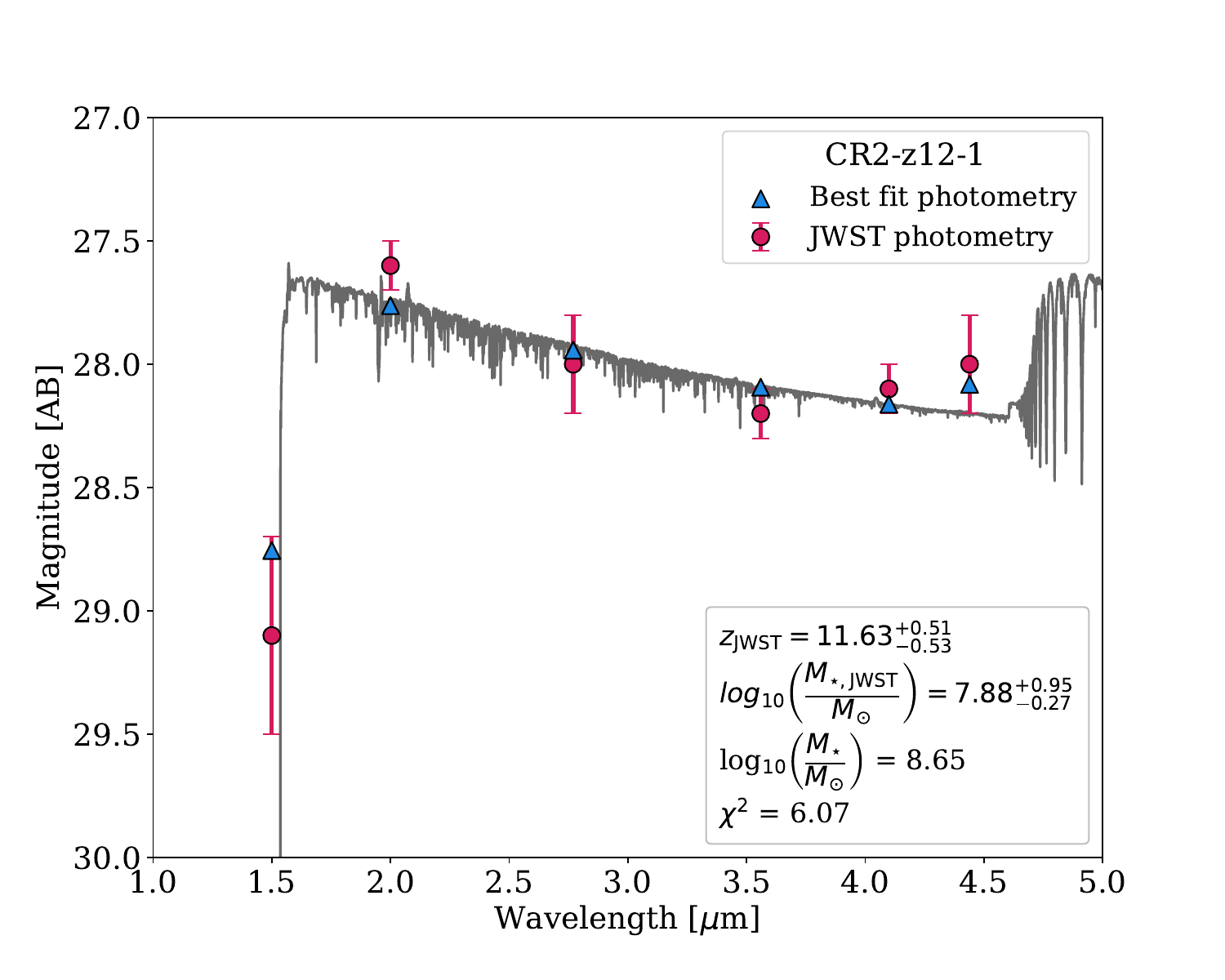}
      \caption{}
      \label{fig:CR2-z12-1}
    \end{subfigure}
    \begin{subfigure}{\columnwidth}
      \centering
      \includegraphics[width=\columnwidth]{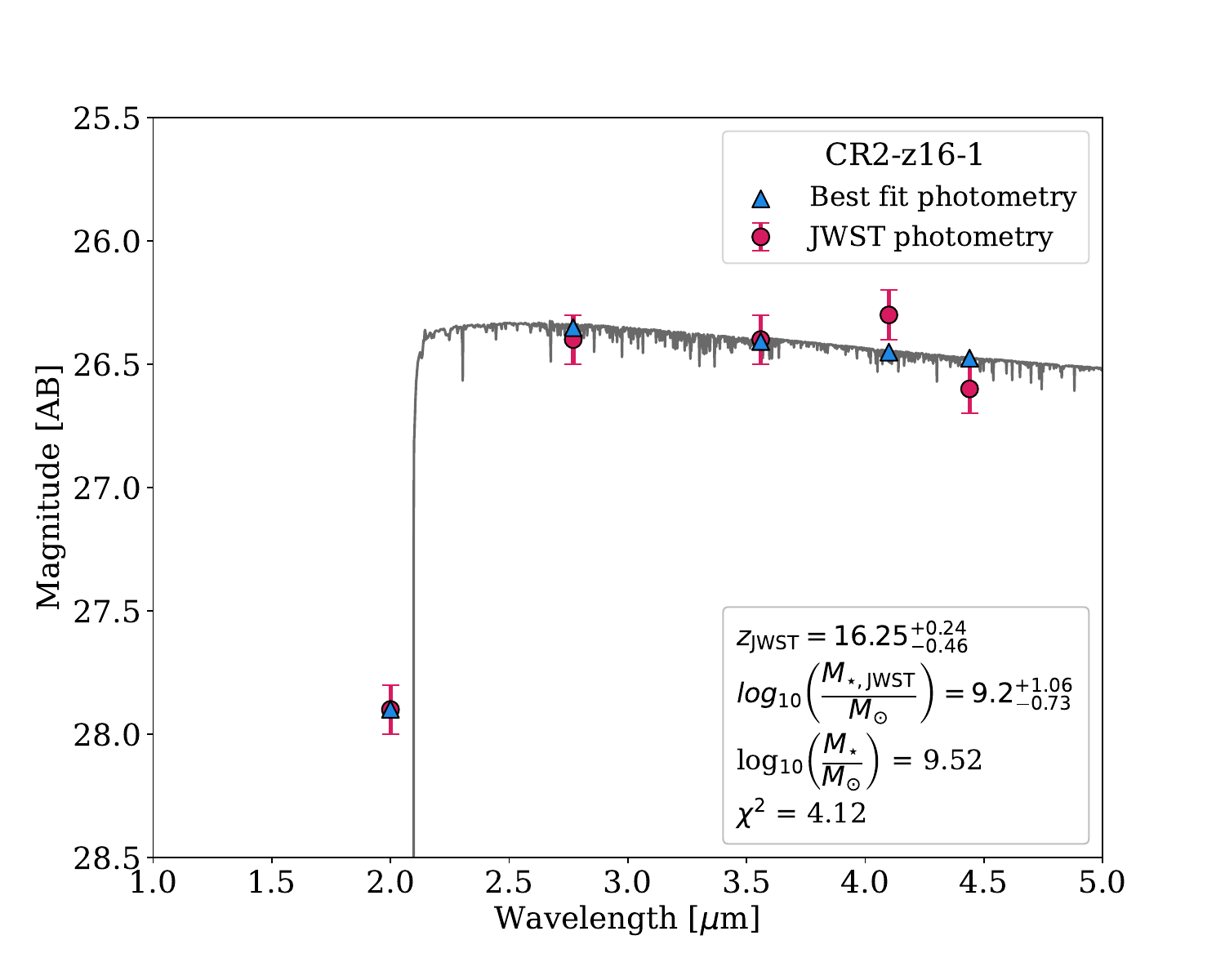}
      \caption{}
      \label{fig:CR2-z16-1}
    \end{subfigure}
    \caption{Fitted SED given the best-fit models for galaxies (a) GL-z9-2, (b) GL-z12-1, (c) CR2-z12-1, and (d) CR2-z16-1. Blue triangles indicate the measured photometry from our fitted spectra and the pink circles are the observed photometry values and errors. Annotated are the measured redshifts and stellar masses from H22, as well as our stellar mass given the best-fit halo mass and our $\chi^{2}$ value.}
    \label{fig:best_spectra_3}
\end{figure*}

A useful property about this approach is that we can estimate a typical star formation history of the galaxies, beginning at its onset, with any scatter being captured by the MCMC.  The full star formation history for each galaxy and their best-fit model can be seen in Figures \ref{fig:adams_history}, \ref{fig:harikane_history} and \ref{fig:all_stellar}. The best fitting spectra for each galaxy is shown in Figures \ref{fig:best_spectra_1}, \ref{fig:best_spectra_2}, and \ref{fig:best_spectra_3}.  We present the best-fit and median parameters for the best-fit model for all galaxies in Tables \ref{tab:best_fits} and \ref{tab:median_params}, respectively.

All galaxies follow a somewhat similar halo growth history, although for all halos, the limit on the SFE can be seen limiting the stellar mass when there is an abrupt knee in the SFR, stellar mass, and SFE plots. The most massive progenitor of the A22 and H22 galaxies reaches the atomic cooling limit, where galaxy formation commences, at redshifts $z = 20-30$ when the universe's age is less than 200~Myr.  Pop III star formation in the progenitors may occur as early as $z \sim 40$ when their halo masses reach $\sim 10^6 \Ms$ as these rare objects undergo an early and rapid assembly.  The galaxies from H22 generally exist at higher redshifts than the A22 galaxies, resulting in shorter star formation sequences with the oldest SSPs having ages of 100--200 Myr and a lower, less diverse metallicity distribution as compared to the lower redshift galaxies from A22. The oldest stars in the A22 galaxies show an age range of 150--300 Myr, shown in Figure \ref{fig:all_stellar}. In general, all galaxies are metal poor, as expected given their stellar masses and high redshifts \citep{Torrey19}.

\subsection{Galaxy Groupings}
There are some important comparisons to make across all the galaxies we modelled. In this section, we group the galaxies based on different aspects of the fitting results. 

\textit{Quality of the spectral fitting:} The best measure of how well our models fit the photometry is $\chi^2$. Objects GL-z9-1 and 6878 have the lowest $\chi^2$ values, with their best-fit models having $\chi^2 < 0.7$. For these objects, the model magnitudes in every filter all lie within the error bars of the photometric magnitudes from A22 and H22 (see Table \ref{tab:magnitudes_delta}). For the other objects, the model magnitudes in at least one filter lie outside the photometric standard error from A22 and H22 and their best-fit $\chi^2$ ranges from 2.28 to 9.55 except for 10234. These objects fit the photometry well in nearly every filter. 10234 has the worst fit by far with $\chi^2 > 78$ for all models. Our pipeline was not able to reproduce the photometry of this galaxy. This is most likely because F200W is quite bright and F350W is quite dim. This means the spectrum must have a steep negative slope in the 2 -- 3.5 $\mu$m range. Our models were not able to find a spectrum with a steep enough slope while also matching filters F150W and F444W. 

We find that for seven of the twelve objects (10234, 1514, 1696, 2462, 2779, 6878, and GL-z9-2) models A\_yggdrasil\_sin and B\_yggdrasil\_sin fit significantly worse than the models which use BPASS. This is typically because the Yggdrasil spectra do not have a pronounced Balmer break like the BPASS spectra do. This makes it difficult for the Yggdrasil models to fit the photometry of F444W at $z \leq 11$. For the higher redshift galaxies, CR2-z12-1 and CR2-z16-1, the BPASS models are still best but $\chi^2$ for the Yggdrasil models is not significantly larger. While the Yggdrasil models typically do not perform as well as the BPASS models, there are two objects (6115 and GL-z12-1) for which the Yggdrasil spectra is the best model; however, in 6115 the AGN component dominates over the stellar one so we cannot conclude that this suggests a top-heavy IMF in this object. Conversely, the stellar component dominates the spectrum of GL-z12-1 so this object may have a top-heavy IMF. We will further discuss these two objects and the IMF variations in Sections \ref{sec:ind_objects} and \ref{sec:variations}.

\textit{Stellar Mass:} The best-fit parameters from our MCMC analysis show that six of the objects have higher best-fit stellar masses than predicted by A22 and H22. These are objects 10234, 1696, 2462, CR2-z12-1, CR2-z16-1, and GL-z12-1. Of these, CR2-z12-1 shows the largest difference in stellar mass than the value from the literature. The best-fit stellar mass of CR2-z12-1 is $\textrm{log}_{10} (\mathrm{M}_{\star} / \Ms) = 8.65$ compared to $\textrm{log}_{10} (\mathrm{M}_{\star} / \Ms) = 7.88_{-0.27}^{0.95}$ from H22. Although this difference is large, our value is still within the H22 error bars. This is true for all H22 objects. Even when our stellar masses differ greatly from the H22 values, the best-fit stellar masses are all within the error bars from H22. 

Four of the objects have comparable stellar masses to the values from A22 and H22. These are 1514, 2779, 6878, and GL-z9-2. All of these objects have $\big|\textrm{log}_{10} (\mathrm{M}_{\star, \mathrm{model}} / \mathrm{M}_{\star, \mathrm{JWST}})\big| < 0.22$. Of these, 2779's stellar mass is closest to the value from A22 with $\textrm{log}_{10} (\mathrm{M}_{\star, \mathrm{model}} / \mathrm{M}_{\star, \mathrm{JWST}}) \approx -0.03$. For objects 1514, 2779, and 6878, their stellar masses, while comparable, are still lower than the values from A22 and H22. Objects 1514 and 6878 may have lower stellar masses because the spectrum is not completely dominated by the stellar component due to the presence of an AGN.

Object 6115 shows the largest difference in stellar mass. The stellar mass of 6115 is $\textrm{log}_{10} (\mathrm{M}_{\star} / \Ms) = 7.53$ compared to $\textrm{log}_{10} (\mathrm{M}_{\star} / \Ms) = 8.40$ from A22. This enormous difference is because the spectrum of 6115 is completely dominated by the AGN component. With a large $f_\mathrm{BH}$, the AGN spectrum fits the photometry of 6115 well with $\chi^2 = 2.28$. This object will be discussed in more detail in Section \ref{sec:ind_objects}.

\textit{Component Spectra and Black Hole Fractions:} We find that the spectra are dominated by the stellar component for most of the objects in our sample. This is the case for 10234, 1696, 2462, 2779, CR2-z12-1, CR2-z16-1, GL-z12-1, and GL-z9-2. For these objects, $\log_{10}(f_\mathrm{BH})$ is near the minimum value of --4, indicating they do not have a bright AGN, or if there is an AGN, it is obscured. There are three objects, 1514, 6878, and GL-z9-1, which have a noticeable contribution from an AGN. These models have larger $\log_{10}(f_\mathrm{BH})$ values ranging from --2.79 to --2.65. Object 6115 is much different from all other objects. It is the only one whose spectrum is completely dominated by a bright AGN with $\log_{10}(f_\mathrm{BH}) = -1.04$ because both BPASS and Yggdrasil stellar models could not provide a good fit to the photometry.

\begin{table*}
	\caption{Best-fit model parameters} \label{tab:best_fits}
	\renewcommand{\arraystretch}{1.2}
	\resizebox{\textwidth}{!}{%
		\begin{tabular}{|l|c|c|c|c|c|c|c|c|}
			\hline \hline
			Galaxy & Model & $\log_{10}(\mathrm{M_{halo}}/\mathrm{M}_\odot$) & $\log_{10}(\mathrm{M_{\star}}/\mathrm{M}_\odot$) & $\log_{10}(f_\mathrm{BH}$) & $\delta_\mathrm{AGN}$ & $\tau_\mathrm{V}$ & $\alpha$ & $\chi^2$ \\ \hline
			10234 & B\_chab100\_bin & 10.09 & 8.44 & --4.00 & --0.94 & 0.00 & 0.80 & 78.75 \\
			1514 & A\_chab100\_sin & 11.10 & 9.58 & --2.65 & --0.97 & 1.07 & -- & 6.35 \\
			1696 & A\_chab100\_sin & 10.68 & 9.16 & --3.99 & --0.99 & 0.24 & -- & 5.35 \\
			2462 & A\_chab100\_sin & 11.47 & 9.95 & --3.99 & --0.98 & 0.76 & -- & 9.55 \\
			2779 & B\_chab100\_bin & 10.37 & 8.67 & --3.95 & --0.80 & 0.19 & 0.40 & 7.04 \\
			6115 & A\_yggdrasil\_sin & 9.70 & 7.53 & --1.04 & --0.84 & 0.00 & -- & 2.28  \\
			6878 & A\_chab100\_sin & 10.71 & 9.18 & --2.79 & --0.88 & 0.41 & -- & 0.66 \\
			CR2-z12-1 & B\_chab100\_bin & 10.18 & 8.65 & --3.82 & 0.73 & 0.02 & 0.41 & 6.07 \\
			CR2-z16-1 & A\_chab100\_sin & 11.04 & 9.52 & --3.53 & --0.77 & 0.36 & -- & 4.12 \\
			GL-z12-1 & B\_yggdrasil\_sin & 10.48 & 8.96 & --3.96 & --0.93 & 0.00 & 0.44 & 4.05 \\
			GL-z9-1 & A\_chab100\_bin & 10.77 & 9.25 & --2.71 & --0.09 & 0.36 & -- & 0.43 \\
			GL-z9-2 & B\_chab100\_bin & 10.34 & 8.72 & --3.91 & --0.81 & 0.01 & 0.42 & 4.47 \\
			\hline
		\end{tabular}
	} 
\end{table*}
\begin{table*}
	\caption{Median of the parameter distributions and their uncertainties for the best-fit models} \label{tab:median_params}
	\renewcommand{\arraystretch}{1.2}
	\resizebox{\textwidth}{!}{%
		\begin{tabular}{|l|c|c|c|c|c|c|c|}
			\hline \hline
			Galaxy & Model & $\log_{10}(\mathrm{M_{halo}}/\mathrm{M}_\odot$) & $\log_{10}(\mathrm{M_{\star}}/\mathrm{M}_\odot$) & $\log_{10}(f_\mathrm{BH}$) & $\delta_\mathrm{AGN}$ & $\tau_\mathrm{V}$ & $\alpha$ \\ 
			\hline
			10234 & B\_chab100\_bin &  $10.11$\raisebox{0.5ex}{\tiny$^{+0.02}_{-0.02}$} &  $8.50$\raisebox{0.5ex}{\tiny$^{+0.04}_{-0.05}$} &  $-3.65$\raisebox{0.5ex}{\tiny$^{+0.22}_{-0.32}$} &  $-0.44$\raisebox{0.5ex}{\tiny$^{+0.38}_{-0.66}$} &  $0.02$\raisebox{0.5ex}{\tiny$^{+0.01}_{-0.02}$} &  $0.73$\raisebox{0.5ex}{\tiny$^{+0.07}_{-0.05}$}\\
			1514 & A\_chab100\_sin &  $11.08$\raisebox{0.5ex}{\tiny$^{+0.10}_{-0.09}$} &  $9.56$\raisebox{0.5ex}{\tiny$^{+0.10}_{-0.09}$} &  $-3.28$\raisebox{0.5ex}{\tiny$^{+0.42}_{-0.44}$} &  $-0.10$\raisebox{0.5ex}{\tiny$^{+0.55}_{-0.63}$} &  $0.99$\raisebox{0.5ex}{\tiny$^{+0.08}_{-0.09}$} &  -- \\
			1696 & A\_chab100\_sin &  $10.67$\raisebox{0.5ex}{\tiny$^{+0.04}_{-0.04}$} &  $9.15$\raisebox{0.5ex}{\tiny$^{+0.04}_{-0.04}$} &  $-3.56$\raisebox{0.5ex}{\tiny$^{+0.27}_{-0.34}$} &  $-0.15$\raisebox{0.5ex}{\tiny$^{+0.55}_{-0.69}$} &  $0.25$\raisebox{0.5ex}{\tiny$^{+0.03}_{-0.04}$} &  -- \\
			2462 & A\_chab100\_sin &  $11.44$\raisebox{0.5ex}{\tiny$^{+0.07}_{-0.07}$} &  $9.92$\raisebox{0.5ex}{\tiny$^{+0.07}_{-0.07}$} &  $-3.59$\raisebox{0.5ex}{\tiny$^{+0.26}_{-0.34}$} &  $-0.09$\raisebox{0.5ex}{\tiny$^{+0.59}_{-0.69}$} &  $0.74$\raisebox{0.5ex}{\tiny$^{+0.07}_{-0.07}$} &  -- \\
			2779 & B\_chab100\_bin &  $10.35$\raisebox{0.5ex}{\tiny$^{+0.04}_{-0.04}$} &  $8.63$\raisebox{0.5ex}{\tiny$^{+0.09}_{-0.07}$} &  $-3.43$\raisebox{0.5ex}{\tiny$^{+0.34}_{-0.40}$} &  $-0.00$\raisebox{0.5ex}{\tiny$^{+0.60}_{-0.62}$} &  $0.29$\raisebox{0.5ex}{\tiny$^{+0.09}_{-0.09}$} &  $0.55$\raisebox{0.5ex}{\tiny$^{+0.10}_{-0.13}$}\\
			6115 & A\_yggdrasil\_sin &  $10.12$\raisebox{0.5ex}{\tiny$^{+0.05}_{-0.03}$} &  $8.39$\raisebox{0.5ex}{\tiny$^{+0.10}_{-0.07}$} &  $-3.19$\raisebox{0.5ex}{\tiny$^{+0.48}_{-0.81}$} &  $-0.30$\raisebox{0.5ex}{\tiny$^{+0.53}_{-0.73}$} &  $0.07$\raisebox{0.5ex}{\tiny$^{+0.04}_{-0.06}$} &  -- \\
			6878 & A\_chab100\_sin &  $10.70$\raisebox{0.5ex}{\tiny$^{+0.05}_{-0.05}$} &  $9.17$\raisebox{0.5ex}{\tiny$^{+0.05}_{-0.05}$} &  $-3.33$\raisebox{0.5ex}{\tiny$^{+0.38}_{-0.40}$} &  $-0.10$\raisebox{0.5ex}{\tiny$^{+0.56}_{-0.65}$} &  $0.37$\raisebox{0.5ex}{\tiny$^{+0.04}_{-0.05}$} &  -- \\
			CR2-z12-1 & B\_chab100\_bin &  $10.21$\raisebox{0.5ex}{\tiny$^{+0.10}_{-0.12}$} &  $8.69$\raisebox{0.5ex}{\tiny$^{+0.14}_{-0.12}$} &  $-3.36$\raisebox{0.5ex}{\tiny$^{+0.39}_{-0.50}$} &  $-0.16$\raisebox{0.5ex}{\tiny$^{+0.53}_{-0.66}$} &  $0.21$\raisebox{0.5ex}{\tiny$^{+0.11}_{-0.12}$} &  $0.62$\raisebox{0.5ex}{\tiny$^{+0.12}_{-0.11}$}\\
			CR2-z16-1 & A\_chab100\_sin &  $10.92$\raisebox{0.5ex}{\tiny$^{+0.23}_{-0.20}$} &  $9.40$\raisebox{0.5ex}{\tiny$^{+0.23}_{-0.20}$} &  $-2.99$\raisebox{0.5ex}{\tiny$^{+0.60}_{-0.72}$} &  $-0.12$\raisebox{0.5ex}{\tiny$^{+0.56}_{-0.66}$} &  $0.29$\raisebox{0.5ex}{\tiny$^{+0.16}_{-0.16}$} &  -- \\
			GL-z12-1 & B\_yggdrasil\_sin &  $10.44$\raisebox{0.5ex}{\tiny$^{+0.10}_{-0.11}$} &  $8.91$\raisebox{0.5ex}{\tiny$^{+0.10}_{-0.11}$} &  $-3.33$\raisebox{0.5ex}{\tiny$^{+0.40}_{-0.51}$} &  $-0.22$\raisebox{0.5ex}{\tiny$^{+0.50}_{-0.66}$} &  $0.09$\raisebox{0.5ex}{\tiny$^{+0.06}_{-0.08}$} &  $0.60$\raisebox{0.5ex}{\tiny$^{+0.12}_{-0.12}$}\\
			GL-z9-1 & A\_chab100\_bin &  $10.84$\raisebox{0.5ex}{\tiny$^{+0.25}_{-0.17}$} &  $9.31$\raisebox{0.5ex}{\tiny$^{+0.25}_{-0.17}$} &  $-2.99$\raisebox{0.5ex}{\tiny$^{+0.59}_{-0.67}$} &  $-0.10$\raisebox{0.5ex}{\tiny$^{+0.55}_{-0.63}$} &  $0.39$\raisebox{0.5ex}{\tiny$^{+0.15}_{-0.14}$} &  -- \\
			GL-z9-2 & B\_chab100\_bin &  $10.27$\raisebox{0.5ex}{\tiny$^{+0.07}_{-0.07}$} &  $8.57$\raisebox{0.5ex}{\tiny$^{+0.14}_{-0.15}$} &  $-3.30$\raisebox{0.5ex}{\tiny$^{+0.42}_{-0.52}$} &  $-0.17$\raisebox{0.5ex}{\tiny$^{+0.53}_{-0.65}$} &  $0.18$\raisebox{0.5ex}{\tiny$^{+0.10}_{-0.14}$} &  $0.64$\raisebox{0.5ex}{\tiny$^{+0.11}_{-0.09}$}\\
			\hline
		\end{tabular}
	}
	\parbox[t]{\textwidth}{\vspace{1em} \textit{Notes:} The upper and lower uncertainties are computed from the 84th and 15th percentiles of the distributions respectively.}
\end{table*}

\subsection{Individual Objects} \label{sec:ind_objects}

While many of the objects can be grouped by their similarities as shown previously, there are a few objects that stand out. Here we describe the most interesting objects in more detail. 

\textbf{6115:} This object has a photometric redshift of $z = 10.94_{-0.15}^{+0.12}$ and a stellar mass of $\textrm{log}_{10} (\mathrm{M}_{\star} / \Ms) = 8.4$ (A22). The best-fits from all models predict lower stellar masses than A22, with Yggdrasil models predicting larger best-fit stellar masses than the BPASS models by 1 dex, and smaller BH masses also by 1 dex. All $\chi^2$ values are below 2.72 for this galaxy, indicating a good reproduction of photometry. Independent of the model choice in the best fit model, the photometry is reproduced with a massive BH with $\log_{10}(f_\mathrm{BH}) \gtrsim -1$; however the median parameters suggest a more typical central BH with $f_{\rm BH} \simeq 10^{-3}$.

\textit{Best-fit parameters:} We find the spectra that best fits this object comes from model A\_yggdrasil\_sin, with a best-fit stellar mass of $\textrm{log}_{10} (\mathrm{M}_{\star} / \Ms) = 7.53$ (see Figure \ref{fig:6115}). This is the lowest stellar mass of all the objects in our sample. The best-fit parameters for all models are shown in Table \ref{tab:6115} and the best-fit spectra and photometry can be seen plotted in Figure \ref{fig:6115}. Our parameters match the photometry in almost every filter except F277W, which is outside the error bounds of A22 by a magnitude of 0.05. The predicted optical depth in every best-fit model was between 0 and 0.02, with four out of the seven models predicting an optical depth of 0.

\begin{figure*}
    \includegraphics[width=\textwidth]{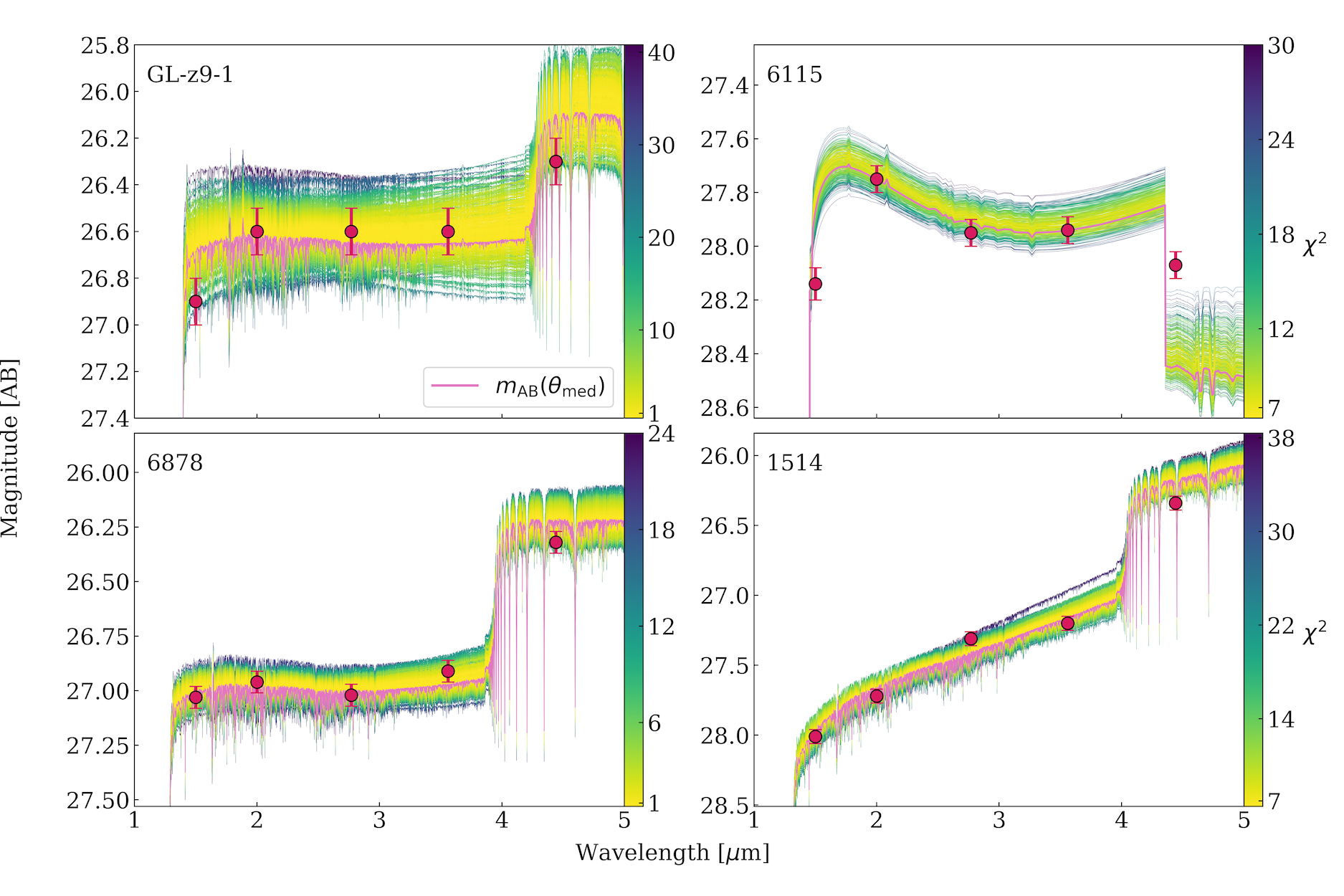}
    \caption{Model spectra colored by $\chi^2$ using the best model from four galaxies: GL-z9-1 with model B\_chab100\_bin (top left), 6115 model A\_yggdrasil\_sin (top right), 6878 model A\_chab100\_sin (bottom left), 1514 model A\_chab100\_sin (bottom right). For each galaxy, the SEDs were generated by randomly sampling 400 sets of model parameters $\theta$ from within the 16th-84th percentile range of each posterior parameter distribution, i.e. within the median +/- standard errors shown in Table \ref{tab:median_params}. The pink line shows the SED computed from the median parameters of the best model. The red markers with error bars show the JWST magnitudes and standard error in filters F150W, F200W, F277W, F365W, and F444W.}
    \label{fig:chi2_spectra}
\end{figure*}

\textit{Median parameters:} We find a median stellar mass of $\textrm{log}_{10}(\mathrm{M}_{\star} / \Ms) = 8.50_{-0.16}^{+0.19}$ from the best model. This larger stellar mass compared to the best-fit value is due to the lower median $\mathrm{log}_{10}(f_\mathrm{BH}) = -3.19_{-1.03}^{+0.49}$. This shows the trade off between a luminous BH and a bright stellar component. The reason for the difference in stellar mass and BH mass is the $f_\mathrm{BH}$ distribution is peaked at around $\mathrm{log}_{10}(f_\mathrm{BH}) \approx -3$ but has a flat tail that extends to larger values where the best fit value is found. The median parameters can be seen in Table \ref{tab:6115}. All models have similar median fits and no significant trend is found. To see a posterior distribution of spectra for the galaxy along with the $\chi^2$ distribution, refer to Figure \ref{fig:chi2_spectra}.

Unlike the other objects, the best-fit corner plot in Figure \ref{fig:corner_6115} shows a noticeable peak in $\delta_\mathrm{AGN}$ while still exploring the full parameter space at the wings of the distribution. This is because when $f_\mathrm{BH}$ is large, the AGN spectrum becomes important and $\delta_\mathrm{AGN}$ can be constrained to fit the photometry. Whereas for other objects with a small $f_\mathrm{BH}$, varying $\delta_\mathrm{AGN}$ has little to no effect on the spectrum. 

\begin{figure*}
    \includegraphics[width=\textwidth]{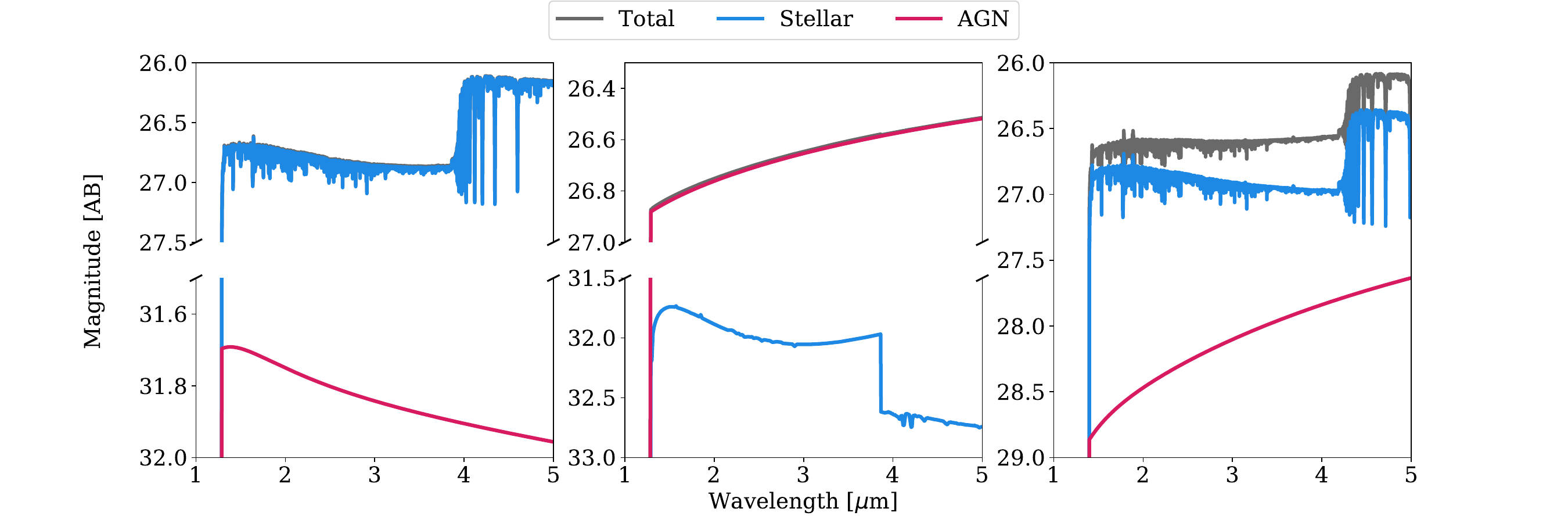}
    \caption{A comparison of SED models showing a stellar dominated SED, an AGN dominated SED, and an SED where the stellar and AGN components are both contributing to the total. The gray lines indicate the total SED, the blue lines indicate the stellar spectrum, and the pink lines indicate the AGN spectrum. Left panel: SED generated for galaxy 1696 from model A\_chab100\_sin. This model shows a completely stellar dominated SED. Middle panel: SED generated for galaxy 1696 for model B\_yggdrasil\_sin. This model shows a completely AGN dominated SED. Right panel: SED generated for galaxy GL-z9-1 for the best fitting model A\_chab100\_bin. This model shows comparable contributions from the stellar and AGN components to the total SED.}
    \label{fig:agn_star_dom}
\end{figure*}

\textbf{GL-z9-1:} 
This galaxy stands out from the rest in our sample because it is the only one that has comparable stellar and AGN luminosities for the non-Yggdrasil spectra. While the stellar component is still brighter, the AGN spectrum makes a significant contribution. This is shown in the rightmost panel of Figure \ref{fig:agn_star_dom}. It is also the object with the best fit to the photometry with $\chi^2 = 0.43$. This object has a photometric redshift of $z = 10.49_{-0.72}^{+0.53}$ and a stellar mass of $\textrm{log}_{10} (\mathrm{M}_{\star} / \Ms) = 9.04_{-0.74}^{+0.61}$ (H22). For all BPASS models, the best-fit parameters and the median parameters give a slightly larger stellar mass than the value from H22. The best-fit spectra match the JWST photometry in every filter. Our magnitudes are all within the error of the photometric magnitudes from H22. This is shown in Figure \ref{fig:GL-z9-1} and Table \ref{tab:magnitudes_delta}. 

\textit{Best-fit parameters:} We find the spectra that best fits this object comes from model A\_chab100\_bin, with a best-fit stellar mass of $\textrm{log}_{10} (\mathrm{M}_{\star} / \Ms) = 9.25$ (see Figure \ref{fig:GL-z9-1}). The best-fit parameters for all models can be seen in the table \ref{tab:GL-z9-1}. All BPASS models fit this galaxy extremely well with $\chi^2$ ranging from 0.43 to 0.45 so there is nothing particularly special about model A\_chab100\_bin. Models using the Yggdrasil spectra for a top-heavy IMF have the largest $\chi^2 \approx 3.6$. Unlike the best models for most other galaxies, GL-z9-1 consistently has a relatively large values of $\log_{10}(f_\mathrm{BH})$ ranging from --2.83 to --2.20.

\textit{Median parameters:} We find a median stellar mass of $\textrm{log}_{10} (\mathrm{M}_{\star} / \Ms) = 9.31_{-0.17}^{+0.25}$ from the best model. The median parameters can be seen in the Table \ref{tab:GL-z9-1}. Compared to the best-fit parameters, the median parameters give larger stellar masses and smaller BH fractions. As with the best-fit parameters, the Yggdrasil models have the largest BHs. As expected, the spectra from the median parameters have larger $\chi^2$ than the best-fit spectra.

The corner plot for GL-z9-1 in Figure \ref{fig:corner_GL-z9-1} has a couple of interesting features. The $M_\mathrm{halo}$ vs. $f_\mathrm{BH}$ plot shows a negative slope with a slight tail that extends to larger $f_\mathrm{BH}$. This shows the trade-off between BH mass and halo mass. To match the photometry, if the AGN is brighter, the stellar component must be dimmer. This results in wider distributions for both $f_\mathrm{BH}$ and $M_\mathrm{halo}$ than when only one component dominates. The $\delta_\mathrm{AGN}$ distribution is skewed toward negative values with a peak at --0.5. Since $f_\mathrm{BH}$ is larger, the AGN component of the spectra is non-negligible, and therefore $\delta_\mathrm{AGN}$ becomes an important parameter. The peak near --0.5 is important because this is consistent with measurements of $\delta_\mathrm{AGN}$ from observations \citep{Yang22}. 

The top-left panel of Figure \ref{fig:chi2_spectra} shows the distribution of spectra within the range of the median parameters and their standard errors, colored by $\chi^2$, for the best model of GL-z9-1. Compared to the other panels, GL-z9-1's spectra have a wider distribution in magnitudes at low $\chi^2$. This is mainly because the observed NIRCam/F277W magnitude is relatively bright causing bump in the photometry, whereas our spectra are more monotonic in this bandwidth. 

\subsection{Variation on models}
\label{sec:variations}

The corner plots for all models (Appendix \ref{appendix:corner}; Figures \ref{fig:corner_10234} -- \ref{fig:corner_GL-z9-2}) reveal how {\sc emcee} constrains the parameters that best match the data. From this, we can see what parameters are most and least important to the model. 

\textit{Host halo mass}: By far and unsurprisingly, the most important parameter that controls the goodness of the fit to the JWST photometry is the halo mass $\mathrm{M}_{\textrm{halo}}$. As will be discussed shortly, the AGN model cannot match the photometry very well on its own while the BPASS stellar spectrum can. Since the halo mass directly controls the stellar mass and the star formation history, $\mathrm{M}_{\textrm{halo}}$ is the parameter which controls the results the most. This can be seen in the corner plots where $\mathrm{M}_{\textrm{halo}}$, and therefore $\mathrm{M}_{\star}$, typically have very peaked distributions. When the Yggdrasil spectrum is used instead of BPASS, {\sc emcee} has a harder time fitting these spectra to the photometry, and the $\mathrm{M}_{\textrm{halo}}$ distribution can sometimes be doubly peaked. In these cases, the AGN spectra will play a more dominant role in fitting the photometry since the Yggdrasil spectra cannot fit the data very well, which we will discuss in Section \ref{sec:discussion}. 

\textit{Halo Growth Models (A or B)}: The exact history of the halos does not have a significant impact. For most of the galaxies in our sample, $\chi^2$ does not differ greatly between models using growth model A or B, indicating that there is not a strong preference for one halo growth model over the other. This can be seen in the distributions of $\alpha$ in our corner plots, where the $\alpha$ parameter tends to have a fairly broad distribution. The value of $\alpha$ that {\sc emcee} lands on is typically around 0.6, which is the median of our range of $\alpha$. This implies that the most recent star formation is what is most important, as the history of the halo does not play a major role. 

\textit{IMF}: Both Yggdrasil models tend to predict higher BH masses, controlled by the $f_{\textrm{BH}}$ parameter. This results in smaller stellar masses since the AGN spectra tends to dominate the fitting, but the Yggdrasil models also tend to have higher $\chi^2$ values than the BPASS models, indicating a systematic issue between matching Yggdrasil spectra to the observed photometry. Some potential causes of this systematic error are discussed in Section \ref{sec:discussion}. There is no significant trend between BPASS models; both the Chabrier and Standard IMF models produce similar results regardless of which one is chosen as the best-fit model. Their $\chi^2$ values are all consistent with each other and most of the BPASS models are able to fit the spectra with low $\chi^2$, showing good fits to photometry.

\textit{Single vs. Binary Stellar Population}: Only BPASS models both single and binary stellar populations, while Yggdrasil only models single stellar populations. Of the 10 galaxies that had best-fit parameters predicted by a BPASS IMF model, six of them have binary stellar population spectra while four have single stellar population spectra. For each galaxy, BPASS IMFs produce similar parameter values regardless of stellar population.

\textit{$\delta_{\textrm{AGN}}$}: The AGN spectrum does not fit the photometry very well by itself, and therefore, it is not typically used by {\sc emcee} as the dominant fitting spectra. Because of this, the slope of the AGN spectrum, controlled by our parameter $\delta_{\textrm{AGN}}$ does not vary that much, and when $f_{\textrm{BH}}$ is small, $\delta_{\textrm{AGN}}$ does not affect the SED at all so the best-fit value can vary greatly. This will be discussed further in Section \ref{sec:discussion}.

\section{Discussion} \label{sec:discussion}
There are a number of reasons why we may be obtaining higher best-fit stellar masses than those predicted by A22 and H22, given our theoretical model. These may include an insensitivity to the halo growth model, a dominant stellar spectra component, and ill-fitting AGN and top-heavy IMF stellar spectra, and the absence of nebular emission in our spectra. We will explore each of these in this section, and focus individually on the AGN spectra and the IMFs in Sections \ref{sec:agnDiscussion} and \ref{sec:stellarimfDiscussion}, respectively. 

A common theme across many of our models and most of the objects analyzed is that the distributions of the exponential halo growth rate $\alpha$ spans the full range of the priors from 0.4 to 0.8.  The median values of $\alpha$ are typically symmetric about 0.6 (see Tables \ref{tab:10234} -- \ref{tab:GL-z9-2}). This indicates that varying $\alpha$ does not strongly affect the probability of our spectra matching the JWST photometry for a given object. Recall that our halos grow exponentially starting from the redshift when they had a mass of $10^7 \Ms$ until the observed redshift, sampled every 5 Myr. The growth rate does not have a significant effect on the spectra because the most recent bursts of star formation dominate the stellar spectrum. For each object, the final burst alone makes up about 35--55 per cent of the total flux of the galaxy in the NIRCam bandwidth.  Although this approach does not capture the individual galaxy progenitors, as in a method based on merger trees, the numerous mergers of low-mass galaxies, which have a large scatter in their star formation histories \citep[e.g.][]{Xu16_fesc, Gutcke22}, average out, producing a more massive galaxy similar to the median values found in this work.  

As we showed in Section \ref{sec:results}, eight out of the 12 objects had $\log_{10}(f_\mathrm{BH})$ near the minimum value of --4. When the BH is this small, the stellar flux dominates the continuum, and therefore the offset in the AGN slope is not an important parameter. This is why we see wide distributions in $f_\mathrm{BH}$ with large uncertainty. We stress that in these cases, $f_\mathrm{BH}$ values have no significant meaning, regardless of their value. This is not the case for the four other objects where $f_\mathrm{BH}$ is larger, and the AGN contributes a noticeable amount to the SED, especially for 6115 where the AGN component is dominant.

\subsection{AGN Spectra} \label{sec:agnDiscussion}

In order to more accurately model the AGN, we attempted to include realistic AGN spectra from the SKIRTOR database of a dusty torus at different inclination angles. Following the parameters given in Figure 4 in \citet{Stalevski12}, we ran {\sc emcee} with the inclination angle as a free parameter. Since the dust-absorbed BPASS spectra match the shape of the photometry well when scaled by stellar mass, the best-fit parameters still trended toward stellar-dominated spectra by predicting best-fit inclinations greater than 40 degrees, at which point a majority of the UV flux is attenuated into the IR, making AGN contributions negligible within our wavelength range of interest. We therefore decided to use the point-source emission model of AGNs and determine the AGN spectral slope in the UV with the $\delta_{\textrm{AGN}}$ parameter, assuming maximum possible contribution from the AGN.

As mentioned earlier to be consistent with previous studies on these objects, we did not include nebular emission lines from star forming regions and AGN.  Photometric detections of $z \geq 6$ galaxies with HST showed that emission lines redward of the Balmer break can contribute a significant amount of flux to broadband filters \citep[e.g.][]{Smit14}.  Emission-line spectra of \hii{} regions can indicate the likelihood of an AGN based on [O\ {\sc iii}]/H$\beta$ and [N {\sc ii}]/H$\alpha$ ratios as shown in BPT diagrams \citep{bpt} and will provide stronger evidence than color-color diagnostics of the existence of an AGN. \citet{Katz_2022} investigated the emission lines of galaxies identified in the JWST Early Release Observations and have used these ratios to determine whether these galaxies were likely to host AGNs. Given a theoretical model that is able to incorporate these elements into the spectra, we may be able to more easily fit a rea listic AGN SED along with our realistic stellar SED and determine whether our high-redshift candidates host AGNs. However, this is highly dependent on the redshift of these galaxies. In practice, if the redshifts from A22 and H22 used in this work are correct, i.e. these galaxies are actually all at $z > 9$, the emission lines mentioned above would be redshifted to wavelengths greater than what is detectable by NIRCam. Therefore these emission lines would have no impact on the photometry and including them in our spectra would not affect our results.

Given that we only have a few best-fit models where an AGN possibly contributes to the continuum, we have too small of a sample to make any general predictions about the nature of the BHs and their seeding \citep{Volonteri22}.  If the AGN contribution is subdominant to the stellar component, color-color diagnostics alone are not sufficient to ascertain the existence of a central BH, and nebular emission lines detected with NIRSpec \citep{Curtis22} will be needed to determine their mass and accretion rate estimates and compare them with their host galaxy properties \citep[e.g.][]{Volonteri22}.  With this additional information, constraints can be placed on the galaxy-BH relation during the Epoch of Reionization and possibly BH seeding mechanisms.

\subsection{Stellar Spectra and IMFs} \label{sec:stellarimfDiscussion}

We explored the possibility of an overabundance of massive stars, i.e. a top-heavy primordial IMF, by using Yggdrasil \citep{yggdrasil}. As discussed in Section \ref{sec:SS}, metal-free stars are thought to be generally massive because of the weak cooling rates associated with molecular hydrogen and from a lack of abundant transitions found in metals. The galaxies that had reasonable fits with Yggdrasil (6115 and GL-z12-1) may be indicative of a top-heavy IMF at high redshifts, however we conclude this is generally not the case for the following reason.

No Balmer break exists in the younger stellar populations in Yggdrasil, but it is still not significant in the older available populations. This may contribute to the poor matching of Yggdrasil spectra by {\sc emcee}, as a bump in emission at approximately 3600 \AA\ in the rest-frame is evident in the observed photometry. While the Pop II stellar populations available in Yggdrasil may have modelled such features more accurately, they do not have a top-heavy IMF and would have been a redundant model, given what we have available in BPASS. There may not be an accurate enough top-heavy IMF database publicly available to us to draw a conclusion whether this discrepancy in spectrum matching is due to the choice of IMF or a systematic error in the chosen spectra. 

Compared to Yggdrasil, BPASS models more metal-enriched binary and single stellar populations and contains a strong Balmer break in its spectra. These features in conjunction with the applied absorption and attenuation functions allowed BPASS SEDs to match well with observed photometry. Almost every Yggdrasil model produced an AGN-dominated SED as shown in the middle panel of Figure \ref{fig:agn_star_dom}. Alternately, almost every BPASS model produced a stellar-dominated SED, except for 6115 which is completely AGN-dominated.

As mentioned before, we did not include nebular emission (lines or continuum) in our SEDs. We recognize that this is a potentially important component of the spectra and we plan to include it in a future version of this work. To do this, we will process the stellar spectra with the radiative transfer code CLOUDY, similarly to how the Flexible Stellar Population Synthesis (FSPS) code does. \cite{Byler17} present useful information on the effect of the nebular emission and how they incorporate it with FSPS using CLOUDY. They found that in the optical and NIR, nebular emission can contribute 30-50 per cent of the total flux at 10 Myr. Given that most of the optical band and NIR (4000 \AA -- 5.3 $\mu m$) is redshifted beyond the NIRCam sensitivity, this would not have a significant effect on our results. At ages greater than 10 Myr, the nebular component for SSPs is negligible (see Figure 12 therein). However, for a constant star formation rate, the nebular emission contributes approximately 20 per cent or less to the total flux at wavelengths of 900-4000 \AA\ (see Figure 13 therein). By not including nebular emission with the BPASS spectra, we are slightly underestimating the flux. Correcting for this would increase the flux for a given stellar mass, and therefore reduce the stellar mass needed to fit the photometry. This may be the primary reason we see slightly higher stellar masses than A22 and H22.

\section{Conclusions} \label{sec:conclusions}

We have developed a galaxy SED model tailored to matching high redshift photometry from JWST.  We use this model to fit spectra to the JWST/NIRCam photometry of 12 high-redshift candidate galaxies from the CEERS and GLASS surveys. In comparison with the early JWST results we compared against (e.g. A22, H22, N22), our model add a few possibly important components to the spectral modeling.

First, we utilize high-redshift star formation history models extending to $z = 15$, which agree well with cosmological simulations. Second, we include AGN spectra in our fitting while many other works assume a stellar-only SED. Third, we test multiple IMFs including a top-heavy IMF that uses the stellar spectra of Pop III.2 stars. The other two IMFs we tested include stellar spectra both with and without stellar binaries. These additions represent a more realistic environment for these very young galaxies assembling during the Epoch of Reionization and could differ from more evolved and massive galaxies at lower redshifts.

Our goal was to determine if the interplay of these features, particularly the inclusion of an AGN, could explain the unexpectedly large stellar masses and bright UV luminosity of galaxies from early JWST photometry. We found the following common trends in our sample:
\begin{enumerate}
    \item Our pipeline was able to fit the JWST photometry moderately well with values of reduced $\chi^2$ comparable to other works. Although almost all our SEDs were dominated by the stellar component while the AGN was of little importance.
    \item Our pipeline was able to fit the JWST photometry moderately well with values of reduced $\chi^2$ comparable to other works. Although almost all our SEDs were dominated by the stellar component while the AGN was of little importance.    
    \item Only the top-heavy IMF given by the Yggdrasil models produced significant AGN contributions, but generally fit the photometry worse than the standard IMF for regular stars. The only exceptions were GL-z12-1 and 6115, where the top-heavy IMF was the best model, but only marginally. In general, models with a top-heavy IMF consistently produced lower stellar masses and larger BH fractions.
    \item There was no significant variation in halo mass, stellar mass, BH mass, or $\chi^2$ when we included stellar binaries. This could indicate that the BPASS spectra do not differ significantly in our wavelength range of interest whether it is a single or binary stellar population. Additionally, because newly formed stars dominate the UV spectrum, the exponential halo growth rate was of little importance.
    \item Ultimately, we predict larger (smaller) best-fit stellar masses than A22 and H22, yet still within the uncertainties, for eight (four) of the 12 galaxies.  In these four cases, this could be caused by the presence of an AGN component, but emission line diagnostics will be needed for their confirmation.
\end{enumerate}

Although we could not confirm our initial hypothesis of an AGN component that moderates the stellar mass estimates, our approach can place constraints on the host halo growth history, star formation history, stellar populations, and central BH mass of these $z \gsim 10$ galaxies observed with JWST in its first data release.  While photometry of these distant galaxies places large uncertainties on the intrinsic galactic properties, the first JWST/NIRSpec detections of $z \gsim 10$ galaxies are just being published, placing stronger constraints on their redshifts, stellar masses, metallicities, and dust attenuation \citep[e.g.][]{Curtis22}.  The first six months of JWST high-redshift observations have been truly groundbreaking, and we, as a community, will continue to push the high-redshift boundary further uncovering the nature and origin of the first galaxies and their central BHs.

\section*{Acknowledgements}

The authors thank Yuichi Harikane and Hakim Atek for providing additional photometric data that were not present in their publications.  The authors thank Sandrine Ferrans and Rohan Srivastava for their contributions to an earlier version of this work.  JHW thanks Raffaella Schneider, Rosa Valiante, and Marta Volonteri for useful discussions.  DS is supported by the NASA FINESST fellowship award 80NSSC20K1540. SS is supported by the NASA FINESST fellowship award 80NSSC22K1589.  This work is supported by NSF grants OAC-1835213 and AST-2108020 and NASA grants 80NSSC20K0520 and 80NSSC21K1053.  The MCMC calculations were performed with NSF's LRAC allocation AST-20007 on the Frontera resource in TACC.  The figures in this paper were constructed with the plotting library {\sc matplotlib} \citep{matplotlib}.

\section*{Data Availability}

 The source code and data underlying this article and the corner plots of the models that are not the best-fit models will be shared on reasonable request to the corresponding author.



\bibliographystyle{mnras}
\bibliography{jwise} 




\appendix

\section{All Model Parameters}
\label{appendix:tables}

In Tables \ref{tab:10234}--\ref{tab:GL-z9-2}, we present our best-fit SED fitting parameters as well as the median parameters and standard error for all models of each object.

\section{Best Model Corner Plots}
\label{appendix:corner}

Figures \ref{fig:corner_10234}--\ref{fig:corner_GL-z9-2} show the corner plots of each galaxy candidate that depicts the posterior parameter distributions for the best models.  Table \ref{tab:magnitudes_delta} lists the apparent magnitudes for the best-fit model parameters and their differences from the observed photometry, along with their uncertainties.

\bsp

\setcounter{table}{0}
\renewcommand{\thetable}{A\arabic{table}}

\begin{table*}
	\caption{Best parameters, median of the parameter distributions and their uncertainties for each model}
    \begin{subtable}{\textwidth}
    \centering
    \renewcommand{\arraystretch}{1.2}
	\resizebox{\textwidth}{!}{%
\begin{tabular}{|l|c|c|c|c|c|c|c|}
	\hline
	    \multicolumn{8}{|c|}{\textbf{10234}} \\  
	    \hline \hline
	    \multicolumn{8}{|c|}{Best Parameters} \\
		\hline 
		Model & $\log_{10}(\mathrm{M_{halo}}/\mathrm{M}_\odot$) & $\log_{10}(\mathrm{M_{\star}}/\mathrm{M}_\odot$) & $\log_{10}(f_\mathrm{BH}$) & $\delta_\mathrm{AGN}$ & $\tau_\mathrm{V}$ & $\alpha$ & $\chi^2$ \\
		\hline
		A\_chab100\_bin & 10.11 & 8.48 & --3.95 & --0.98 & 0.00 & -- & 80.63 \\
		A\_chab100\_sin & 10.12 & 8.51 & --3.97 & --1.00 & 0.00 & -- & 96.31 \\
		A\_135\_100\_bin & 10.14 & 8.55 & --3.99 & --0.98 & 0.00 & -- & 81.35 \\
		B\_chab100\_bin & 10.09 & 8.44 & --4.00 & --0.94 & 0.00 & 0.80 & 78.75 \\
		B\_135\_100\_bin & 10.12 & 8.51 & --3.95 & --0.95 & 0.00 & 0.79 & 79.21 \\
		B\_yggdrasil\_sin & 9.30 & 6.83 & --0.16 & --1.00 & 0.00 & 0.56 & 126.93 \\
		A\_yggdrasil\_sin & 9.29 & 6.80 & --0.14 & --1.00 & 0.00 & -- & 127.20 \\
		\hline		
	\end{tabular}%
	}
	\bigskip
	\resizebox{\textwidth}{!}{%
	\begin{tabular}{|l|c|c|c|c|c|c|c|c|}
	    \hline 
	    \multicolumn{9}{|c|}{Median Parameters} \\
		\hline
		Model & $\log_{10}(\mathrm{M_{halo}}/\mathrm{M}_\odot$) & $\log_{10}(\mathrm{M_{\star}}/\mathrm{M}_\odot$) & $\log_{10}(f_\mathrm{BH}$) & $\delta_\mathrm{AGN}$ & $\tau_\mathrm{V}$ & $\alpha$ & $\chi^2$ & $\chi^2(\theta_{\mathrm{med}})$ \\
		\hline
		A\_chab100\_bin &  $10.11$\raisebox{0.5ex}{\tiny$^{+0.01}_{-0.01}$} &  $8.49$\raisebox{0.5ex}{\tiny$^{+0.02}_{-0.02}$} &  $-3.64$\raisebox{0.5ex}{\tiny$^{+0.23}_{-0.33}$} &  $-0.45$\raisebox{0.5ex}{\tiny$^{+0.37}_{-0.67}$} &  $0.02$\raisebox{0.5ex}{\tiny$^{+0.01}_{-0.02}$} & -- &  $90.94$\raisebox{0.5ex}{\tiny$^{+4.86}_{-8.20}$} &  $87.20$\\
		A\_chab100\_sin &  $10.12$\raisebox{0.5ex}{\tiny$^{+0.01}_{-0.01}$} &  $8.52$\raisebox{0.5ex}{\tiny$^{+0.02}_{-0.02}$} &  $-3.64$\raisebox{0.5ex}{\tiny$^{+0.22}_{-0.34}$} &  $-0.51$\raisebox{0.5ex}{\tiny$^{+0.34}_{-0.63}$} &  $0.01$\raisebox{0.5ex}{\tiny$^{+0.01}_{-0.02}$} & -- &  $106.88$\raisebox{0.5ex}{\tiny$^{+4.97}_{-8.34}$} &  $103.03$\\
		A\_135\_100\_bin &  $10.14$\raisebox{0.5ex}{\tiny$^{+0.01}_{-0.01}$} &  $8.56$\raisebox{0.5ex}{\tiny$^{+0.02}_{-0.02}$} &  $-3.66$\raisebox{0.5ex}{\tiny$^{+0.22}_{-0.32}$} &  $-0.48$\raisebox{0.5ex}{\tiny$^{+0.35}_{-0.65}$} &  $0.02$\raisebox{0.5ex}{\tiny$^{+0.01}_{-0.02}$} & -- &  $91.67$\raisebox{0.5ex}{\tiny$^{+4.82}_{-8.20}$} &  $87.87$\\
		B\_chab100\_bin &  $10.11$\raisebox{0.5ex}{\tiny$^{+0.02}_{-0.02}$} &  $8.50$\raisebox{0.5ex}{\tiny$^{+0.04}_{-0.05}$} &  $-3.65$\raisebox{0.5ex}{\tiny$^{+0.22}_{-0.32}$} &  $-0.44$\raisebox{0.5ex}{\tiny$^{+0.38}_{-0.66}$} &  $0.02$\raisebox{0.5ex}{\tiny$^{+0.01}_{-0.02}$} &  $0.73$\raisebox{0.5ex}{\tiny$^{+0.07}_{-0.05}$} &  $91.88$\raisebox{0.5ex}{\tiny$^{+5.47}_{-8.71}$} &  $87.54$\\
\hline
	\end{tabular}%
	}
    \vspace{-1em}
    \caption{} \label{tab:10234}
    \end{subtable}
\end{table*}
\begin{table}
\ContinuedFloat
    \begin{subtable}{\textwidth}
    \centering
    \renewcommand{\arraystretch}{1.2}
	\resizebox{\textwidth}{!}{%
		\begin{tabular}{|l|c|c|c|c|c|c|c|}
		\hline
			\multicolumn{8}{|c|}{\textbf{1514}} \\ 
			\hline \hline
			\multicolumn{8}{|c|}{Best Parameters} \\
			\hline 
			Model & $\log_{10}(\mathrm{M_{halo}}/\mathrm{M}_\odot$) & $\log_{10}(\mathrm{M_{\star}}/\mathrm{M}_\odot$) & $\log_{10}(f_\mathrm{BH}$) & $\delta_\mathrm{AGN}$ & $\tau_\mathrm{V}$ & $\alpha$ & $\chi^2$ \\
			\hline
			A\_chab100\_bin & 11.33 & 9.80 & --3.97 & --0.94 & 1.22 & -- & 7.58 \\
			A\_chab100\_sin & 11.10 & 9.58 & --2.65 & --0.97 & 1.07 & -- & 6.35 \\
			A\_135\_100\_bin & 11.37 & 9.85 & --3.98 & --0.94 & 1.20 & -- & 7.37 \\
			B\_chab100\_bin & 11.40 & 9.88 & --3.95 & --0.95 & 1.13 & 0.41 & 6.42 \\
			B\_135\_100\_bin & 11.44 & 9.91 & --4.00 & --0.94 & 1.14 & 0.45 & 6.47 \\
			B\_yggdrasil\_sin & 9.89 & 7.68 & --0.97 & 1.00 & 0.08 & 0.79 & 64.77 \\
			A\_yggdrasil\_sin & 9.92 & 7.74 & --1.00 & 1.00 & 0.11 & -- & 64.77 \\			
			\hline
		\end{tabular}%
	}
	\bigskip
	\resizebox{\textwidth}{!}{%
	\begin{tabular}{|l|c|c|c|c|c|c|c|c|}
	    \hline 
	    \multicolumn{9}{|c|}{Median Parameters} \\
		\hline
		Model & $\log_{10}(\mathrm{M_{halo}}/\mathrm{M}_\odot$) & $\log_{10}(\mathrm{M_{\star}}/\mathrm{M}_\odot$) & $\log_{10}(f_\mathrm{BH}$) & $\delta_\mathrm{AGN}$ & $\tau_\mathrm{V}$ & $\alpha$ & $\chi^2$ & $\chi^2(\theta_{\mathrm{med}})$ \\
		\hline
		A\_chab100\_bin &  $11.25$\raisebox{0.5ex}{\tiny$^{+0.10}_{-0.08}$} &  $9.73$\raisebox{0.5ex}{\tiny$^{+0.10}_{-0.08}$} &  $-3.43$\raisebox{0.5ex}{\tiny$^{+0.34}_{-0.41}$} &  $-0.03$\raisebox{0.5ex}{\tiny$^{+0.60}_{-0.64}$} &  $1.17$\raisebox{0.5ex}{\tiny$^{+0.09}_{-0.08}$} & -- &  $13.65$\raisebox{0.5ex}{\tiny$^{+3.46}_{-7.22}$} &  $9.80$\\
		A\_chab100\_sin &  $11.08$\raisebox{0.5ex}{\tiny$^{+0.10}_{-0.09}$} &  $9.56$\raisebox{0.5ex}{\tiny$^{+0.10}_{-0.09}$} &  $-3.28$\raisebox{0.5ex}{\tiny$^{+0.42}_{-0.44}$} &  $-0.10$\raisebox{0.5ex}{\tiny$^{+0.55}_{-0.63}$} &  $0.99$\raisebox{0.5ex}{\tiny$^{+0.08}_{-0.09}$} & -- &  $11.76$\raisebox{0.5ex}{\tiny$^{+3.20}_{-7.30}$} &  $7.41$\\
		A\_135\_100\_bin &  $11.31$\raisebox{0.5ex}{\tiny$^{+0.10}_{-0.08}$} &  $9.79$\raisebox{0.5ex}{\tiny$^{+0.10}_{-0.08}$} &  $-3.47$\raisebox{0.5ex}{\tiny$^{+0.32}_{-0.40}$} &  $-0.05$\raisebox{0.5ex}{\tiny$^{+0.59}_{-0.64}$} &  $1.17$\raisebox{0.5ex}{\tiny$^{+0.09}_{-0.07}$} & -- &  $13.66$\raisebox{0.5ex}{\tiny$^{+3.76}_{-7.42}$} &  $9.18$\\
		B\_chab100\_bin &  $11.28$\raisebox{0.5ex}{\tiny$^{+0.10}_{-0.08}$} &  $9.76$\raisebox{0.5ex}{\tiny$^{+0.10}_{-0.08}$} &  $-3.47$\raisebox{0.5ex}{\tiny$^{+0.32}_{-0.39}$} &  $-0.07$\raisebox{0.5ex}{\tiny$^{+0.58}_{-0.64}$} &  $1.15$\raisebox{0.5ex}{\tiny$^{+0.09}_{-0.08}$} &  $0.60$\raisebox{0.5ex}{\tiny$^{+0.12}_{-0.12}$} &  $12.89$\raisebox{0.5ex}{\tiny$^{+3.52}_{-7.31}$} &  $8.52$\\
\hline		
	\end{tabular}%
	}
    \vspace{-1em}
    \caption{} \label{tab:1514}
    \end{subtable}
\end{table}
\clearpage
\begin{table}
\ContinuedFloat
    \begin{subtable}{\textwidth}
        \centering
    \renewcommand{\arraystretch}{1.2}
	\resizebox{\textwidth}{!}{%
		\begin{tabular}{|l|c|c|c|c|c|c|c|}
		\hline
			\multicolumn{8}{|c|}{\textbf{1696}} \\ 
			\hline \hline
			\multicolumn{8}{|c|}{Best Parameters} \\
			\hline 
			Model & $\log_{10}(\mathrm{M_{halo}}/\mathrm{M}_\odot$) & $\log_{10}(\mathrm{M_{\star}}/\mathrm{M}_\odot$) & $\log_{10}(f_\mathrm{BH}$) & $\delta_\mathrm{AGN}$ & $\tau_\mathrm{V}$ & $\alpha$ & $\chi^2$ \\
			\hline
			A\_chab100\_bin & 10.78 & 9.26 & --3.98 & 0.99 & 0.34 & -- & 14.64 \\
			A\_chab100\_sin & 10.68 & 9.16 & --3.99 & --0.99 & 0.24 & -- & 5.35 \\
			A\_135\_100\_bin & 10.84 & 9.32 & --4.00 & 0.99 & 0.33 & -- & 14.43 \\
			B\_chab100\_bin & 10.86 & 9.34 & --3.97 & 0.99 & 0.22 & 0.41 & 8.78 \\
			B\_135\_100\_bin & 10.92 & 9.39 & --3.92 & 0.99 & 0.22 & 0.41 & 8.70 \\
			B\_yggdrasil\_sin & 9.48 & 6.84 & --0.02 & --0.27 & 0.01 & 0.42 & 115.00 \\
			A\_yggdrasil\_sin & 9.47 & 6.81 & --0.01 & --0.25 & 0.00 & -- & 115.40 \\			
			\hline
			\end{tabular}
		}
		\bigskip
	\resizebox{\textwidth}{!}{%
	\begin{tabular}{|l|c|c|c|c|c|c|c|c|}
	    \hline 
	    \multicolumn{9}{|c|}{Median Parameters} \\
		\hline
		Model & $\log_{10}(\mathrm{M_{halo}}/\mathrm{M}_\odot$) & $\log_{10}(\mathrm{M_{\star}}/\mathrm{M}_\odot$) & $\log_{10}(f_\mathrm{BH}$) & $\delta_\mathrm{AGN}$ & $\tau_\mathrm{V}$ & $\alpha$ & $\chi^2$ & $\chi^2(\theta_{\mathrm{med}})$ \\
		\hline
		A\_chab100\_bin &  $10.78$\raisebox{0.5ex}{\tiny$^{+0.05}_{-0.06}$} &  $9.26$\raisebox{0.5ex}{\tiny$^{+0.05}_{-0.06}$} &  $-3.60$\raisebox{0.5ex}{\tiny$^{+0.25}_{-0.32}$} &  $-0.11$\raisebox{0.5ex}{\tiny$^{+0.58}_{-0.69}$} &  $0.36$\raisebox{0.5ex}{\tiny$^{+0.05}_{-0.06}$} & -- &  $20.67$\raisebox{0.5ex}{\tiny$^{+3.55}_{-7.69}$} &  $16.69$\\
		A\_chab100\_sin &  $10.67$\raisebox{0.5ex}{\tiny$^{+0.04}_{-0.04}$} &  $9.15$\raisebox{0.5ex}{\tiny$^{+0.04}_{-0.04}$} &  $-3.56$\raisebox{0.5ex}{\tiny$^{+0.27}_{-0.34}$} &  $-0.15$\raisebox{0.5ex}{\tiny$^{+0.55}_{-0.69}$} &  $0.25$\raisebox{0.5ex}{\tiny$^{+0.03}_{-0.04}$} & -- &  $11.65$\raisebox{0.5ex}{\tiny$^{+3.62}_{-7.42}$} &  $7.92$\\
		A\_135\_100\_bin &  $10.83$\raisebox{0.5ex}{\tiny$^{+0.06}_{-0.06}$} &  $9.30$\raisebox{0.5ex}{\tiny$^{+0.06}_{-0.06}$} &  $-3.60$\raisebox{0.5ex}{\tiny$^{+0.25}_{-0.32}$} &  $-0.07$\raisebox{0.5ex}{\tiny$^{+0.60}_{-0.67}$} &  $0.34$\raisebox{0.5ex}{\tiny$^{+0.06}_{-0.07}$} & -- &  $20.53$\raisebox{0.5ex}{\tiny$^{+3.43}_{-7.35}$} &  $16.61$\\
		B\_chab100\_bin &  $10.81$\raisebox{0.5ex}{\tiny$^{+0.07}_{-0.07}$} &  $9.28$\raisebox{0.5ex}{\tiny$^{+0.07}_{-0.07}$} &  $-3.58$\raisebox{0.5ex}{\tiny$^{+0.26}_{-0.33}$} &  $-0.05$\raisebox{0.5ex}{\tiny$^{+0.60}_{-0.66}$} &  $0.28$\raisebox{0.5ex}{\tiny$^{+0.08}_{-0.08}$} &  $0.55$\raisebox{0.5ex}{\tiny$^{+0.09}_{-0.12}$} &  $17.66$\raisebox{0.5ex}{\tiny$^{+3.95}_{-7.37}$} &  $13.68$\\
\hline
		\end{tabular}
		}
        \vspace{-1em}
        \caption{} \label{tab:1696}
    \end{subtable} 
\end{table}

\begin{table}
\ContinuedFloat
    \begin{subtable}{\textwidth}
        \centering
    \renewcommand{\arraystretch}{1.2}
	\resizebox{\textwidth}{!}{%
	\begin{tabular}{|l|c|c|c|c|c|c|c|}
	\hline
	    \multicolumn{8}{|c|}{\textbf{2462}} \\ 
	    \hline \hline
	    \multicolumn{8}{|c|}{Best Parameters} \\
		\hline 
		Model & $\log_{10}(\mathrm{M_{halo}}/\mathrm{M}_\odot$) & $\log_{10}(\mathrm{M_{\star}}/\mathrm{M}_\odot$) & $\log_{10}(f_\mathrm{BH}$) & $\delta_\mathrm{AGN}$ & $\tau_\mathrm{V}$ & $\alpha$ & $\chi^2$ \\
		\hline
		A\_chab100\_bin & 11.58 & 10.05 & --3.82 & 1.00 & 0.89 & -- & 20.86 \\
		A\_chab100\_sin & 11.47 & 9.95 & --3.99 & --0.98 & 0.76 & -- & 9.55 \\
		A\_135\_100\_bin & 11.64 & 10.12 & --3.89 & 0.99 & 0.89 & -- & 21.70 \\
		B\_chab100\_bin & 11.66 & 10.14 & --3.93 & 1.00 & 0.86 & 0.46 & 19.64 \\
		B\_135\_100\_bin & 11.70 & 10.18 & --3.98 & 0.98 & 0.84 & 0.47 & 19.68 \\
		B\_yggdrasil\_sin & 10.57 & 9.04 & --1.84 & 1.00 & 0.16 & 0.43 & 116.38 \\
		A\_yggdrasil\_sin & 10.59 & 9.07 & --1.79 & 1.00 & 0.32 & -- & 119.27 \\
		\hline
	\end{tabular}
	}
	\bigskip
	\resizebox{\textwidth}{!}{%
	\begin{tabular}{|l|c|c|c|c|c|c|c|c|}
	    \hline 
	    \multicolumn{9}{|c|}{Median Parameters} \\
		\hline
		Model & $\log_{10}(\mathrm{M_{halo}}/\mathrm{M}_\odot$) & $\log_{10}(\mathrm{M_{\star}}/\mathrm{M}_\odot$) & $\log_{10}(f_\mathrm{BH}$) & $\delta_\mathrm{AGN}$ & $\tau_\mathrm{V}$ & $\alpha$ & $\chi^2$ & $\chi^2(\theta_{\mathrm{med}})$ \\
		\hline
		A\_chab100\_bin &  $11.60$\raisebox{0.5ex}{\tiny$^{+0.05}_{-0.06}$} &  $10.07$\raisebox{0.5ex}{\tiny$^{+0.05}_{-0.06}$} &  $-3.63$\raisebox{0.5ex}{\tiny$^{+0.23}_{-0.31}$} &  $-0.10$\raisebox{0.5ex}{\tiny$^{+0.59}_{-0.70}$} &  $0.92$\raisebox{0.5ex}{\tiny$^{+0.06}_{-0.06}$} & -- &  $27.41$\raisebox{0.5ex}{\tiny$^{+3.61}_{-9.00}$} &  $23.68$\\
		A\_chab100\_sin &  $11.44$\raisebox{0.5ex}{\tiny$^{+0.07}_{-0.07}$} &  $9.92$\raisebox{0.5ex}{\tiny$^{+0.07}_{-0.07}$} &  $-3.59$\raisebox{0.5ex}{\tiny$^{+0.26}_{-0.34}$} &  $-0.09$\raisebox{0.5ex}{\tiny$^{+0.59}_{-0.69}$} &  $0.74$\raisebox{0.5ex}{\tiny$^{+0.07}_{-0.07}$} & -- &  $16.05$\raisebox{0.5ex}{\tiny$^{+3.66}_{-7.69}$} &  $11.87$\\
		A\_135\_100\_bin &  $11.63$\raisebox{0.5ex}{\tiny$^{+0.06}_{-0.07}$} &  $10.11$\raisebox{0.5ex}{\tiny$^{+0.06}_{-0.07}$} &  $-3.63$\raisebox{0.5ex}{\tiny$^{+0.23}_{-0.30}$} &  $-0.06$\raisebox{0.5ex}{\tiny$^{+0.61}_{-0.68}$} &  $0.90$\raisebox{0.5ex}{\tiny$^{+0.07}_{-0.07}$} & -- &  $27.94$\raisebox{0.5ex}{\tiny$^{+3.34}_{-7.68}$} &  $24.38$\\
		B\_chab100\_bin &  $11.61$\raisebox{0.5ex}{\tiny$^{+0.06}_{-0.07}$} &  $10.09$\raisebox{0.5ex}{\tiny$^{+0.06}_{-0.07}$} &  $-3.63$\raisebox{0.5ex}{\tiny$^{+0.23}_{-0.30}$} &  $-0.07$\raisebox{0.5ex}{\tiny$^{+0.60}_{-0.68}$} &  $0.89$\raisebox{0.5ex}{\tiny$^{+0.07}_{-0.07}$} &  $0.58$\raisebox{0.5ex}{\tiny$^{+0.10}_{-0.11}$} &  $27.19$\raisebox{0.5ex}{\tiny$^{+3.72}_{-8.48}$} &  $23.20$\\
\hline
	\end{tabular}
	}
        \vspace{-1em}
        \caption{} \label{tab:2462}
    \end{subtable}
\end{table}
\clearpage
\begin{table}
\ContinuedFloat
    \begin{subtable}{\textwidth}
        \centering
    \renewcommand{\arraystretch}{1.2}
	\resizebox{\textwidth}{!}{%
	\begin{tabular}{|l|c|c|c|c|c|c|c|}
	\hline
	    \multicolumn{8}{|c|}{\textbf{2779}} \\ 
	    \hline \hline
	    \multicolumn{8}{|c|}{Best Parameters} \\
		\hline 
		Model & $\log_{10}(\mathrm{M_{halo}}/\mathrm{M}_\odot$) & $\log_{10}(\mathrm{M_{\star}}/\mathrm{M}_\odot$) & $\log_{10}(f_\mathrm{BH}$) & $\delta_\mathrm{AGN}$ & $\tau_\mathrm{V}$ & $\alpha$ & $\chi^2$ \\
		\hline
		A\_chab100\_bin & 10.32 & 8.56 & --3.27 & 0.99 & 0.34 & -- & 14.96 \\
		A\_chab100\_sin & 10.32 & 8.57 & --3.94 & 1.00 & 0.30 & -- & 12.91 \\
		A\_135\_100\_bin & 10.36 & 8.63 & --3.41 & 0.99 & 0.35 & -- & 15.24 \\
		B\_chab100\_bin & 10.37 & 8.67 & --3.95 & --0.80 & 0.19 & 0.40 & 7.04 \\
		B\_135\_100\_bin & 10.40 & 8.73 & --3.98 & --0.74 & 0.18 & 0.40 & 7.10 \\
		B\_yggdrasil\_sin & 9.28 & 6.42 & --0.03 & --0.49 & 0.01 & 0.40 & 77.38 \\
		A\_yggdrasil\_sin & 9.25 & 6.38 & --0.00 & --0.48 & 0.00 & -- & 77.92 \\
		\hline
	\end{tabular}
	}
	\bigskip
	\resizebox{\textwidth}{!}{%
	\begin{tabular}{|l|c|c|c|c|c|c|c|c|}
	    \hline 
	    \multicolumn{9}{|c|}{Median Parameters} \\
		\hline
		Model & $\log_{10}(\mathrm{M_{halo}}/\mathrm{M}_\odot$) & $\log_{10}(\mathrm{M_{\star}}/\mathrm{M}_\odot$) & $\log_{10}(f_\mathrm{BH}$) & $\delta_\mathrm{AGN}$ & $\tau_\mathrm{V}$ & $\alpha$ & $\chi^2$ & $\chi^2(\theta_{\mathrm{med}})$ \\
		\hline
		A\_chab100\_bin &  $10.32$\raisebox{0.5ex}{\tiny$^{+0.04}_{-0.04}$} &  $8.57$\raisebox{0.5ex}{\tiny$^{+0.09}_{-0.08}$} &  $-3.38$\raisebox{0.5ex}{\tiny$^{+0.37}_{-0.45}$} &  $0.03$\raisebox{0.5ex}{\tiny$^{+0.63}_{-0.61}$} &  $0.35$\raisebox{0.5ex}{\tiny$^{+0.09}_{-0.08}$} & -- &  $20.60$\raisebox{0.5ex}{\tiny$^{+3.48}_{-7.44}$} &  $16.39$\\
		A\_chab100\_sin &  $10.31$\raisebox{0.5ex}{\tiny$^{+0.04}_{-0.04}$} &  $8.53$\raisebox{0.5ex}{\tiny$^{+0.09}_{-0.07}$} &  $-3.40$\raisebox{0.5ex}{\tiny$^{+0.36}_{-0.43}$} &  $-0.00$\raisebox{0.5ex}{\tiny$^{+0.62}_{-0.63}$} &  $0.29$\raisebox{0.5ex}{\tiny$^{+0.08}_{-0.08}$} & -- &  $18.48$\raisebox{0.5ex}{\tiny$^{+3.47}_{-7.38}$} &  $14.31$\\
		A\_135\_100\_bin &  $10.35$\raisebox{0.5ex}{\tiny$^{+0.05}_{-0.04}$} &  $8.63$\raisebox{0.5ex}{\tiny$^{+0.09}_{-0.08}$} &  $-3.41$\raisebox{0.5ex}{\tiny$^{+0.36}_{-0.43}$} &  $0.04$\raisebox{0.5ex}{\tiny$^{+0.64}_{-0.61}$} &  $0.35$\raisebox{0.5ex}{\tiny$^{+0.09}_{-0.08}$} & -- &  $20.99$\raisebox{0.5ex}{\tiny$^{+3.50}_{-7.40}$} &  $16.77$\\
		B\_chab100\_bin &  $10.35$\raisebox{0.5ex}{\tiny$^{+0.04}_{-0.04}$} &  $8.63$\raisebox{0.5ex}{\tiny$^{+0.09}_{-0.07}$} &  $-3.43$\raisebox{0.5ex}{\tiny$^{+0.34}_{-0.40}$} &  $-0.00$\raisebox{0.5ex}{\tiny$^{+0.60}_{-0.62}$} &  $0.29$\raisebox{0.5ex}{\tiny$^{+0.09}_{-0.09}$} &  $0.55$\raisebox{0.5ex}{\tiny$^{+0.10}_{-0.13}$} &  $18.02$\raisebox{0.5ex}{\tiny$^{+4.26}_{-7.33}$} &  $13.57$\\
\hline
	\end{tabular}
	}
        \vspace{-1em}
        \caption{} \label{tab:2779}
    \end{subtable}
\end{table}

\begin{table}
\ContinuedFloat
    \begin{subtable}{\textwidth}
        \centering
    \renewcommand{\arraystretch}{1.2}
	\resizebox{\textwidth}{!}{%
	\begin{tabular}{|l|c|c|c|c|c|c|c|}
	\hline
	    \multicolumn{8}{|c|}{\textbf{6115}} \\ 
	    \hline \hline
	    \multicolumn{8}{|c|}{Best Parameters} \\
		\hline 
		Model & $\log_{10}(\mathrm{M_{halo}}/\mathrm{M}_\odot$) & $\log_{10}(\mathrm{M_{\star}}/\mathrm{M}_\odot$) & $\log_{10}(f_\mathrm{BH}$) & $\delta_\mathrm{AGN}$ & $\tau_\mathrm{V}$ & $\alpha$ & $\chi^2$ \\
		\hline
		A\_chab100\_bin & 9.25 & 6.60 & --0.03 & --0.88 & 0.02 & -- & 2.72 \\
		A\_chab100\_sin & 9.25 & 6.59 & --0.03 & --0.86 & 0.00 & -- & 2.67 \\
		A\_135\_100\_bin & 9.27 & 6.64 & --0.06 & --0.88 & 0.02 & -- & 2.69 \\
		B\_chab100\_bin & 9.26 & 6.61 & --0.04 & --0.87 & 0.01 & 0.79 & 2.68 \\
		B\_135\_100\_bin & 9.27 & 6.64 & --0.09 & --0.85 & 0.00 & 0.65 & 2.68 \\
		B\_yggdrasil\_sin & 9.80 & 7.73 & --1.25 & --0.84 & 0.00 & 0.63 & 2.30 \\
		A\_yggdrasil\_sin & 9.70 & 7.53 & --1.04 & --0.84 & 0.00 & -- & 2.28 \\
		\hline
	\end{tabular}
	}
	\bigskip
	\resizebox{\textwidth}{!}{%
	\begin{tabular}{|l|c|c|c|c|c|c|c|c|}
	    \hline 
	    \multicolumn{9}{|c|}{Median Parameters} \\
		\hline
		Model & $\log_{10}(\mathrm{M_{halo}}/\mathrm{M}_\odot$) & $\log_{10}(\mathrm{M_{\star}}/\mathrm{M}_\odot$) & $\log_{10}(f_\mathrm{BH}$) & $\delta_\mathrm{AGN}$ & $\tau_\mathrm{V}$ & $\alpha$ & $\chi^2$ & $\chi^2(\theta_{\mathrm{med}})$ \\
		\hline
		A\_chab100\_bin &  $10.17$\raisebox{0.5ex}{\tiny$^{+0.07}_{-0.05}$} &  $8.49$\raisebox{0.5ex}{\tiny$^{+0.14}_{-0.10}$} &  $-3.09$\raisebox{0.5ex}{\tiny$^{+0.53}_{-0.71}$} &  $-0.28$\raisebox{0.5ex}{\tiny$^{+0.51}_{-0.69}$} &  $0.12$\raisebox{0.5ex}{\tiny$^{+0.07}_{-0.08}$} & -- &  $12.89$\raisebox{0.5ex}{\tiny$^{+3.11}_{-7.12}$} &  $10.80$\\
		A\_chab100\_sin &  $10.17$\raisebox{0.5ex}{\tiny$^{+0.11}_{-0.04}$} &  $8.48$\raisebox{0.5ex}{\tiny$^{+0.22}_{-0.08}$} &  $-3.13$\raisebox{0.5ex}{\tiny$^{+0.52}_{-1.09}$} &  $-0.39$\raisebox{0.5ex}{\tiny$^{+0.46}_{-0.73}$} &  $0.09$\raisebox{0.5ex}{\tiny$^{+0.05}_{-0.07}$} & -- &  $13.21$\raisebox{0.5ex}{\tiny$^{+3.11}_{-7.14}$} &  $13.08$\\
		A\_135\_100\_bin &  $10.20$\raisebox{0.5ex}{\tiny$^{+0.08}_{-0.05}$} &  $8.55$\raisebox{0.5ex}{\tiny$^{+0.17}_{-0.10}$} &  $-3.13$\raisebox{0.5ex}{\tiny$^{+0.52}_{-0.82}$} &  $-0.32$\raisebox{0.5ex}{\tiny$^{+0.50}_{-0.71}$} &  $0.12$\raisebox{0.5ex}{\tiny$^{+0.07}_{-0.08}$} & -- &  $12.68$\raisebox{0.5ex}{\tiny$^{+3.19}_{-7.36}$} &  $10.94$\\
		B\_chab100\_bin &  $10.18$\raisebox{0.5ex}{\tiny$^{+0.15}_{-0.05}$} &  $8.52$\raisebox{0.5ex}{\tiny$^{+0.31}_{-0.11}$} &  $-3.05$\raisebox{0.5ex}{\tiny$^{+0.56}_{-1.21}$} &  $-0.40$\raisebox{0.5ex}{\tiny$^{+0.46}_{-0.72}$} &  $0.10$\raisebox{0.5ex}{\tiny$^{+0.06}_{-0.08}$} &  $0.67$\raisebox{0.5ex}{\tiny$^{+0.09}_{-0.08}$} &  $12.74$\raisebox{0.5ex}{\tiny$^{+3.52}_{-7.43}$} &  $11.87$\\
\hline
	\end{tabular}
	}
        \vspace{-1em}
        \caption{} \label{tab:6115}
    \end{subtable}
\end{table}
\clearpage
\begin{table}
\ContinuedFloat
    \begin{subtable}{\textwidth}
        \centering
    \renewcommand{\arraystretch}{1.2}
	\resizebox{\textwidth}{!}{%
	\begin{tabular}{|l|c|c|c|c|c|c|c|}
	\hline
	    \multicolumn{8}{|c|}{\textbf{6878}} \\ 
	    \hline \hline
	    \multicolumn{8}{|c|}{Best Parameters} \\
		\hline 
		Model & $\log_{10}(\mathrm{M_{halo}}/\mathrm{M}_\odot$) & $\log_{10}(\mathrm{M_{\star}}/\mathrm{M}_\odot$) & $\log_{10}(f_\mathrm{BH}$) & $\delta_\mathrm{AGN}$ & $\tau_\mathrm{V}$ & $\alpha$ & $\chi^2$ \\
		\hline
		A\_chab100\_bin & 10.87 & 9.35 & --3.98 & 1.00 & 0.52 & -- & 1.56 \\
		A\_chab100\_sin & 10.71 & 9.18 & --2.79 & --0.88 & 0.41 & -- & 0.66 \\
		A\_135\_100\_bin & 10.94 & 9.41 & --3.97 & 0.99 & 0.52 & -- & 1.57 \\
		B\_chab100\_bin & 10.98 & 9.46 & --3.99 & --0.98 & 0.44 & 0.41 & 0.70 \\
		B\_135\_100\_bin & 11.02 & 9.50 & --3.99 & --0.89 & 0.44 & 0.44 & 0.74 \\
		B\_yggdrasil\_sin & 10.29 & 8.51 & --1.94 & 0.98 & 0.01 & 0.50 & 63.96 \\
		A\_yggdrasil\_sin & 10.17 & 8.26 & --1.66 & 0.98 & 0.00 & -- & 63.58 \\
		\hline
	\end{tabular}
	}
	\bigskip
	\resizebox{\textwidth}{!}{%
	\begin{tabular}{|l|c|c|c|c|c|c|c|c|}
	    \hline 
	    \multicolumn{9}{|c|}{Median Parameters} \\
		\hline
		Model & $\log_{10}(\mathrm{M_{halo}}/\mathrm{M}_\odot$) & $\log_{10}(\mathrm{M_{\star}}/\mathrm{M}_\odot$) & $\log_{10}(f_\mathrm{BH}$) & $\delta_\mathrm{AGN}$ & $\tau_\mathrm{V}$ & $\alpha$ & $\chi^2$ & $\chi^2(\theta_{\mathrm{med}})$ \\
		\hline
		A\_chab100\_bin &  $10.84$\raisebox{0.5ex}{\tiny$^{+0.07}_{-0.07}$} &  $9.32$\raisebox{0.5ex}{\tiny$^{+0.07}_{-0.07}$} &  $-3.47$\raisebox{0.5ex}{\tiny$^{+0.32}_{-0.38}$} &  $-0.09$\raisebox{0.5ex}{\tiny$^{+0.58}_{-0.66}$} &  $0.52$\raisebox{0.5ex}{\tiny$^{+0.07}_{-0.07}$} & -- &  $6.59$\raisebox{0.5ex}{\tiny$^{+3.21}_{-7.14}$} &  $2.66$\\
		A\_chab100\_sin &  $10.70$\raisebox{0.5ex}{\tiny$^{+0.05}_{-0.05}$} &  $9.17$\raisebox{0.5ex}{\tiny$^{+0.05}_{-0.05}$} &  $-3.33$\raisebox{0.5ex}{\tiny$^{+0.38}_{-0.40}$} &  $-0.10$\raisebox{0.5ex}{\tiny$^{+0.56}_{-0.65}$} &  $0.37$\raisebox{0.5ex}{\tiny$^{+0.04}_{-0.05}$} & -- &  $5.63$\raisebox{0.5ex}{\tiny$^{+3.36}_{-7.33}$} &  $1.69$\\
		A\_135\_100\_bin &  $10.89$\raisebox{0.5ex}{\tiny$^{+0.08}_{-0.07}$} &  $9.37$\raisebox{0.5ex}{\tiny$^{+0.08}_{-0.07}$} &  $-3.48$\raisebox{0.5ex}{\tiny$^{+0.31}_{-0.37}$} &  $-0.07$\raisebox{0.5ex}{\tiny$^{+0.59}_{-0.65}$} &  $0.50$\raisebox{0.5ex}{\tiny$^{+0.07}_{-0.07}$} & -- &  $6.75$\raisebox{0.5ex}{\tiny$^{+3.30}_{-7.22}$} &  $2.66$\\
		B\_chab100\_bin &  $10.87$\raisebox{0.5ex}{\tiny$^{+0.08}_{-0.08}$} &  $9.35$\raisebox{0.5ex}{\tiny$^{+0.08}_{-0.08}$} &  $-3.45$\raisebox{0.5ex}{\tiny$^{+0.33}_{-0.38}$} &  $-0.06$\raisebox{0.5ex}{\tiny$^{+0.58}_{-0.63}$} &  $0.46$\raisebox{0.5ex}{\tiny$^{+0.08}_{-0.08}$} &  $0.59$\raisebox{0.5ex}{\tiny$^{+0.11}_{-0.12}$} &  $6.11$\raisebox{0.5ex}{\tiny$^{+3.29}_{-7.14}$} &  $1.85$\\
\hline
	\end{tabular}
	}
        \vspace{-1em}
        \caption{} \label{tab:6878}
    \end{subtable}
\end{table}

\begin{table}
\ContinuedFloat
    \begin{subtable}{\textwidth}
        \centering
    \renewcommand{\arraystretch}{1.2}
	\resizebox{\textwidth}{!}{%
	\begin{tabular}{|l|c|c|c|c|c|c|c|}
	\hline
	    \multicolumn{8}{|c|}{\textbf{CR2-z12-1}} \\ 
	    \hline \hline
	    \multicolumn{8}{|c|}{Best Parameters} \\
		\hline 
		Model & $\log_{10}(\mathrm{M_{halo}}/\mathrm{M}_\odot$) & $\log_{10}(\mathrm{M_{\star}}/\mathrm{M}_\odot$) & $\log_{10}(f_\mathrm{BH}$) & $\delta_\mathrm{AGN}$ & $\tau_\mathrm{V}$ & $\alpha$ & $\chi^2$ \\
		\hline
		A\_chab100\_bin & 10.02 & 8.35 & --2.68 & 0.99 & 0.00 & -- & 6.24 \\
		A\_chab100\_sin & 10.04 & 8.40 & --2.88 & 0.97 & 0.00 & -- & 6.35 \\
		A\_135\_100\_bin & 10.06 & 8.44 & --2.81 & 0.92 & 0.02 & -- & 6.27 \\
		B\_chab100\_bin & 10.18 & 8.65 & --3.82 & 0.73 & 0.02 & 0.41 & 6.07 \\
		B\_135\_100\_bin & 10.22 & 8.70 & --3.68 & 0.97 & 0.00 & 0.40 & 6.07 \\
		B\_yggdrasil\_sin & 10.14 & 8.61 & --3.10 & --0.96 & 0.01 & 0.47 & 6.75 \\
		A\_yggdrasil\_sin & 9.97 & 8.25 & --2.54 & --1.00 & 0.00 & -- & 6.96 \\
		\hline
	\end{tabular}
	}
	\bigskip
	\resizebox{\textwidth}{!}{%
	\begin{tabular}{|l|c|c|c|c|c|c|c|c|}
	    \hline 
	    \multicolumn{9}{|c|}{Median Parameters} \\
		\hline
		Model & $\log_{10}(\mathrm{M_{halo}}/\mathrm{M}_\odot$) & $\log_{10}(\mathrm{M_{\star}}/\mathrm{M}_\odot$) & $\log_{10}(f_\mathrm{BH}$) & $\delta_\mathrm{AGN}$ & $\tau_\mathrm{V}$ & $\alpha$ & $\chi^2$ & $\chi^2(\theta_{\mathrm{med}})$ \\
		\hline
		A\_chab100\_bin &  $10.18$\raisebox{0.5ex}{\tiny$^{+0.10}_{-0.13}$} &  $8.66$\raisebox{0.5ex}{\tiny$^{+0.19}_{-0.13}$} &  $-3.33$\raisebox{0.5ex}{\tiny$^{+0.41}_{-0.53}$} &  $-0.15$\raisebox{0.5ex}{\tiny$^{+0.54}_{-0.66}$} &  $0.24$\raisebox{0.5ex}{\tiny$^{+0.12}_{-0.12}$} & -- &  $10.43$\raisebox{0.5ex}{\tiny$^{+2.93}_{-6.95}$} &  $7.37$\\
		A\_chab100\_sin &  $10.18$\raisebox{0.5ex}{\tiny$^{+0.10}_{-0.12}$} &  $8.66$\raisebox{0.5ex}{\tiny$^{+0.17}_{-0.12}$} &  $-3.36$\raisebox{0.5ex}{\tiny$^{+0.40}_{-0.52}$} &  $-0.17$\raisebox{0.5ex}{\tiny$^{+0.53}_{-0.66}$} &  $0.20$\raisebox{0.5ex}{\tiny$^{+0.11}_{-0.11}$} & -- &  $10.76$\raisebox{0.5ex}{\tiny$^{+3.09}_{-7.01}$} &  $7.71$\\
		A\_135\_100\_bin &  $10.22$\raisebox{0.5ex}{\tiny$^{+0.11}_{-0.13}$} &  $8.70$\raisebox{0.5ex}{\tiny$^{+0.15}_{-0.13}$} &  $-3.35$\raisebox{0.5ex}{\tiny$^{+0.40}_{-0.50}$} &  $-0.15$\raisebox{0.5ex}{\tiny$^{+0.54}_{-0.66}$} &  $0.21$\raisebox{0.5ex}{\tiny$^{+0.10}_{-0.12}$} & -- &  $10.08$\raisebox{0.5ex}{\tiny$^{+2.77}_{-6.73}$} &  $7.34$\\
		B\_chab100\_bin &  $10.21$\raisebox{0.5ex}{\tiny$^{+0.10}_{-0.12}$} &  $8.69$\raisebox{0.5ex}{\tiny$^{+0.14}_{-0.12}$} &  $-3.36$\raisebox{0.5ex}{\tiny$^{+0.39}_{-0.50}$} &  $-0.16$\raisebox{0.5ex}{\tiny$^{+0.53}_{-0.66}$} &  $0.21$\raisebox{0.5ex}{\tiny$^{+0.11}_{-0.12}$} &  $0.62$\raisebox{0.5ex}{\tiny$^{+0.12}_{-0.11}$} &  $10.32$\raisebox{0.5ex}{\tiny$^{+2.92}_{-6.89}$} &  $6.99$\\
\hline
	\end{tabular}
	}
        \vspace{-1em}
        \caption{} \label{tab:CR2-z12-1}
    \end{subtable}
\end{table}
\clearpage
\begin{table}
\ContinuedFloat
    \begin{subtable}{\textwidth}
        \centering
    \renewcommand{\arraystretch}{1.2}
	\resizebox{\textwidth}{!}{%
	\begin{tabular}{|l|c|c|c|c|c|c|c|}
	\hline
	    \multicolumn{8}{|c|}{\textbf{CR2-z16-1}} \\ 
	    \hline \hline
	    \multicolumn{8}{|c|}{Best Parameters} \\
		\hline 
		Model & $\log_{10}(\mathrm{M_{halo}}/\mathrm{M}_\odot$) & $\log_{10}(\mathrm{M_{\star}}/\mathrm{M}_\odot$) & $\log_{10}(f_\mathrm{BH}$) & $\delta_\mathrm{AGN}$ & $\tau_\mathrm{V}$ & $\alpha$ & $\chi^2$ \\
		\hline
		A\_chab100\_bin & 11.04 & 9.52 & --3.16 & --0.70 & 0.40 & -- & 4.15 \\
		A\_chab100\_sin & 11.04 & 9.52 & --3.53 & --0.77 & 0.36 & -- & 4.12 \\
		A\_135\_100\_bin & 11.10 & 9.58 & --3.48 & --0.65 & 0.38 & -- & 4.17 \\
		B\_chab100\_bin & 11.00 & 9.48 & --3.95 & --0.82 & 0.44 & 0.64 & 4.13 \\
		B\_135\_100\_bin & 11.03 & 9.50 & --3.95 & --0.68 & 0.45 & 0.79 & 4.13 \\
		B\_yggdrasil\_sin & 10.51 & 8.98 & --1.73 & --0.86 & 0.20 & 0.43 & 4.58 \\
		A\_yggdrasil\_sin & 10.47 & 8.94 & --1.62 & --0.96 & 0.23 & -- & 4.57 \\
		\hline
	\end{tabular}
	}
	\bigskip
	\resizebox{\textwidth}{!}{%
	\begin{tabular}{|l|c|c|c|c|c|c|c|c|}
	    \hline 
	    \multicolumn{9}{|c|}{Median Parameters} \\
		\hline
		Model & $\log_{10}(\mathrm{M_{halo}}/\mathrm{M}_\odot$) & $\log_{10}(\mathrm{M_{\star}}/\mathrm{M}_\odot$) & $\log_{10}(f_\mathrm{BH}$) & $\delta_\mathrm{AGN}$ & $\tau_\mathrm{V}$ & $\alpha$ & $\chi^2$ & $\chi^2(\theta_{\mathrm{med}})$ \\
		\hline
		A\_chab100\_bin &  $10.93$\raisebox{0.5ex}{\tiny$^{+0.25}_{-0.21}$} &  $9.41$\raisebox{0.5ex}{\tiny$^{+0.25}_{-0.21}$} &  $-2.97$\raisebox{0.5ex}{\tiny$^{+0.62}_{-0.73}$} &  $-0.11$\raisebox{0.5ex}{\tiny$^{+0.57}_{-0.65}$} &  $0.33$\raisebox{0.5ex}{\tiny$^{+0.17}_{-0.17}$} & -- &  $8.52$\raisebox{0.5ex}{\tiny$^{+3.14}_{-7.09}$} &  $5.75$\\
		A\_chab100\_sin &  $10.92$\raisebox{0.5ex}{\tiny$^{+0.23}_{-0.20}$} &  $9.40$\raisebox{0.5ex}{\tiny$^{+0.23}_{-0.20}$} &  $-2.99$\raisebox{0.5ex}{\tiny$^{+0.60}_{-0.72}$} &  $-0.12$\raisebox{0.5ex}{\tiny$^{+0.56}_{-0.66}$} &  $0.29$\raisebox{0.5ex}{\tiny$^{+0.16}_{-0.16}$} & -- &  $8.23$\raisebox{0.5ex}{\tiny$^{+2.99}_{-6.92}$} &  $5.82$\\
		A\_135\_100\_bin &  $10.98$\raisebox{0.5ex}{\tiny$^{+0.26}_{-0.21}$} &  $9.46$\raisebox{0.5ex}{\tiny$^{+0.26}_{-0.21}$} &  $-2.99$\raisebox{0.5ex}{\tiny$^{+0.61}_{-0.73}$} &  $-0.12$\raisebox{0.5ex}{\tiny$^{+0.56}_{-0.65}$} &  $0.32$\raisebox{0.5ex}{\tiny$^{+0.17}_{-0.17}$} & -- &  $8.55$\raisebox{0.5ex}{\tiny$^{+3.13}_{-7.07}$} &  $5.88$\\
		B\_chab100\_bin &  $10.88$\raisebox{0.5ex}{\tiny$^{+0.25}_{-0.21}$} &  $9.36$\raisebox{0.5ex}{\tiny$^{+0.25}_{-0.21}$} &  $-2.94$\raisebox{0.5ex}{\tiny$^{+0.62}_{-0.74}$} &  $-0.10$\raisebox{0.5ex}{\tiny$^{+0.56}_{-0.64}$} &  $0.35$\raisebox{0.5ex}{\tiny$^{+0.18}_{-0.18}$} &  $0.61$\raisebox{0.5ex}{\tiny$^{+0.12}_{-0.11}$} &  $8.40$\raisebox{0.5ex}{\tiny$^{+3.07}_{-7.05}$} &  $5.00$\\
\hline
	\end{tabular}
	}
        \vspace{-1em}
        \caption{} \label{tab:CR2-z16-1}
    \end{subtable}
\end{table}
\vspace{-2em}
\begin{table}
\ContinuedFloat
    \begin{subtable}{\textwidth}
        \centering
    \renewcommand{\arraystretch}{1.2}
	\resizebox{\textwidth}{!}{%
	\begin{tabular}{|l|c|c|c|c|c|c|c|}
	\hline
	    \multicolumn{8}{|c|}{\textbf{GL-z12-1}} \\ 
	    \hline \hline
	    \multicolumn{8}{|c|}{Best Parameters} \\
		\hline 
		Model & $\log_{10}(\mathrm{M_{halo}}/\mathrm{M}_\odot$) & $\log_{10}(\mathrm{M_{\star}}/\mathrm{M}_\odot$) & $\log_{10}(f_\mathrm{BH}$) & $\delta_\mathrm{AGN}$ & $\tau_\mathrm{V}$ & $\alpha$ & $\chi^2$ \\
		\hline
		A\_chab100\_bin & 10.61 & 9.09 & --3.96 & --0.89 & 0.23 & -- & 6.45 \\
		A\_chab100\_sin & 10.58 & 9.06 & --4.00 & --0.82 & 0.17 & -- & 6.64 \\
		A\_135\_100\_bin & 10.66 & 9.13 & --3.93 & --0.94 & 0.21 & -- & 6.45 \\
		B\_chab100\_bin & 10.65 & 9.12 & --3.94 & --0.87 & 0.14 & 0.43 & 6.14 \\
		B\_135\_100\_bin & 10.71 & 9.19 & --3.96 & --0.88 & 0.13 & 0.41 & 6.04 \\
		B\_yggdrasil\_sin & 10.48 & 8.96 & --3.96 & --0.93 & 0.00 & 0.44 & 4.05 \\
		A\_yggdrasil\_sin & 10.33 & 8.81 & --3.95 & --0.90 & 0.00 & -- & 4.07 \\
		\hline
	\end{tabular}
	}
	\bigskip
	\resizebox{\textwidth}{!}{%
	\begin{tabular}{|l|c|c|c|c|c|c|c|c|}
	    \hline 
	    \multicolumn{9}{|c|}{Median Parameters} \\
		\hline
		Model & $\log_{10}(\mathrm{M_{halo}}/\mathrm{M}_\odot$) & $\log_{10}(\mathrm{M_{\star}}/\mathrm{M}_\odot$) & $\log_{10}(f_\mathrm{BH}$) & $\delta_\mathrm{AGN}$ & $\tau_\mathrm{V}$ & $\alpha$ & $\chi^2$ & $\chi^2(\theta_{\mathrm{med}})$ \\
		\hline
		A\_chab100\_bin &  $10.54$\raisebox{0.5ex}{\tiny$^{+0.13}_{-0.13}$} &  $9.02$\raisebox{0.5ex}{\tiny$^{+0.13}_{-0.13}$} &  $-3.25$\raisebox{0.5ex}{\tiny$^{+0.45}_{-0.54}$} &  $-0.13$\raisebox{0.5ex}{\tiny$^{+0.55}_{-0.65}$} &  $0.20$\raisebox{0.5ex}{\tiny$^{+0.10}_{-0.11}$} & -- &  $11.23$\raisebox{0.5ex}{\tiny$^{+2.95}_{-6.52}$} &  $7.23$\\
		A\_chab100\_sin &  $10.52$\raisebox{0.5ex}{\tiny$^{+0.10}_{-0.11}$} &  $9.00$\raisebox{0.5ex}{\tiny$^{+0.10}_{-0.11}$} &  $-3.25$\raisebox{0.5ex}{\tiny$^{+0.45}_{-0.53}$} &  $-0.15$\raisebox{0.5ex}{\tiny$^{+0.54}_{-0.65}$} &  $0.15$\raisebox{0.5ex}{\tiny$^{+0.08}_{-0.09}$} & -- &  $10.68$\raisebox{0.5ex}{\tiny$^{+2.71}_{-6.70}$} &  $7.26$\\
		A\_135\_100\_bin &  $10.60$\raisebox{0.5ex}{\tiny$^{+0.13}_{-0.13}$} &  $9.07$\raisebox{0.5ex}{\tiny$^{+0.13}_{-0.13}$} &  $-3.29$\raisebox{0.5ex}{\tiny$^{+0.43}_{-0.53}$} &  $-0.15$\raisebox{0.5ex}{\tiny$^{+0.54}_{-0.66}$} &  $0.19$\raisebox{0.5ex}{\tiny$^{+0.10}_{-0.11}$} & -- &  $11.13$\raisebox{0.5ex}{\tiny$^{+3.02}_{-6.50}$} &  $7.34$\\
		B\_chab100\_bin &  $10.56$\raisebox{0.5ex}{\tiny$^{+0.13}_{-0.13}$} &  $9.04$\raisebox{0.5ex}{\tiny$^{+0.13}_{-0.13}$} &  $-3.27$\raisebox{0.5ex}{\tiny$^{+0.44}_{-0.52}$} &  $-0.14$\raisebox{0.5ex}{\tiny$^{+0.54}_{-0.64}$} &  $0.19$\raisebox{0.5ex}{\tiny$^{+0.10}_{-0.11}$} &  $0.61$\raisebox{0.5ex}{\tiny$^{+0.12}_{-0.11}$} &  $10.93$\raisebox{0.5ex}{\tiny$^{+2.85}_{-6.48}$} &  $6.96$\\
\hline
	\end{tabular}
	}
        \vspace{-1em}
        \caption{} \label{tab:GL-z12-1}
    \end{subtable}
\end{table}
\clearpage
\begin{table}
\ContinuedFloat
    \begin{subtable}{\textwidth}
        \centering
    \renewcommand{\arraystretch}{1.2}
	\resizebox{\textwidth}{!}{%
	\begin{tabular}{|l|c|c|c|c|c|c|c|}
	\hline
	    \multicolumn{8}{|c|}{\textbf{GL-z9-1}} \\ 
	    \hline \hline
	    \multicolumn{8}{|c|}{Best Parameters} \\
		\hline 
		Model & $\log_{10}(\mathrm{M_{halo}}/\mathrm{M}_\odot$) & $\log_{10}(\mathrm{M_{\star}}/\mathrm{M}_\odot$) & $\log_{10}(f_\mathrm{BH}$) & $\delta_\mathrm{AGN}$ & $\tau_\mathrm{V}$ & $\alpha$ & $\chi^2$ \\
		\hline
		A\_chab100\_bin & 10.77 & 9.25 & --2.71 & --0.09 & 0.36 & -- & 0.43 \\
		A\_chab100\_sin & 10.60 & 9.08 & --2.20 & --0.55 & 0.30 & -- & 0.45 \\
		A\_135\_100\_bin & 10.82 & 9.30 & --2.74 & --0.14 & 0.35 & -- & 0.44 \\
		B\_chab100\_bin & 10.80 & 9.28 & --2.77 & --0.16 & 0.39 & 0.74 & 0.43 \\
		B\_135\_100\_bin & 10.82 & 9.30 & --2.83 & 0.01 & 0.36 & 0.80 & 0.44 \\
		B\_yggdrasil\_sin & 9.48 & 6.97 & --0.02 & --0.27 & 0.02 & 0.44 & 3.61 \\
		A\_yggdrasil\_sin & 9.53 & 7.06 & --0.14 & --0.25 & 0.01 & -- & 3.62 \\
		\hline
	\end{tabular}
	}
	\bigskip
	\resizebox{\textwidth}{!}{%
	\begin{tabular}{|l|c|c|c|c|c|c|c|c|}
	    \hline 
	    \multicolumn{9}{|c|}{Median Parameters} \\
		\hline
		Model & $\log_{10}(\mathrm{M_{halo}}/\mathrm{M}_\odot$) & $\log_{10}(\mathrm{M_{\star}}/\mathrm{M}_\odot$) & $\log_{10}(f_\mathrm{BH}$) & $\delta_\mathrm{AGN}$ & $\tau_\mathrm{V}$ & $\alpha$ & $\chi^2$ & $\chi^2(\theta_{\mathrm{med}})$ \\
		\hline
		A\_chab100\_bin &  $10.84$\raisebox{0.5ex}{\tiny$^{+0.25}_{-0.17}$} &  $9.31$\raisebox{0.5ex}{\tiny$^{+0.25}_{-0.17}$} &  $-2.99$\raisebox{0.5ex}{\tiny$^{+0.59}_{-0.67}$} &  $-0.10$\raisebox{0.5ex}{\tiny$^{+0.55}_{-0.63}$} &  $0.39$\raisebox{0.5ex}{\tiny$^{+0.15}_{-0.14}$} & -- &  $5.14$\raisebox{0.5ex}{\tiny$^{+3.38}_{-7.68}$} &  $1.58$\\
		A\_chab100\_sin &  $10.71$\raisebox{0.5ex}{\tiny$^{+0.21}_{-0.16}$} &  $9.19$\raisebox{0.5ex}{\tiny$^{+0.21}_{-0.16}$} &  $-2.88$\raisebox{0.5ex}{\tiny$^{+0.64}_{-0.68}$} &  $-0.14$\raisebox{0.5ex}{\tiny$^{+0.53}_{-0.64}$} &  $0.27$\raisebox{0.5ex}{\tiny$^{+0.12}_{-0.13}$} & -- &  $6.08$\raisebox{0.5ex}{\tiny$^{+3.30}_{-7.53}$} &  $3.82$\\
		A\_135\_100\_bin &  $10.88$\raisebox{0.5ex}{\tiny$^{+0.27}_{-0.18}$} &  $9.36$\raisebox{0.5ex}{\tiny$^{+0.27}_{-0.18}$} &  $-2.99$\raisebox{0.5ex}{\tiny$^{+0.60}_{-0.67}$} &  $-0.11$\raisebox{0.5ex}{\tiny$^{+0.54}_{-0.63}$} &  $0.38$\raisebox{0.5ex}{\tiny$^{+0.15}_{-0.14}$} & -- &  $5.19$\raisebox{0.5ex}{\tiny$^{+3.40}_{-7.68}$} &  $1.70$\\
		B\_chab100\_bin &  $10.86$\raisebox{0.5ex}{\tiny$^{+0.24}_{-0.17}$} &  $9.34$\raisebox{0.5ex}{\tiny$^{+0.24}_{-0.17}$} &  $-3.01$\raisebox{0.5ex}{\tiny$^{+0.57}_{-0.66}$} &  $-0.12$\raisebox{0.5ex}{\tiny$^{+0.53}_{-0.62}$} &  $0.37$\raisebox{0.5ex}{\tiny$^{+0.15}_{-0.15}$} &  $0.60$\raisebox{0.5ex}{\tiny$^{+0.12}_{-0.12}$} &  $5.04$\raisebox{0.5ex}{\tiny$^{+3.28}_{-7.46}$} &  $1.57$\\
\hline
	\end{tabular}
	}
        \vspace{-1em}
        \caption{} \label{tab:GL-z9-1}
    \end{subtable}
\end{table}
\clearpage
\begin{table}
\ContinuedFloat
    \begin{subtable}{\textwidth}
        \centering
    \renewcommand{\arraystretch}{1.2}
	\resizebox{\textwidth}{!}{%
	\begin{tabular}{|l|c|c|c|c|c|c|c|}
	\hline
	    \multicolumn{8}{|c|}{\textbf{GL-z9-2}} \\ 
	    \hline \hline
	    \multicolumn{8}{|c|}{Best Parameters} \\
		\hline 
		Model & $\log_{10}(\mathrm{M_{halo}}/\mathrm{M}_\odot$) & $\log_{10}(\mathrm{M_{\star}}/\mathrm{M}_\odot$) & $\log_{10}(f_\mathrm{BH}$) & $\delta_\mathrm{AGN}$ & $\tau_\mathrm{V}$ & $\alpha$ & $\chi^2$ \\
		\hline
		A\_chab100\_bin & 10.26 & 8.55 & --3.95 & --0.95 & 0.24 & -- & 4.68 \\
		A\_chab100\_sin & 10.25 & 8.52 & --3.91 & --0.97 & 0.18 & -- & 4.62 \\
		A\_135\_100\_bin & 10.29 & 8.62 & --3.97 & --0.99 & 0.24 & -- & 4.71 \\
		B\_chab100\_bin & 10.34 & 8.72 & --3.91 & --0.81 & 0.01 & 0.42 & 4.47 \\
		B\_135\_100\_bin & 10.35 & 8.74 & --3.74 & --0.86 & 0.00 & 0.42 & 4.49 \\
		B\_yggdrasil\_sin & 9.34 & 6.67 & --0.08 & --0.99 & 0.12 & 0.69 & 9.71 \\
		A\_yggdrasil\_sin & 9.36 & 6.70 & --0.11 & --0.99 & 0.13 & -- & 9.74 \\
		\hline
	\end{tabular}
	}
	\bigskip
	\resizebox{\textwidth}{!}{%
	\begin{tabular}{|l|c|c|c|c|c|c|c|c|}
	    \hline 
	    \multicolumn{9}{|c|}{Median Parameters} \\
		\hline
		Model & $\log_{10}(\mathrm{M_{halo}}/\mathrm{M}_\odot$) & $\log_{10}(\mathrm{M_{\star}}/\mathrm{M}_\odot$) & $\log_{10}(f_\mathrm{BH}$) & $\delta_\mathrm{AGN}$ & $\tau_\mathrm{V}$ & $\alpha$ & $\chi^2$ & $\chi^2(\theta_{\mathrm{med}})$ \\
		\hline
		A\_chab100\_bin &  $10.23$\raisebox{0.5ex}{\tiny$^{+0.07}_{-0.08}$} &  $8.49$\raisebox{0.5ex}{\tiny$^{+0.14}_{-0.16}$} &  $-3.25$\raisebox{0.5ex}{\tiny$^{+0.45}_{-0.53}$} &  $-0.12$\raisebox{0.5ex}{\tiny$^{+0.56}_{-0.65}$} &  $0.21$\raisebox{0.5ex}{\tiny$^{+0.12}_{-0.14}$} & -- &  $8.52$\raisebox{0.5ex}{\tiny$^{+2.63}_{-6.54}$} &  $5.37$\\
		A\_chab100\_sin &  $10.23$\raisebox{0.5ex}{\tiny$^{+0.06}_{-0.07}$} &  $8.49$\raisebox{0.5ex}{\tiny$^{+0.12}_{-0.15}$} &  $-3.28$\raisebox{0.5ex}{\tiny$^{+0.43}_{-0.53}$} &  $-0.14$\raisebox{0.5ex}{\tiny$^{+0.55}_{-0.66}$} &  $0.18$\raisebox{0.5ex}{\tiny$^{+0.10}_{-0.13}$} & -- &  $8.40$\raisebox{0.5ex}{\tiny$^{+2.64}_{-6.82}$} &  $5.45$\\
		A\_135\_100\_bin &  $10.26$\raisebox{0.5ex}{\tiny$^{+0.07}_{-0.08}$} &  $8.55$\raisebox{0.5ex}{\tiny$^{+0.14}_{-0.16}$} &  $-3.28$\raisebox{0.5ex}{\tiny$^{+0.43}_{-0.52}$} &  $-0.12$\raisebox{0.5ex}{\tiny$^{+0.56}_{-0.65}$} &  $0.21$\raisebox{0.5ex}{\tiny$^{+0.12}_{-0.14}$} & -- &  $8.55$\raisebox{0.5ex}{\tiny$^{+2.63}_{-6.55}$} &  $5.37$\\
		B\_chab100\_bin &  $10.27$\raisebox{0.5ex}{\tiny$^{+0.07}_{-0.07}$} &  $8.57$\raisebox{0.5ex}{\tiny$^{+0.14}_{-0.15}$} &  $-3.30$\raisebox{0.5ex}{\tiny$^{+0.42}_{-0.52}$} &  $-0.17$\raisebox{0.5ex}{\tiny$^{+0.53}_{-0.65}$} &  $0.18$\raisebox{0.5ex}{\tiny$^{+0.10}_{-0.14}$} &  $0.64$\raisebox{0.5ex}{\tiny$^{+0.11}_{-0.09}$} &  $8.74$\raisebox{0.5ex}{\tiny$^{+2.87}_{-7.12}$} &  $6.02$\\
\hline
	\end{tabular}
	}
    \vspace{-1em}
    \caption{} \label{tab:GL-z9-2}
    \end{subtable}
	\parbox[t]{\textwidth}{\vspace{1em} \textit{Notes:} The upper and lower uncertainties are computed from the 84th and 16th percentiles of the distributions respectively. $\chi^2(\theta_{\mathrm{med}})$ is $\chi^2$ for the spectra computed from the median parameters in columns 3-8 rather than the median of the $\chi^2$ distribution.}
\end{table}

\clearpage


\setcounter{table}{0}
\renewcommand{\thetable}{B\arabic{table}}

\begin{table*}
	\caption{Magnitudes for the best-fit model parameters and their differences with observed magnitudes} \label{tab:best_fits_mags}
	\label{tab:magnitudes_delta}
	\renewcommand{\arraystretch}{1.2}
	\resizebox{\textwidth}{!}{%
		\begin{tabular}{|l|c|c|c|c|c|c|c|c|}
			\hline \hline
			\multirow{2}{*}{Galaxy} & \multirow{2}{*}{Model} & F150W & F200W & F277W & F356W & F410M & F444W & $M_\mathrm{UV}$ \\ 
			 &  & $\Delta$F150W ($\sigma_{\mathrm{obs}}$)& $\Delta$F200W ($\sigma_{\mathrm{obs}}$)& $\Delta$F277W ($\sigma_{\mathrm{obs}}$)& $\Delta$F356W ($\sigma_{\mathrm{obs}}$)& $\Delta$F410M ($\sigma_{\mathrm{obs}}$)& $\Delta$F444W ($\sigma_{\mathrm{obs}}$) & $\Delta M_\mathrm{UV}$ ($\sigma_{\mathrm{obs}}$) \\ 
			\hline
			\multirow{2}{*}{10234} & \multirow{2}{*}{B\_chab100\_bin} &  28.22 &  \textit{27.49} &  27.75 &  \textit\textit{27.98} & \multirow{2}{*}{--} &  \textit{28.04} &  \multirow{2}{*}{--} \\
			 &  &  --0.019 (0.07) &  \textit{0.364} (0.05) &  --0.047 (0.05) &  \textit{--0.241} (0.05) &  &  \textit{--0.064} (0.05) &   \\
			 \arrayrulecolor{lightgray}\hline
			 \arrayrulecolor{black}
			\multirow{2}{*}{1514} & \multirow{2}{*}{A\_chab100\_sin} &  28.01 &  27.68 &  \textit{27.41} &  \textit{27.14} & \multirow{2}{*}{--} &  26.35 & \multirow{2}{*}{--}  \\
			 &  &  --0.004 (0.05) &  --0.037 (0.05) &  \textit{0.101} (0.05) &  \textit{--0.064} (0.05) &  &  0.009 (0.05) &   \\
			\arrayrulecolor{lightgray}\hline
			 \arrayrulecolor{black}
			\multirow{2}{*}{1696} & \multirow{2}{*}{A\_chab100\_sin} &  26.71 &  26.76 &  \textit{26.85} &  26.84 &   &  26.23 &    \\
			 &  &  0.044 (0.05) &  0.021 (0.05) &  \textit{--0.099} (0.05) &  --0.012 (0.05) &   &  0.033 (0.05) &   \\
			\arrayrulecolor{lightgray}\hline
			 \arrayrulecolor{black}
			\multirow{2}{*}{2462} & \multirow{2}{*}{A\_chab100\_sin} & \textit{26.43} &  26.24 &  \textit{26.09} &  25.84 & \multirow{2}{*}{--} &  \textit{25.06} & \multirow{2}{*}{--}  \\
			 &  &  \textit{0.070} (0.05) &  --0.025 (0.05) &  \textit{--0.106} (0.05) &  --0.027 (0.05) &   &  \textit{0.080} (0.05) &   \\
			\arrayrulecolor{lightgray}\hline
			 \arrayrulecolor{black}
			\multirow{2}{*}{2779} & \multirow{2}{*}{B\_chab100\_bin} & 27.78 &  \textit{27.87} &  \textit{28.03} &  28.07 & \multirow{2}{*}{--} &  27.62 & \multirow{2}{*}{--}  \\
			 &  &  --0.017 (0.05) &  \textit{0.084} (0.05) &  \textit{--0.095} (0.05) &  0.003 (0.05) &   &  0.035 (0.05) &   \\
			\arrayrulecolor{lightgray}\hline
			 \arrayrulecolor{black}
			\multirow{2}{*}{6115} & \multirow{2}{*}{A\_yggdrasil\_sin} &  28.17 &  27.75 &  \textit{27.89} &  27.97 & \multirow{2}{*}{--} &  28.08 & \multirow{2}{*}{--} \\
			 &  &  0.025 (0.06) &  0.004 (0.05) &  \textit{--0.064} (0.05) &  0.034 (0.05) &   &  0.006 (0.05) &   \\
			\arrayrulecolor{lightgray}\hline
			 \arrayrulecolor{black}
			\multirow{2}{*}{6878} & \multirow{2}{*}{A\_chab100\_sin} &   27.02 &  26.99 &  26.99 &  26.92 & \multirow{2}{*}{--} &  26.32 & \multirow{2}{*}{--}   \\
			 &  &  --0.011 (0.05) &  0.027 (0.05) &  --0.027 (0.05) &  0.010 (0.05) &   &  0.002 (0.05) &   \\
			\arrayrulecolor{lightgray}\hline
			 \arrayrulecolor{black}
			\multirow{2}{*}{CR2-z12-1} & \multirow{2}{*}{B\_chab100\_bin} &  28.76 &  \textit{27.76} &  27.94 &  \textit{28.09} &  28.16 &  28.08 &  --20.00  \\
			 &  &  --0.345 (0.4) &  \textit{0.162} (0.1) &  --0.057 (0.2) &  \textit{--0.107} (0.1) &  0.062 (0.1) &  0.081 (0.2) &  --0.096 (0.1) \\
			\arrayrulecolor{lightgray}\hline
			 \arrayrulecolor{black}
			\multirow{2}{*}{CR2-z16-1} & \multirow{2}{*}{A\_chab100\_sin} & \multirow{2}{*}{--} &  27.90 &  26.35 &  26.41 &  \textit{26.45} &  \textit{26.48} &  --21.88 \\
			 &  &   &  --0.001 (0.1) &  --0.048 (0.1) &  0.008 (0.1) &  \textit{0.151} (0.1) &  \textit{--0.124} (0.1) &  0.024 (0.1) \\
			\arrayrulecolor{lightgray}\hline
			 \arrayrulecolor{black}
			\multirow{2}{*}{GL-z12-1} & \multirow{2}{*}{B\_yggdrasil\_sin} &  29.01 &  \textit{26.81} &  \textit{27.00} &  27.12 & \multirow{2}{*}{--} &  27.16 &  --21.02 \\
			 &  &  --0.091 (0.1) &  \textit{0.111} (0.1) &  \textit{0.103} (0.1) &  --0.083 (0.1) &   &  --0.043 (0.1) &  --0.025 (0.1) \\
			\arrayrulecolor{lightgray}\hline
			 \arrayrulecolor{black}
			\multirow{2}{*}{GL-z9-1} & \multirow{2}{*}{B\_chab100\_bin} &  26.93 &  26.62 &  26.61 &  26.59 & \multirow{2}{*}{--} &  26.30 &  --20.95 \\
			 &  &  0.025 (0.1) &  0.023 (0.1) &  0.009 (0.1) &  --0.011 (0.1) &   &  0.004 (0.1) &  --0.054 (0.1) \\
			\arrayrulecolor{lightgray}\hline
			 \arrayrulecolor{black}
			\multirow{2}{*}{GL-z9-2} & \multirow{2}{*}{B\_chab100\_bin} &  28.11 &  \textit{27.97} &  28.14 &  28.25 & \multirow{2}{*}{--} &  27.95 &  --19.68 \\
			 &  & --0.086 (0.1) &  \textit{0.175} (0.1) &  --0.062 (0.1) &  --0.046 (0.1) &   &  0.050 (0.2) &  0.018 (0.1) \\
			\hline
		\end{tabular}
	}
	\parbox[t]{\textwidth}{\vspace{1em} \textit{Notes:} First row: Magnitudes for best-fit. Second-row: Difference between model magnitudes and observed magnitudes and the observed standard error. Observed values are from H22 and A22. Italicized numbers indicate where our filter magnitudes lie outside the observed standard error.}
\end{table*}

\setcounter{figure}{0}
\renewcommand{\thefigure}{\thesection\arabic{figure}}    

 \begin{figure*}
    \centering
     \caption{Top: Posterior distributions of the parameters $\theta$ determined from the MCMC chains for the best models of each object, 10234 (\ref{fig:corner_10234}), 1514 (\ref{fig:corner_1514}), 1696 (\ref{fig:corner_1696}), 2462 (\ref{fig:corner_2462}), 2779 (\ref{fig:corner_2779}), 6115 (\ref{fig:corner_6115}), 6878 (\ref{fig:corner_6878}), CR2-z12-1 (\ref{fig:corner_CR2-z12-1}), CR2-z16-1 (\ref{fig:corner_CR2-z16-1}), GL-z12-1 (\ref{fig:corner_GL-z12-1}), GL-z9-1 (\ref{fig:corner_GL-z9-1}), GL-z9-2 (\ref{fig:corner_GL-z9-2}). The plots at the top of each column show the 1d histograms for each parameter $\delta_\mathrm{AGN}$, $f_\mathrm{BH}$, $M_\mathrm{halo}$, $\tau_\mathrm{V}$ and $\alpha$ for models using halo growth model B. The dashed lines on each histogram mark the median value and the 16th and 84th percentiles. The titles on top of each histogram show the median value and standard error as determined from the 16th and 84th percentiles. The contour plots below the histograms show the 2-dimensional distributions for each pair of parameters and the colormap represents the probability density. The contours outline the 0.5$\sigma$, 1$\sigma$, 1.5$\sigma$, and 2$\sigma$ confidence levels. Bottom row: Posterior distributions of the magnitudes in each of the JWST/NIRCam filters computed from the SEDs of the MCMC samples. The colored bands show the range of the observed  photometry from A22 and H22 within the standard error, $m_\mathrm{obs} \pm \sigma_\mathrm{obs}$. The dotted lines show the AB magnitudes $m_\mathrm{AB}(\theta_\mathrm{best})$ for the best parameters which minimize $\chi^2$ (see Table \ref{tab:best_fits} and \ref{tab:magnitudes_delta}).}
     \centering
    \begin{subfigure}{\textwidth}
        \centering
        \includegraphics[width=0.825\textwidth]{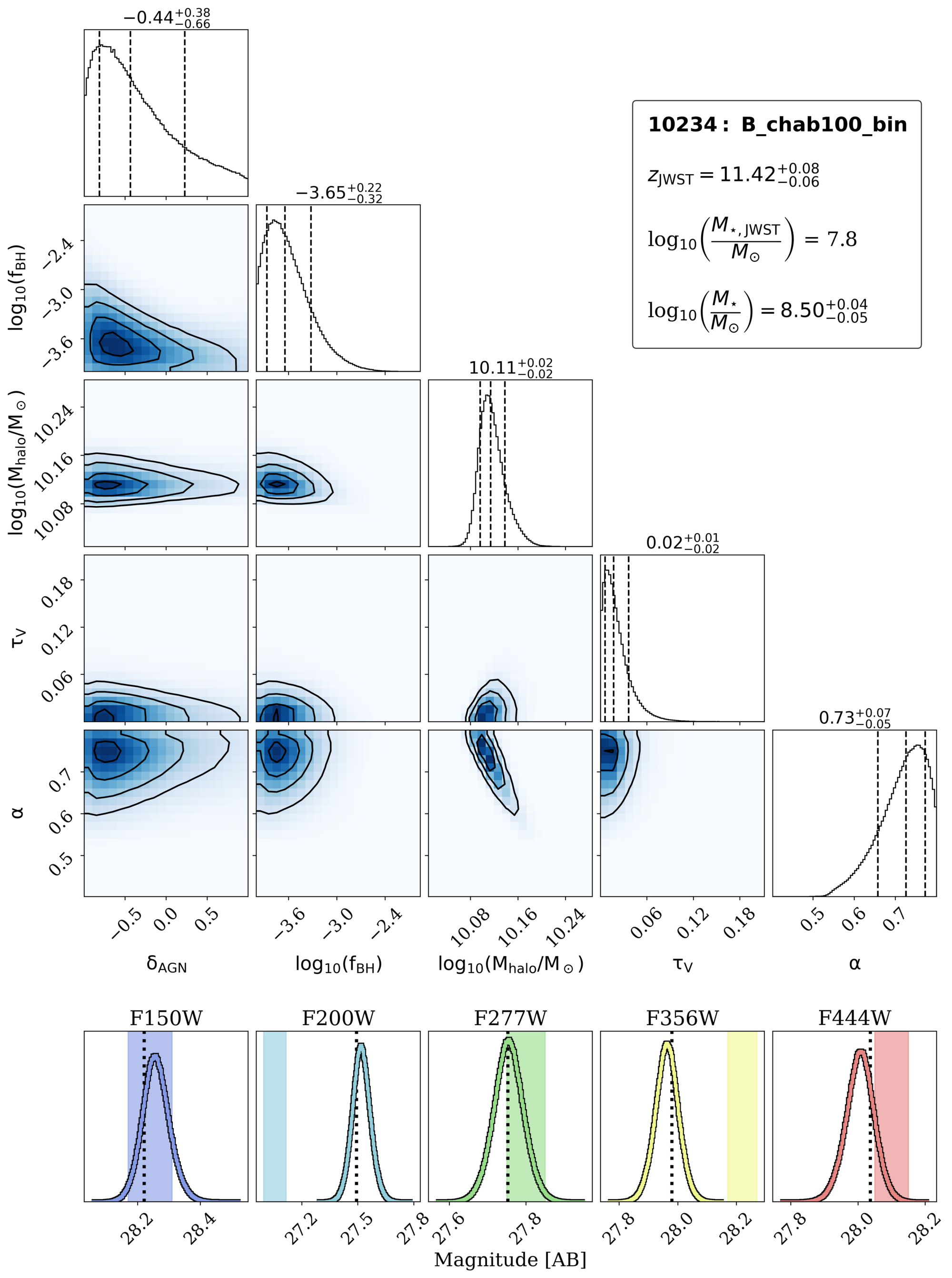}
        \caption{}
        \label{fig:corner_10234}
    \end{subfigure}
 \end{figure*}
 \clearpage
 \begin{figure*}\ContinuedFloat  
    \centering
    \begin{subfigure}{\textwidth}
        \centering
        \includegraphics[width=\textwidth]{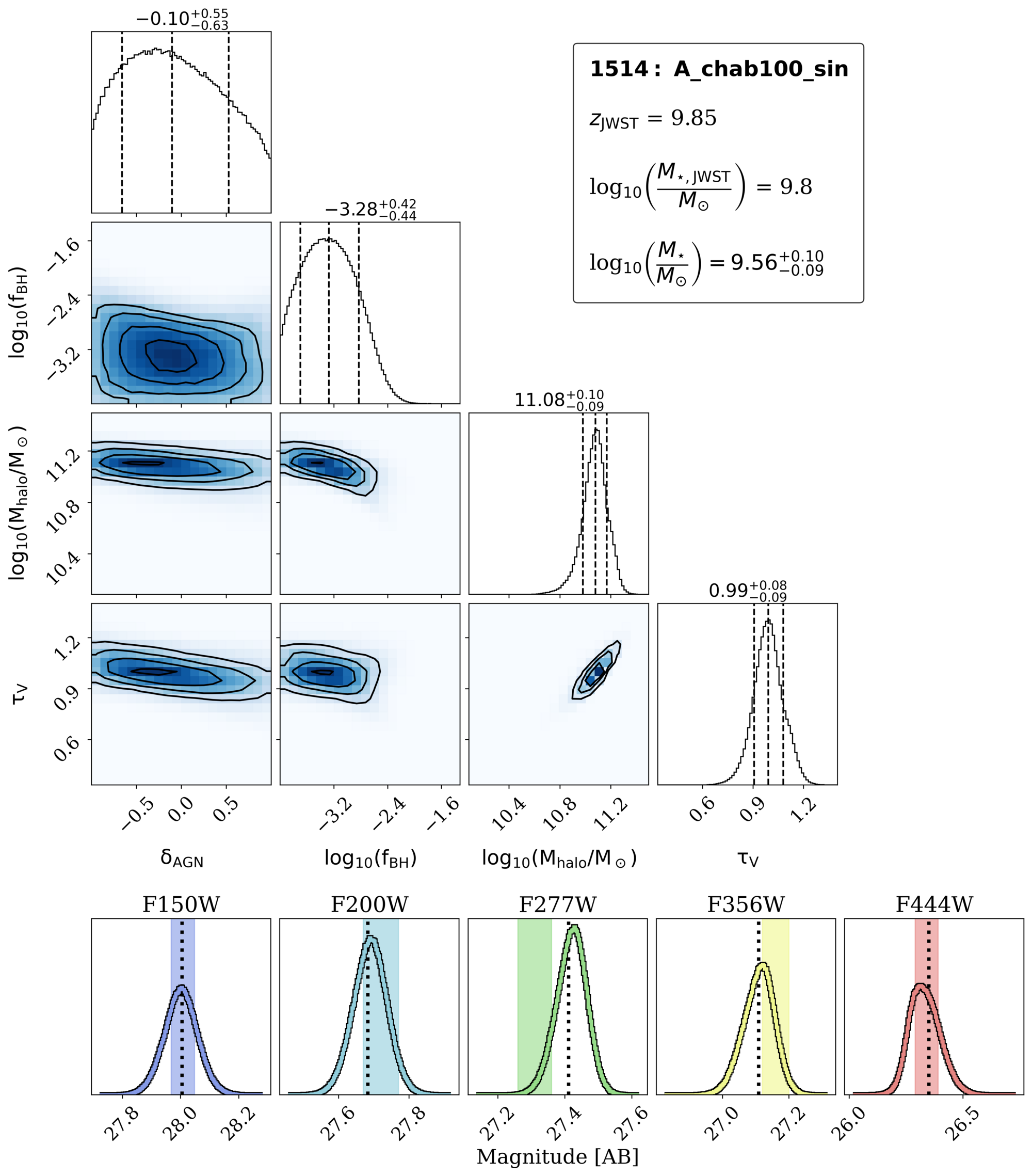}
        \caption{}
        \label{fig:corner_1514}
    \end{subfigure}
 \end{figure*}
 \clearpage
 \begin{figure*}\ContinuedFloat 
    \centering
    \begin{subfigure}{\textwidth}
        \includegraphics[width=\textwidth]{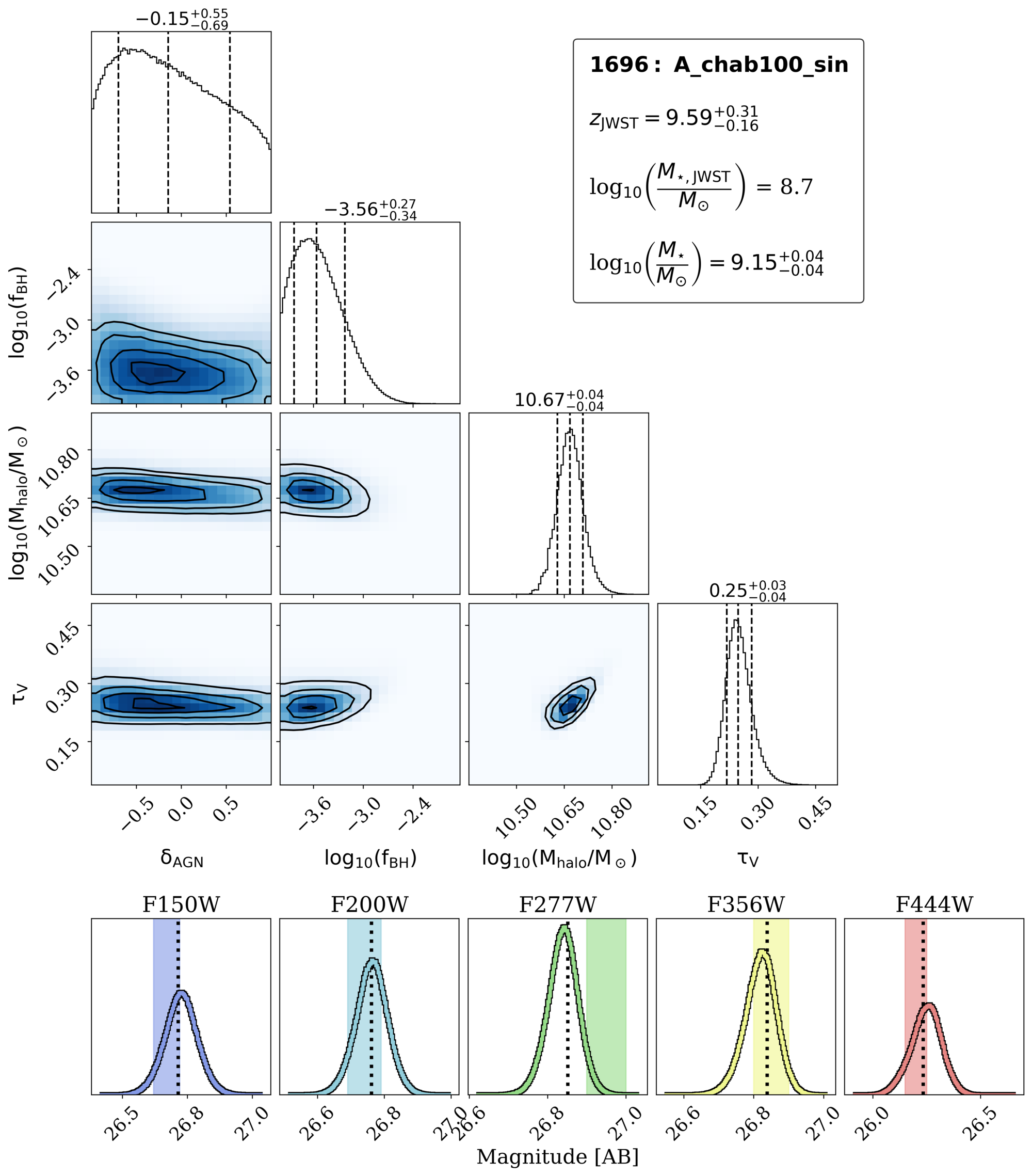}
              \caption{}
        \label{fig:corner_1696}
    \end{subfigure}
 \end{figure*}
 \clearpage
 \begin{figure*}\ContinuedFloat 
    \centering
    \begin{subfigure}{\textwidth}
        \includegraphics[width=\textwidth]{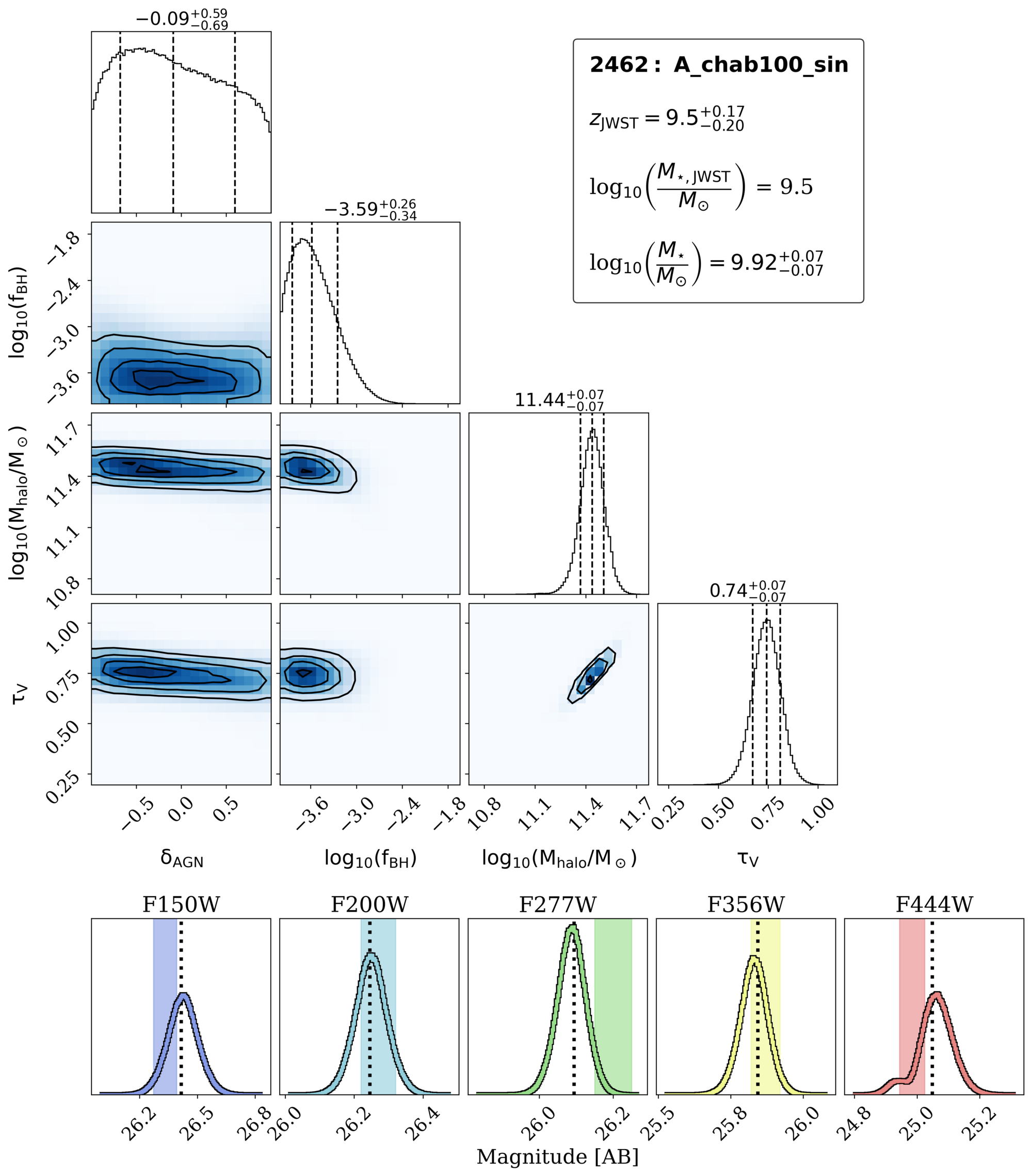}
          \caption{}
        \label{fig:corner_2462}
    \end{subfigure}
 \end{figure*}
 \clearpage
 \begin{figure*}\ContinuedFloat 
    \centering
    \begin{subfigure}{\textwidth}
        \includegraphics[width=0.98\textwidth]{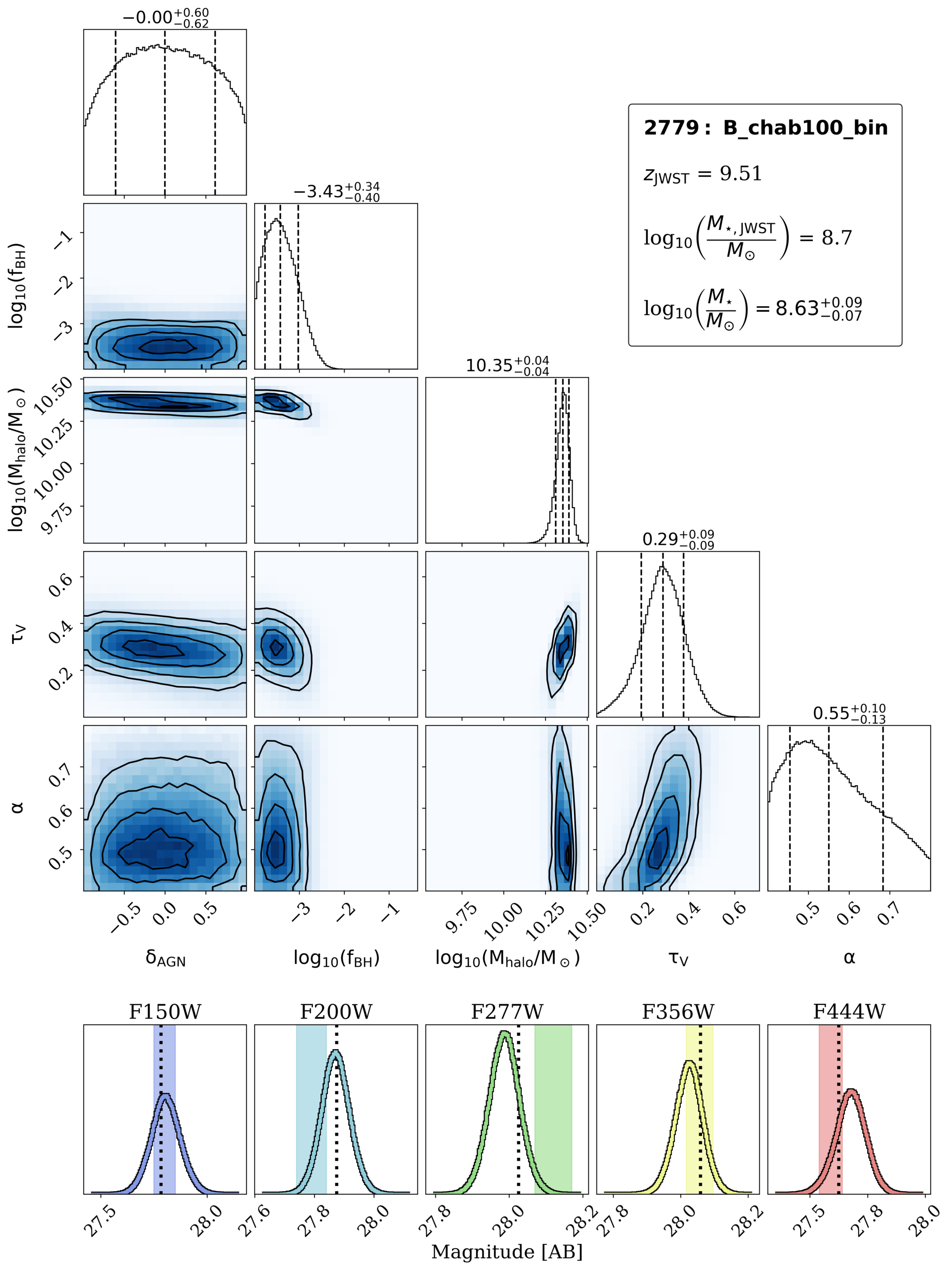}
          \caption{}
        \label{fig:corner_2779}
    \end{subfigure}
 \end{figure*}
 \clearpage
 \begin{figure*}\ContinuedFloat 
    \centering
    \begin{subfigure}{\textwidth}
        \includegraphics[width=\textwidth]{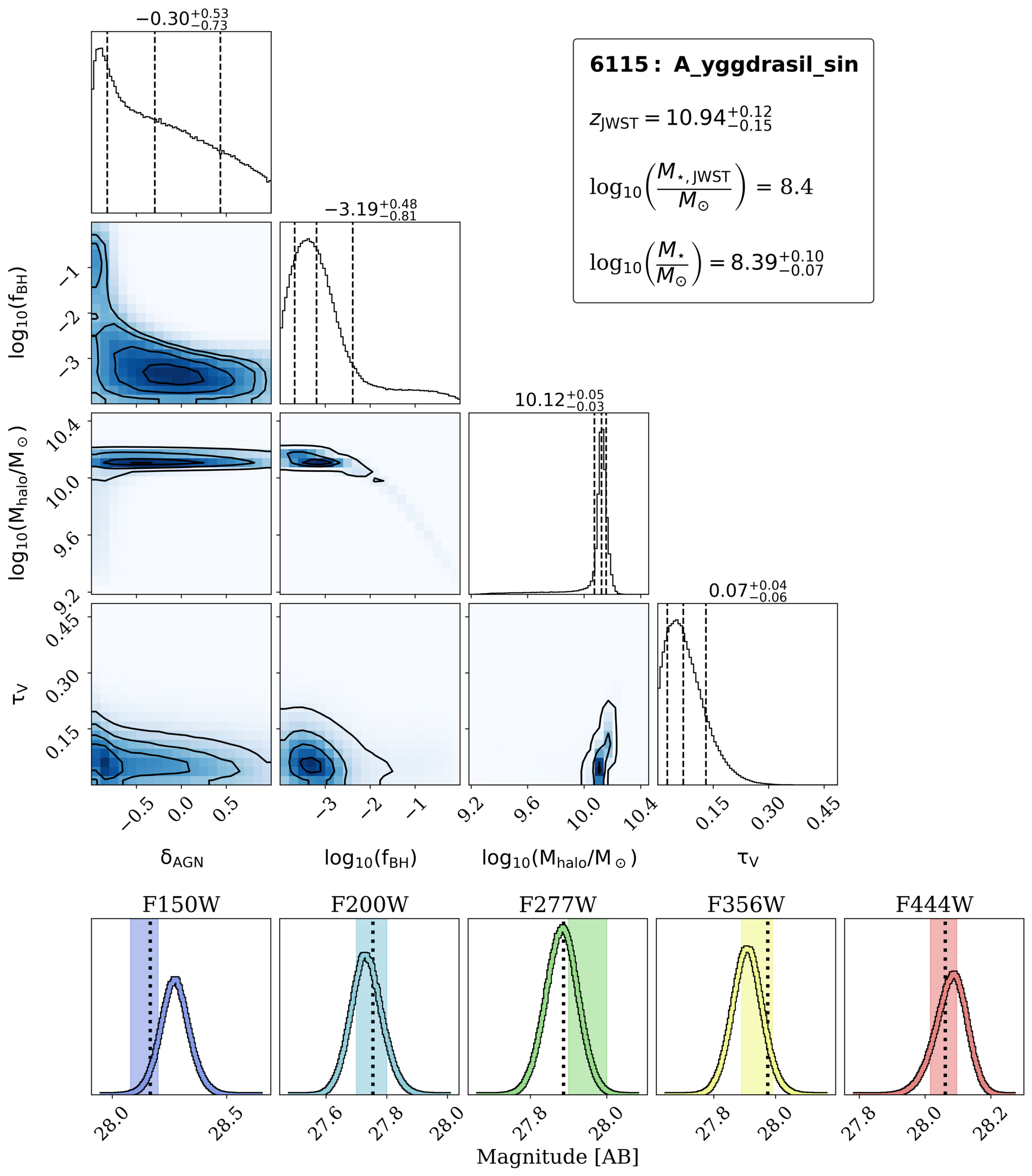}
          \caption{}
        \label{fig:corner_6115}
    \end{subfigure}
 \end{figure*}
 \clearpage
 \begin{figure*}\ContinuedFloat 
    \centering
    \begin{subfigure}{\textwidth}
        \includegraphics[width=\textwidth]{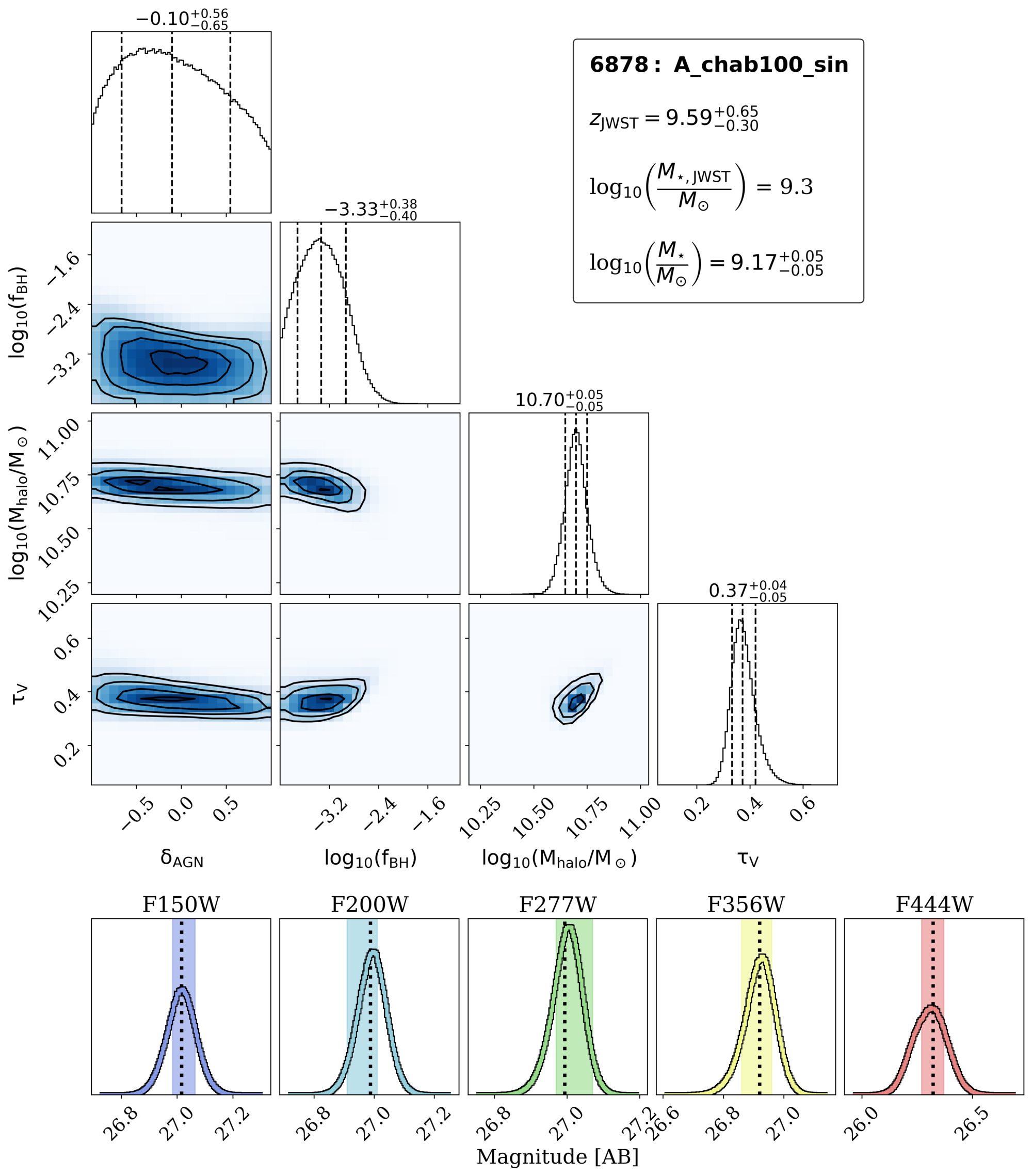}
          \caption{}
        \label{fig:corner_6878}
    \end{subfigure}
 \end{figure*}
 \clearpage
 \begin{figure*}\ContinuedFloat 
    \centering
    \begin{subfigure}{\textwidth}
        \includegraphics[width=\textwidth]{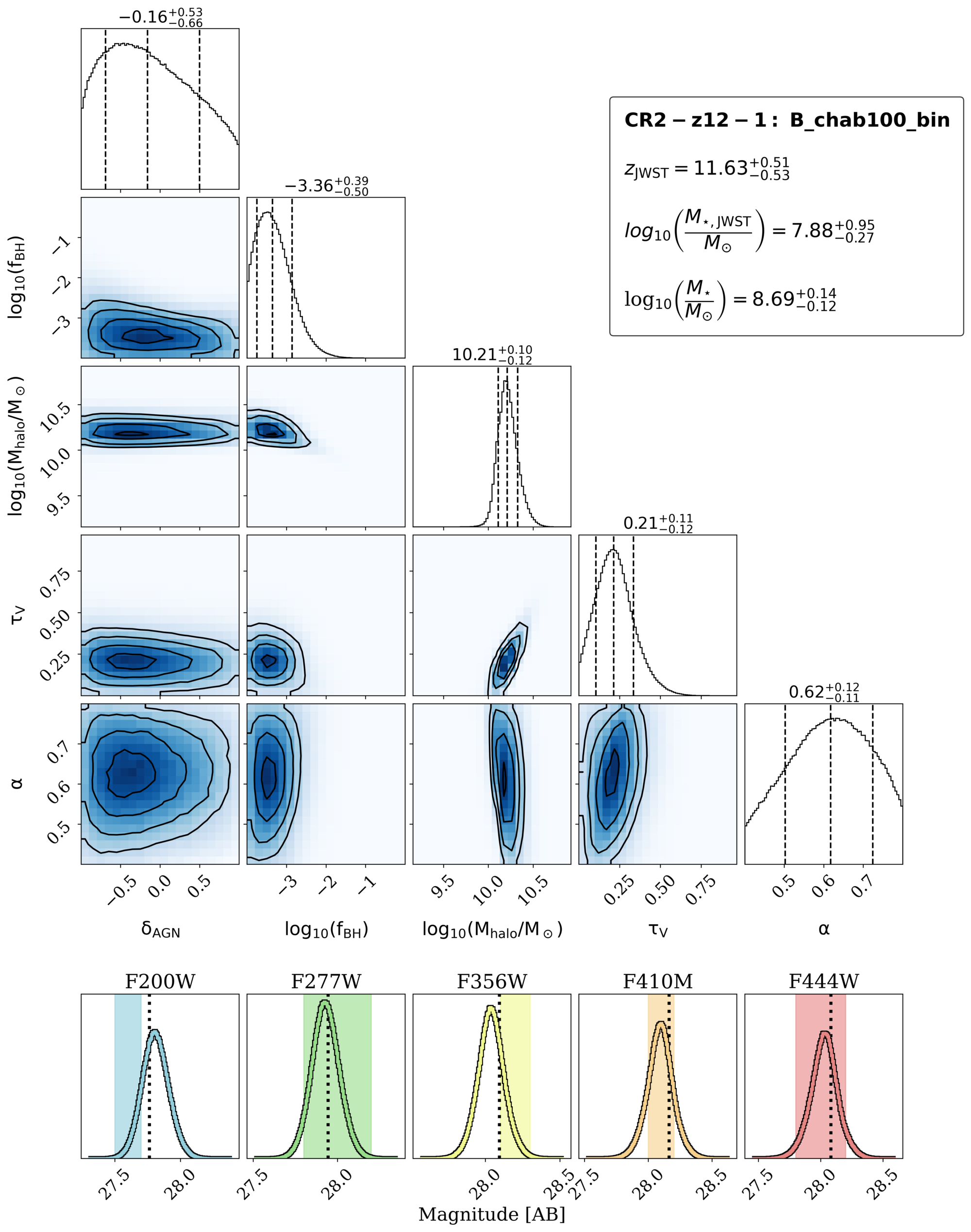}
          \caption{}
        \label{fig:corner_CR2-z12-1}
    \end{subfigure}
 \end{figure*}
 \clearpage
 \begin{figure*}\ContinuedFloat 
    \centering
    \begin{subfigure}{\textwidth}
        \includegraphics[width=\textwidth]{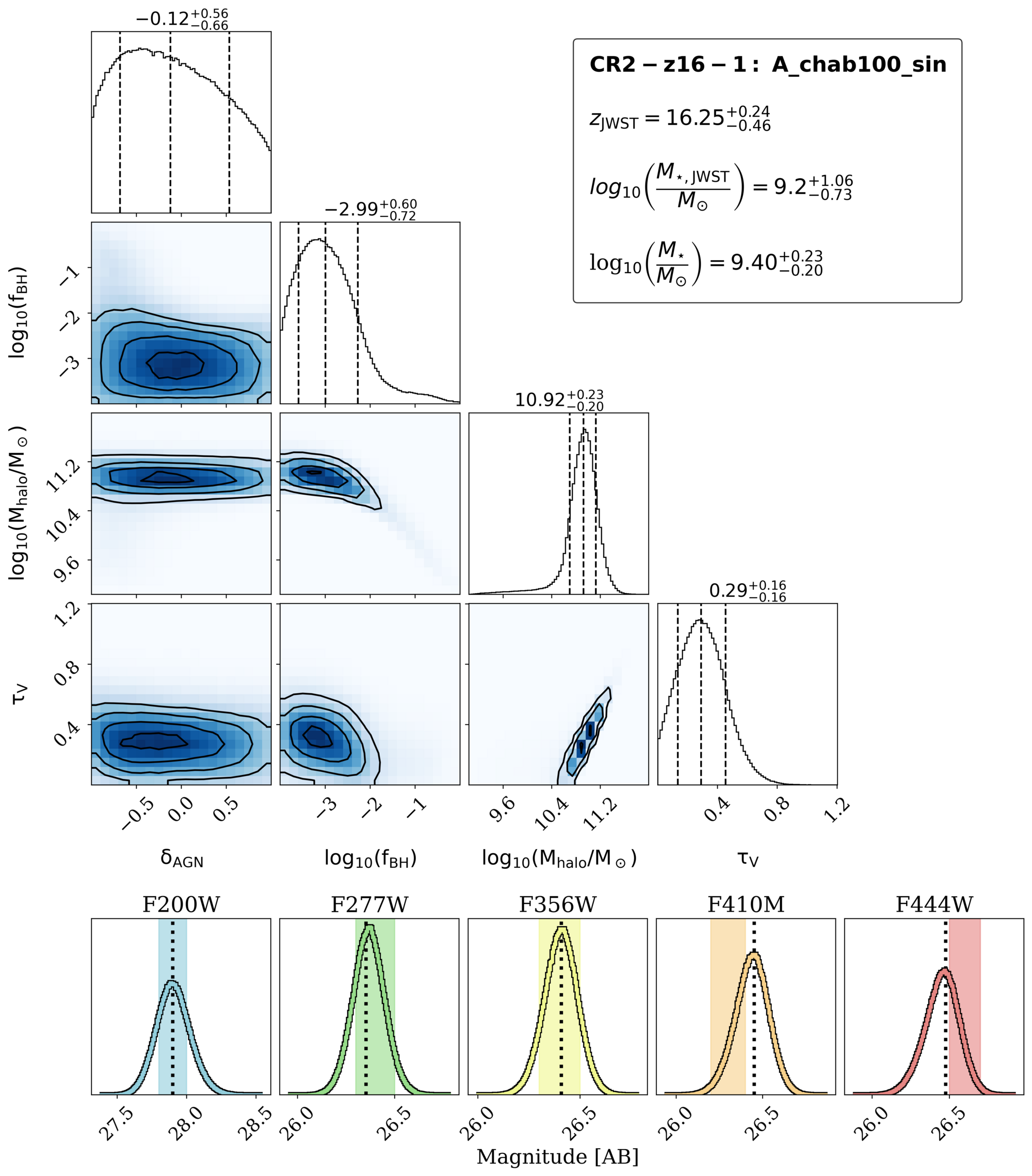}
          \caption{}
        \label{fig:corner_CR2-z16-1}
    \end{subfigure}
 \end{figure*}
 \clearpage
 \begin{figure*}\ContinuedFloat 
    \centering
    \begin{subfigure}{\textwidth}
        \includegraphics[width=\textwidth]{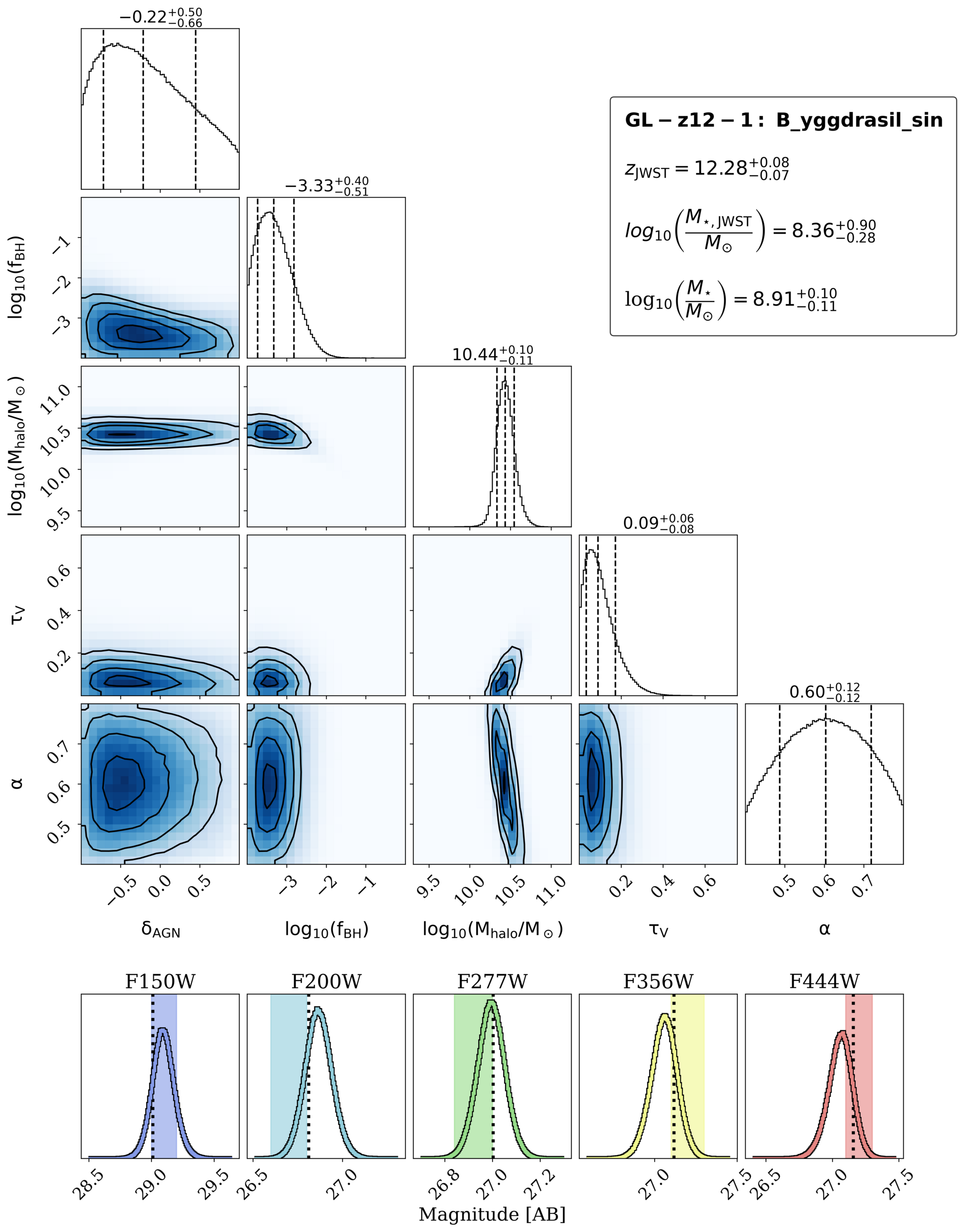}
          \caption{}
        \label{fig:corner_GL-z12-1}
    \end{subfigure}
 \end{figure*}
 \clearpage
 \begin{figure*}\ContinuedFloat 
    \centering
    \begin{subfigure}{\textwidth}
        \includegraphics[width=\textwidth]{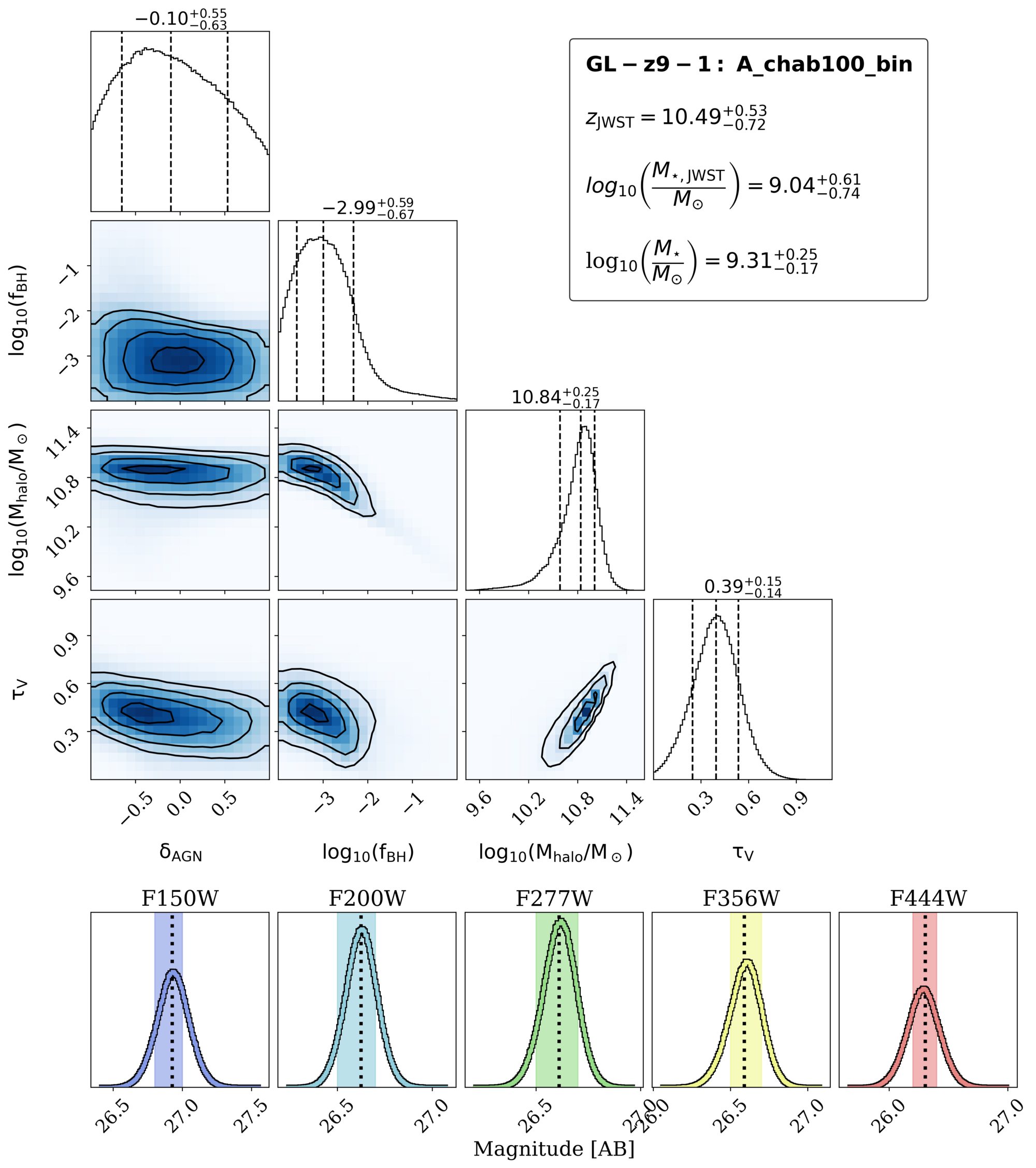}
          \caption{}
        \label{fig:corner_GL-z9-1}
    \end{subfigure}
 \end{figure*}
 \clearpage
 \begin{figure*}\ContinuedFloat 
    \centering
    \begin{subfigure}{\textwidth}
        \includegraphics[width=\textwidth]{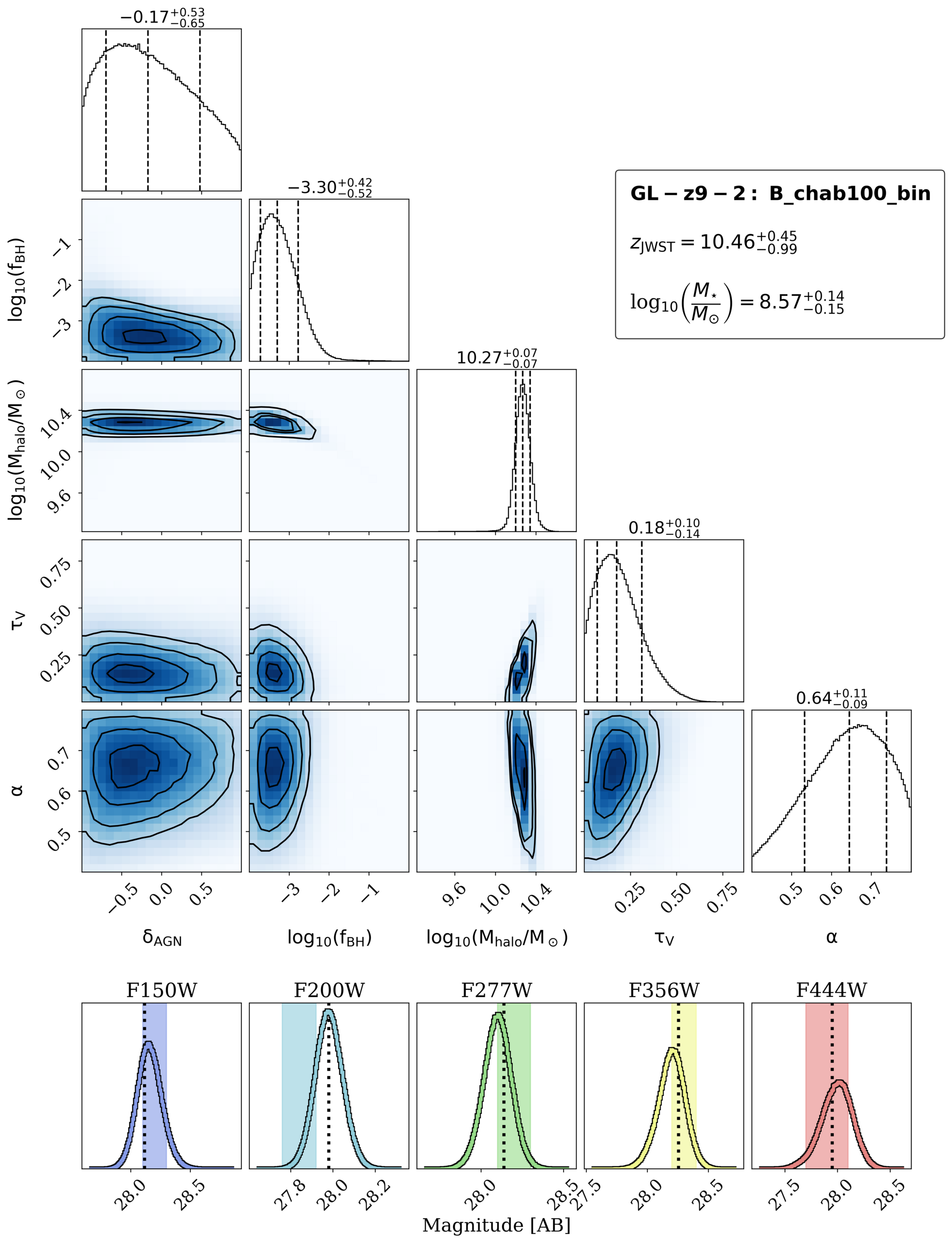}
          \caption{}
        \label{fig:corner_GL-z9-2}
    \end{subfigure}
    \label{lastpage}
\end{figure*}




\end{document}